\documentclass[11pt,a4paper]{article}
\pdfoutput=1
\usepackage[english]{babel}
\usepackage[utf8]{inputenc}
\usepackage{jheppub}
\usepackage{amsmath}
\usepackage{nccmath}
\usepackage{mathtools}
\usepackage{amssymb}
\usepackage{amsfonts}
\usepackage{nameref}
\usepackage{latexsym}
\numberwithin{equation}{section}
\allowdisplaybreaks

\def\be{\begin{equation}}
\def\ee{\end{equation}}
\def\bea{\begin{eqnarray}}
\def\eea{\end{eqnarray}}
\def\bequ{\begin{equation}}
\def\eequ{\end{equation}}

\def\del{\partial}

\title{A String Dual for Partially Topological Chern-Simons-Matter Theories}
\author[a]{Ofer Aharony,}
\author[a]{Andrey Feldman}
\author[b]{and Masazumi Honda}
\affiliation[a]{Department of Particle Physics and Astrophysics, \\
	Weizmann Institute of Science, Rehovot 7610001, Israel}
\affiliation[b]{Department of Applied Mathematics and Theoretical Physics, \\ Centre for Mathematical Sciences, Wilberforce Road, Cambridge, CB3 0WA, UK}
\emailAdd{ofer.aharony@weizmann.ac.il}
\emailAdd{andrey.feldman@weizmann.ac.il}
\emailAdd{mh974@damtp.cam.ac.uk}
\abstract{%In this paper 
We consider a string dual of a partially topological $\mathrm{U} (N)$ Chern-Simons-matter (PTCSM) theory recently introduced by Aganagic, Costello, McNamara and Vafa.
In this theory, 
fundamental matter fields are coupled to the Chern-Simons theory in a way that depends only on a transverse holomorphic structure on a manifold; they are not fully dynamical, but the theory is also not fully topological.
One description of this theory arises 
from topological strings on the deformed conifold $T^* S^3$ with $N$ Lagrangian 3-branes and additional coisotropic `flavor' 5-branes. 
Applying the idea of the Gopakumar-Vafa duality to this setup, we suggest that this has a dual description as a topological string on the resolved conifold ${\cal O} \left( - 1 \right) \oplus {\cal O} \left( - 1 \right) \rightarrow \mathbb{CP}^1$, in the presence of coisotropic 5-branes. We test this duality by computing the annulus amplitude on the deformed conifold and the disc amplitude on the resolved conifold via equivariant localization, and we find an agreement between the two. 
We find a small discrepancy between the topological string results 
and the large $N$ limit of the partition function of the PTCSM theory arising from the deformed conifold,
computed via field theory localization by a method proposed by Aganagic et al. 
We discuss
possible origins of the mismatch.}

\begin{document}

\keywords{Chern-Simons-matter theory, topological string theory, branes, holographic duality}

\maketitle

\flushbottom

\selectlanguage{english}

\section{Introduction} \label{sec:intro}
Holographic duality is one of the key avenues to understanding string theory and quantum gravity.  
The idea that a gauge theory may be dual to a closed string theory was first suggested in \cite{tHooft}. 
Since then, such a gauge/gravity duality was constructed for many pairs of physical systems. Among the most prominent examples are the duality between certain zero-dimensional gauge theories and two-dimensional pure topological gravity \cite{KontsevichAiry, WittenIntersection}, the equivalence between 2d QCD and 2d string theory \cite{GrossTaylor}, the description of M-theory in terms of the $N=\infty$ limit of $\mathrm{U} (N)$ matrix quantum mechanics \cite{BanksFischlerShenkerSusskind}, and the AdS/CFT correspondence, relating Type IIB superstring theory in an $AdS_5 \times S^5$ background to the 4d ${\cal N}=4$ super Yang-Mills (SYM) theory \cite{MaldacenaHol, WittenHol, GKPHol}. The latter duality has the richest physical content, but is too complicated to be proven by currently available techniques, due to non-trivial local dynamics on the gauge theory side of the correspondence.\footnote{However, some progress in this direction has been achieved in \cite{BerkovitsOoguriVafa, BerkovitsVafa}.}

A non-trivial but still exactly soluble example of gauge/gravity duality, known as the Gopakumar-Vafa (GV) duality, was given in \cite{GV}, where it was shown at the level of partition functions that a $\mathrm{U} (N)$ Chern-Simons (CS) gauge theory on $S^3$ is equivalent to the topological string A-model on the total space of the ${\cal O}(-1) \oplus {\cal O}(-1) \rightarrow \mathbb{CP}^1$ bundle, which can also be conveniently thought of as the resolved $S^2 \times S^3$ conifold. Here `resolved' means that the $S^2 = \mathbb{CP}^1$ does not shrink to zero size at the tip of the conifold, while the $S^3$ does. The duality was later derived in \cite{OV}, using the GLSM techniques developed in \cite{2DPhases}.
However, it is fair to say that
this setup is far from typical examples of gauge/gravity duality,
in the sense that both sides are topological and do not have propagating degrees of freedom.

In this paper we propose a gauge/gravity duality between two partially topological theories,
which can be regarded as something between the Gopakumar-Vafa duality and a typical gauge/gravity duality. 
%Our goal is to add matter fields to this duality. 
This is achieved by adding matter fields to the Gopakumar-Vafa duality.
The pure CS theory is equivalent \cite{WittenChernString} to topological strings on the deformed conifold with $N$ 3-branes. 
%(the word `deformed' means that the $S^2$ shrinks to zero size at the tip of the conifold, while the $S^3$ does not). 
Adding 5-branes to this, which can be treated as probes at leading order in the string coupling, gives rise to matter fields from 3-5 strings. The duality then maps this configuration to 5-brane probes on the resolved conifold, which add to this closed topological string theory an open string sector. These are `flavor' branes capturing the dynamics of the matter fields, as usual when coupling fundamental matter to dualities between gauge theories and closed string theories.

Ideally, we would like to do this for dynamical matter fields, in order to obtain a dual for the standard CS-matter (CSM) theories that have been extensively studied in the past few years. These theories have a dual description at large $N$ by a theory of higher spin gravity 
%\cite{FradkinVasiliev,KP}, 
(see \cite{GiombiReview} for a review),
so this would give a string theory completion for this theory. However, we do not yet know how to do this. What we can do, following \cite{AganagicCostelloMcNamaraVafa}, is to add partially topological matter fields to the CS theory, which arise from adding specific coisotropic 5-branes to the deformed conifold side of the duality. In this paper we suggest a dual for the resulting PTCSM theory, and test it by comparing the partition functions on both sides.

The easiest way to understand the GV duality is to represent a $\mathrm{U} (N)$ CS theory on $S^3$ as the topological A-model on $T^{*} S^3$ with $N$ Lagrangian 3-branes wrapping the minimal $S^3$. As shown in \cite{WittenChernString}, the full string field theory action of the topological A-model
\begin{equation}
S = \frac{1}{g_s} \int \mathrm{Tr} \left[ \frac{1}{2} \Psi \star Q_B \Psi + \frac{1}{3} \Psi \star \Psi \star \Psi \right]
\end{equation}
in this setup is reduced to the CS action for a $\mathrm{U} (N)$ gauge field $A$ living on the 3-branes
\begin{equation} \label{CSaction}
S_{CS} = \mathrm{i} \frac{k}{2 \pi} \int_{S^3} \mathrm{Tr} \left[ \frac{1}{2} A \wedge \textrm{d}A + \frac{1}{3} A \wedge A \wedge A \right],
\end{equation}
where $k$ is an integer and is finitely renormalized as $k \rightarrow k+N$ for the $\mathrm{SU} (N)$ factor. For more general topological string backgrounds than a stack of Lagrangian 3-branes wrapping the zero section of $T^*M$, the open string action is corrected by holomorphic instantons
\begin{equation}
S = S_{CS} + \sum_{\beta} \mathrm{e}^{-\int_{\beta} \omega} \mathrm{Tr} \left( U_{{\cal K} \left( \beta \right)} \right),
\end{equation}
which play a crucial role in the GV duality. Here $\omega$ is the complexified K\"{a}hler form, which measures the symplectic area of the worldsheet instanton $\beta$ in units of the string length, ${\cal K} \left( \beta \right)$ is a knot, on which the worldsheet instanton ends, $U_{{\cal K} \left( \beta \right)}$ is the holonomy of the 3-brane gauge field along this knot, and the sum runs over all holomorphic worldsheet instantons.

The cotangent bundle $T^* S^3$ is a deformed $S^2 \times S^3$ conifold, and is known to be related to the resolved conifold via a geometric transition, in which the $S^2$ shrinks to zero at the tip, and the $S^3$ blows up \cite{CandelasdelaOssa}. The GV duality is an equivalence between the topological A-model in two different backgrounds --- the resolved conifold with no branes, where the worldvolume of the closed topological string can wrap the non-vanishing $S^2$, and the deformed conifold in the presence of a stack of Lagrangian branes, on which the open topological strings may end. More precisely, the resolved conifold appearing here is geometrically singular, but with a non-zero flux of the NS-NS 2-form B-field through $S^2$. Then, the integral of the complexified K\"{a}hler class $\kappa = \omega + \mathrm{i} B$ over the non-trivial $S^2$ cycle is purely imaginary. It is related to the number of Lagrangian 3-branes $N$ on the deformed conifold side of the duality as
\begin{equation}
\mathrm{t}_r \equiv \int_{S^2} \kappa = N g_s = - \mathrm{i} \frac{2 \pi N}{k+N},
\end{equation}
where the string coupling $g_s$ is the same on both sides.
The case of a real string coupling is physically most sensible, but in principle, one can analytically continue $g_s$ to any complex value. One often does it in computation, even if at the end the string coupling is put to be real. In topological string computations on the resolved conifold side of the GV duality, the volume $\mathrm{t}_r$ is generally complex.

The CS theory, originally solved in \cite{WittenChernSimons}, is a topological theory, so it does not have any local degrees of freedom, and the only observables are the partition function and the Wilson loops. Wilson loops can be included in the framework of the GV duality by adding a stack of non-compact Lagrangian 3-branes to the deformed conifold background.
%\footnote{There is also an alternative description of Wilson loops via strings ending on the appropriate loop at the boundary of space, as in other holographic dualities \cite{GV}.}
These branes are of $\mathbb{R}^2 \times S^1$ topology, and intersect the zero section of $T^* S^3$ along the contour of the Wilson loop of the CS theory. These 3-branes go through the geometric transition, and one can then extract the value of the Wilson loop from the partition function of the topological A-model in the resolved conifold background with a stack of non-compact 3-branes. The simplest Wilson loop representing a trivial knot was included in the GV duality in \cite{KnotInvariantsAndTopologicalStrings}, and this was later generalized to links \cite{LabastidaMarinoVafa}, framed knots \cite{FramedKnotsAtLargeN}, and algebraic knots \cite{VafaAlgebraicKnots}.\footnote{A pedagogical review of the GV duality can be found in \cite{MarinoReview}.}

Recently, the string theory construction of the CS theory given in \cite{WittenChernSimons} was generalized to the CS theory coupled to partially topological matter, by putting coisotropic 5-branes on the deformed conifold in addition to the stack of $N$ 3-branes \cite{AganagicCostelloMcNamaraVafa}. The procedure of coupling matter to a gauge theory living on a brane by adding a higher-dimensional brane is familiar in conventional non-topological string theory. Exactly in this way one can couple fundamental quarks to ${\cal N}=4$ SYM describing the low-energy dynamics of D3-branes --- one puts additional D7-branes in the background of a large stack of D3-branes, adding 3-7 and 7-3 strings to the theory \cite{D3D7us,D3D7}. In the case of the A-model construction of the PTCSM theory, the fundamental matter dynamics is highly restricted by the topologicity of the theory. The only structure on a 3-fold, besides its topology, that this field theory is sensitive to, is the transverse holomorphic structure determined by a transverse holomorphic foliation (THF) \cite{AganagicCostelloMcNamaraVafa, ClossetDumitrescuFestucciaKomargodski1, GeometrySUSYPartition, ClossetDumitrescuFestucciaKomargodski2, DumitrescuReview}. We will review the construction of this theory below.

In this paper we show how these coisotropic 5-branes go through the GV duality, leading to a dual description of this PTCSM theory. We test the resulting duality by computing the topological string vacuum amplitudes at the first subleading order in the perturbative expansion in the string coupling $g_s$, both on the resolved and the deformed sides. The amplitude on the resolved conifold at the given order in $g_s$ corresponds to the disc amplitude with a coisotropic boundary condition for the strings ending on the 5-branes. On the deformed conifold, it corresponds to the annulus amplitude, with a Lagrangian condition on one boundary of the annulus (ending on the 3-branes), and a coisotropic one on the other (ending on the 5-branes). We find an agreement between our results on both sides, consistent with the duality. 

In \cite{AganagicCostelloMcNamaraVafa} a specific way to evaluate the partition function of the PTCSM theory on $S^3$ was suggested, by mapping it to a computation of a 3d ${\cal N}=2$ CSM partition function. We compare our topological string results to a direct evaluation of this partition function, and we find similar results but with a small discrepancy, which we do not yet understand.

The organization of the paper is the following. In section \ref{sec:topstrings} we give a brief review of the topological A-model, both closed and open (with Lagrangian and coisotropic boundary conditions). Section \ref{sec:equivariantlocalization} reviews the main computational tool we use in our paper, which is the equivariant localization technique for computing open Gromov-Witten (GW) invariants (which gives the amplitudes we are interested in). In section \ref{sec:GVduality} we review the GV duality with Lagrangian 3-branes. These sections review known material, and readers who are familiar with it, or are not interested in the technicalities of the computations we present, may skip them. In section \ref{sec:addingmatter} we propose a generalization of the GV duality by adding the 5-branes into the formalism, and describe the geometry of the brane setups on both sides of the duality. Using the equivariant localization, we compute the annulus amplitude on the deformed conifold, and the disc amplitude in the resolved conifold background, in sections \ref{sec:annulus} and \ref{sec:disc}, respectively. In section \ref{sec:higher} we briefly discuss higher order corrections to this. Independently of the topological string computations, in section \ref{sec:fieldtheory} we briefly review the PTCSM theory constructed in \cite{AganagicCostelloMcNamaraVafa}, and we then compute the subleading term in the large $N$ expansion of the partition function of this theory living on $S^3$. In section \ref{sec:conclusions}, we discuss our results and propose directions in which our project may be continued.

%%%%%%%%%%%%%%%%%%%%%%%%%%%%%%%
%%%%%%%%%%%%%%%%%%%%%%%%%%%%%%%
%%%%%%%%%%%%%%%%%%%%%%%%%%%%%%%
\section{A review of topological string theory} \label{sec:topstrings}
%%%%%%%%%%%%%%%%%%%%%%%%%%%%%%%
%%%%%%%%%%%%%%%%%%%%%%%%%%%%%%%
%%%%%%%%%%%%%%%%%%%%%%%%%%%%%%%
\subsection{Closed A-model} \label{sec:closedAmodel}
As was mentioned in the \nameref{sec:intro}, it is most convenient to consider the GV duality as a geometric transition of the A-model topological string background. In this section, we briefly review the closed A-model, following mostly \cite{KlemmBook, MarinoReview, NeitzkeVafa, Vonk}.

The worldsheet theory of the closed topological A-model is a two-dimensional topological sigma-model \cite{WittenTopologicalSigma} coupled to two-dimensional topological gravity \cite{WittenTopologicalGravity, WittenIntersection}. The sigma-model has a K\"{a}hler target space $X$ with coordinates $x^I$, where the spacetime index $I$ can be divided into holomorphic and anti-holomorphic parts $I=(i,\overline{i})$.
The only bosonic field of the model is $\Phi^I$, which is the embedding of the worldsheet into the target spacetime $X$. The fermionic fields are a section $\chi$ of $\Phi^*TX$, and a one-form $\rho$, also taking values in $\Phi^*TX$. The worldsheet Grassmann one-form $\rho$ is a self-dual field obeying the condition that its $(1,0)$ part takes values in $\Phi^*T^{(0,1)}X$, while the $(0,1)$ part takes values in $\Phi^*T^{(1,0)}X$. The dynamics of these fields is governed by the action \cite{WittenTopologicalSigma, WittenMirrorManifolds} \footnote{One can also add a superpotential term to this action, but we will not need it in what follows.}
\begin{equation}  \label{sigmamodelaction}
S=2 \mathrm{T} \int_{\Sigma_g} \mathrm{d}^2 z \left[ \frac{1}{2} g_{IJ} \partial_{z} \Phi^I \partial_{\overline{z}} \Phi^J + \mathrm{i} g_{i \overline{j}} \rho^{i}_{\overline{z}} \mathrm{D}_{z} \chi^{\overline{j}} + \mathrm{i} g_{\overline{i} j} \rho^{\overline{i}}_{z} \mathrm{D}_{\overline{z}} \chi^j - R_{i \overline{i} j \overline{j}} \rho^{i}_{\overline{z}} \rho^{\overline{i}}_{z} \chi^j \chi^{\overline{j}} \right].
\end{equation}
Here $\Sigma_{g}$ is a Riemann surface of genus $g$, $R_{i \overline{i} j \overline{j}}$ is the target space Riemann tensor, and $\mathrm{T}$ is the string tension. This parameter introduces the mass dimension in the theory, and in what follows will be usually put to $1$. 
The covariant derivatives are the pullbacks of their spacetime cousins
\begin{align}
\begin{split}
\mathrm{D}_z \chi^{\overline{i}} = \partial_z \chi^{\overline{i}}+ \partial_z \Phi^{\overline{j}} \Gamma^{\overline{i}}_{\overline{j} \overline{k}} \chi^{\overline{k}}, \qquad
\mathrm{D}_{\overline{z}} \chi^i =\partial_{\overline{z}} \chi^i+\partial_{\overline{z}} \Phi^j \Gamma^i_{jk} \chi^k.
\end{split}
\end{align}

The action (\ref{sigmamodelaction}) has the following BRST-like nilpotent symmetry ${\cal Q}$ with a Grassmann parameter $\alpha$:
\begin{align} \label{BRSTAction}
\begin{split}
\delta \Phi^I &= \mathrm{i} \alpha \left\{ {\cal Q} , \Phi^I \right\} = \mathrm{i} \alpha \chi^I, \\
\delta \chi^I &= \mathrm{i} \alpha \left\{ {\cal Q} , \chi^I \right\} = 0, \\
\delta \rho^{\overline{i}}_z &= \mathrm{i} \alpha \left\{ {\cal Q} , \rho^{\overline{i}}_z \right\} = - \alpha \left[ \partial_z \Phi^{\overline{i}}+\mathrm{i} \chi^{\overline{j}} \Gamma^{\overline{i}}_{\overline{j} \overline{m}} \rho^{\overline{m}}_z \right], \\
\delta \rho^i_{\overline{z}} &= \mathrm{i} \alpha \left\{ {\cal Q} , \rho^i_{\overline{z}} \right\} = - \alpha \left[ \partial_{\overline{z}} \Phi^i+\mathrm{i} \chi^j \Gamma^i_{jm} \rho^m_{\overline{z}} \right].
\end{split}
\end{align}
There is also a $\mathrm{U} (1)$ ghost number symmetry, under which the fields $\Phi$, $\chi$ and $\rho$ have charges $0$, $1$ and $-1$ respectively. The BRST operator ${\cal Q}$ has charge 1.

All observables of the A-model are restricted to the cohomology of ${\cal Q}$
\begin{equation}
\begin{gathered}
\left\{ {\cal Q}, \mathfrak{L} \right\}=0, \qquad 
\mathfrak{L} \sim \mathfrak{L} + \left\{ {\cal Q}, \mathfrak{l} \right\},
\end{gathered}
\end{equation}
and the vacuum is ${\cal Q}$-invariant, so the theory is a topological field theory of cohomological type (or Witten type). The ${\cal Q}$-cohomology is equivalent to the de Rham cohomology of the target space $X$ with the following identification:
\begin{equation}
{\cal O}_{\Lambda_n} = \mathfrak{L}_{I_{1} \cdots I_{n}} \left( \Phi \right) \chi^{I_1} \cdots \chi^{I_{n}} \sim \mathfrak{L}_{I_{1} \cdots I_{n}} \left( x \right) \mathrm{d} x^{I_1} \wedge \cdots \wedge \mathrm{d} x^{I_{n}}.
\end{equation}
Thus, the Hilbert space of the theory is finite-dimensional. The fact that there is no local dynamics can be understood easily, if one rewrites the action (\ref{sigmamodelaction}) in the following form:
\begin{equation} \label{actionintopologicalform}
S=\left\{{\cal Q}, \mathfrak{V} \right\} + \mathrm{T} \int \Phi^{*} \omega,
\end{equation}
where $\omega$ is the K\"{a}hler class of $X$, and
\begin{equation} \label{vinaction}
\mathfrak{V} = \mathrm{i} \mathrm{T} \int \mathrm{d}^2 z g_{i \overline{j}} \left(\rho_z^{\overline{j}} \partial_{\overline{z}} \Phi^i+\rho_{\overline{z}}^i \partial_z \Phi^{\overline{j}} \right).
\end{equation}

We see that the local dynamics is ${\cal Q}$-exact, and the only non-exact term in the action (\ref{actionintopologicalform}) is topological, depending only on the cohomology class $H^2 (X, \mathbb{Z})$. In particular, the stress-energy tensor of the theory is
\begin{equation} \label{ExactnessofT}
T_{\alpha \beta} = \left\{ {\cal Q} , b_{\alpha \beta} \right\} \equiv \left\{ {\cal Q} , \frac{\delta}{\delta h^{\alpha \beta}} \mathfrak{V} \right\},
\end{equation}
where $h_{\alpha \beta}$ is the worldsheet metric, which we put previously to be the unit matrix.
The ${\cal Q}$-exactness of the non-topological term leads to its independence on $\mathrm{T}$. Applying the usual considerations of supersymmetric localization \cite{Pestun}, one can see that the path integral localizes to the semiclassical minima of the action, namely to the holomorphic maps
\begin{equation}
\partial_{\overline{z}} \Phi^i = \partial_z \Phi^{\overline{i}} = 0.
\end{equation}

The elementary field $\chi^I$ and the composite field $b_{\alpha \beta}$ play the same role in the A-model as the ghosts $c^{\alpha}$ and $b_{\alpha \beta}$ in the usual bosonic string theory \cite{WittenChernString}. The ghost $\mathrm{U} (1)$ symmetry is in general anomalous, which means that the genus $g$ correlation function $\left< {\cal O}_{\mathfrak{L}_1} \cdots {\cal O}_{\mathfrak{L}_n} \right>_g$ vanishes unless\footnote{See \cite{NeitzkeVafa, Vonk, BigMirrorBook} for a detailed discussion of the subject.}
\begin{equation} \label{SelectionRule}
\sum\limits_{i=1}^{n} \mathrm{deg} \left( {\cal O}_{\mathfrak{L}_i} \right) = 2 D \left( 1 - g \right) + 2 \int_{\Sigma_g} \Phi^* c_1 \left( X \right),
\end{equation}
where $\mathrm{deg} \left( {\cal O}_{\mathfrak{L}_i} \right)$ is the degree of the differential form corresponding to the operator ${\cal O}_{\mathfrak{L}_i}$, $D$ is the complex dimension of the target space, and $c_1 \left( X \right)$ is its first Chern class. Physically, the most interesting target space is a Calabi-Yau 3-fold, which means that $c_1 \left( X \right) = 0$. In what follows, we restrict ourselves only to this case. For these manifolds, the selection rule (\ref{SelectionRule}) states that the correlation function of three operators, corresponding to the target space 2-forms, is in general non-vanishing at genus zero, and is of the form
\begin{equation}
\left< {\cal O}_{\mathfrak{L}_1} {\cal O}_{\mathfrak{L}_2} {\cal O}_{\mathfrak{L}_3} \right>_{g=0} = \left( D_1 \cap D_2 \cap D_3 \right) + \sum_{\beta} I_{0,3,\beta} \left( \mathfrak{L}_1, \mathfrak{L}_2, \mathfrak{L}_3 \right) \mathfrak{T}^{\beta}.
\end{equation}
The first term corresponds to the trivial instanton sector, where the image of the worldsheet sphere is a point in spacetime, and it represents a classical intersection number of the divisors $D_i$, which are in a one-to-one correspondence with the differential forms $\mathfrak{L}_i$. We also introduced the notation $\mathfrak{T}^{\beta} = \prod_i \mathrm{e}^{-n_i \mathrm{t}_i}$, where $\beta = \sum_i n_i \left[ h_i \right]$, $h_i$ form a basis of the homology group $H_2 \left( X \right)$, and $\mathrm{t}_i$ are the symplectic volumes of the corresponding cycles $h_i$ measured in units of $\mathrm{T}$
\begin{equation}
\mathrm{t}_i = \int_{h_i} \left( \omega + \mathrm{i} B \right), \quad i = 1 , \cdots , b_2 \left( X \right).
\end{equation}
The coefficients $I_{0,3,\beta} \left( \mathfrak{L}_1, \mathfrak{L}_2, \mathfrak{L}_3 \right)$ count the number of holomorphic maps from the sphere to the target space, wrapping the cycle $\beta$, such that the insertion point of ${\cal O}_{\mathfrak{L}_i}$ is mapped to $D_i$, and can be written as
\begin{equation}
I_{0,3, \beta} = \mathrm{GW}_{0, \beta} \int_{\beta} \mathfrak{L}_1 \int_{\beta} \mathfrak{L}_2 \int_{\beta} \mathfrak{L}_3.
\end{equation}
Here $\mathrm{GW}_{0, \beta}$ are the so-called Gromov-Witten (GW) invariants at genus 0, in the homology class $\beta$. It is common to combine them into a free energy of the theory
\begin{equation}
\mathrm{F}_0 (\mathrm{t}) = \sum_{\beta} \mathrm{GW}_{0, \beta} \mathfrak{T}^{\beta}.
\end{equation}
The selection rule (\ref{SelectionRule}) allows for a non-zero free energy $\mathrm{F}_1 (\mathrm{t})$ with no insertions for $g=1$, but all the higher-point correlation functions vanish. If $g \geq 2$, all the correlators are zero, including the free energy. This is because there is no holomorphic map from $\Sigma_{g \geq 2}$ to the target space for a general worldsheet metric. In topological string theory, one integrates over all the worldsheet metrics, coupling the sigma-model to topological gravity, thus allowing for non-zero correlation functions for arbitrary $g \geq 0$.

The coupling to worldsheet gravity in topological string theory is almost the same as in the usual bosonic string theory \cite{TopologicalGravityReview, WittenChernString, BCOVLong}, so we just follow a well-known procedure. One defines the genus $g$ free energy as
\begin{equation} \label{FreeEnergyg}
\mathrm{F}_g = \int\limits_{\overline{\cal M}_g} \left< \prod\limits_{k=1}^{6 g - 6} \left( b , \mu_k \right) \right>,
\end{equation}
where the vacuum expectation value (VEV) is taken with respect to the fields of the topological sigma-model, $\overline{\cal M}_g$ is the Deligne-Mumford compactification of the moduli space of Riemann surfaces, $\mu_k$ are the Beltrami differentials (anti-holomorphic 1-forms on $\Sigma_g$ with values in $T^{(1,0)} \Sigma_g$, spanning the space of infinitesimal deformations of the $\overline{\partial}$-operator on $\Sigma_g$, which is the tangent
space to $\overline{M}_g$), and
\begin{equation}
\left( b, \mu_k \right) \equiv \int_{\Sigma_g} \mathrm{d}^2 z \left[ b_{zz} \left( \mu_k \right)_{\overline{z}}^{\ z} + b_{\overline{zz}} \left( \overline{\mu}_k \right)_z^{\ \overline{z}} \right].
\end{equation}

More geometrically, $\left( b , \mu_k \right)$ is an operator-valued $(1,0) \oplus (0,1)$ differential form on $\overline{\cal M}_g$, so the expectation value of the product of $(6 g - 6)$ copies of $\left( b , \mu_k \right)$ corresponds to integrating a top form over the moduli space. The ghost number of the aforementioned top form is $(6 g -6)$, so its insertion exactly cancels the first term on the right-hand side of (\ref{SelectionRule}),\footnote{The case of $g \leq 1$ is special, because the virtual dimension of the moduli space does not coincide with the actual one for these genera.} allowing for non-zero correlation functions at higher genera.
Remember that, as we showed in $\left( \ref{ExactnessofT} \right)$, the field $b_{\alpha \beta}$ is a composite field in topological string theory.
Since in (\ref{FreeEnergyg}) one integrates also over embeddings of the worldsheet into the target space, it is convenient to represent the expression for the free energy or some correlation function with $n$ insertions, not in terms of integrals over the moduli space of curves (possibly with $n$ marked points corresponding to the insertions) $\overline{\cal M}_{g,n}$, but in terms of their generalizations, namely integrals over the moduli space of maps to the homology cycle $\beta$ of the target space $X$
\begin{equation}
\overline{\cal M}_{g,n} \rightarrow \overline{\cal M}_{g,n} \left( X, \beta \right).
\end{equation}

It is convenient to define a total free energy of the topological A-model with coupling $g_s$ as
\begin{equation} \label{ClosedFreeEnergy}
\mathrm{F} (g_s, \mathrm{t}) = \sum\limits_{g=0}^{\infty} g_s^{2 g - 2} \mathrm{F}_g (\mathrm{t}) = \sum\limits_{g=0}^{\infty} g_s^{2 g - 2} \mathrm{GW}_{g, \beta} \mathfrak{T}^{\beta},
\end{equation}
where we introduced the higher genus generalization of the GW invariants $\mathrm{GW}_{g, \beta}$.

\noindent These invariants are defined as \cite{WittenMirrorManifolds, AspinwallMorrison}
\begin{equation} \label{ClosedGWinvariants}
\mathrm{GW}_{g, \beta} = \int\limits_{\overline{\cal M}_g \left( X, \beta \right)} e \left( {\cal V} \right).
\end{equation}
Here $e \left( {\cal V} \right)$ is the top Euler class of the bundle ${\cal V}$, formed by the zero modes of $\rho$
\begin{equation}
\mathrm{D}_z \rho^i_{\overline{z}} = \mathrm{D}_{\overline{z}} \rho^{\overline{i}}_z = 0
\end{equation}
over the moduli space $\overline{\cal M}_{g,n=0} \left( X, \beta \right)$.

\subsection{Open strings, branes and boundary conditions} \label{sec:openAmodel}

Let us next review the open topological A-model and the corresponding boundary conditions, which we need in order to formulate the Gopakumar-Vafa duality, following \cite{OoguriOzYin, KapustinOrlovPaper, KapustinOrlovLectures, N=2, Herbst}.

\subsubsection{Open topological sigma-model} \label{sec:opensigmamodel}

A generalization of the sigma-model considered above is a worldsheet theory with boundaries, corresponding to a target space with D-branes. In order to take boundaries of the worldsheet into account, one must replace the covariant derivatives in the action (\ref{sigmamodelaction}) as $\mathrm{D} \rightarrow \frac{1}{2} \mathrm{\overset{\leftrightarrow}D}$, and couple the theory to the gauge field on the brane via \cite{WittenChernString}
\begin{equation} \label{coupledaction}
\int D \Phi D \chi D \rho \ \mathrm{exp} \left(- S \left[\Phi, \chi, \rho \right]\right) \prod_{i}\mathrm{Tr} \left[ \mathrm{\hat{P}} \mathrm{exp} \left( \oint_{{\cal K}_i} \Phi^{*} A_i \right) \right],
\end{equation}
where $A_i$ is a $\mathrm{U} (N_i)$ gauge connection living on the $i$'th stack of $N_i$ D-branes, ${\cal K}_i$ is an $i$'th boundary of the worldsheet, and $\mathrm{\hat{P}}$ is the path-ordering operator.
One can also couple the theory to the NS-NS two-form field $B$ by complexifying the K\"{a}hler class in (\ref{actionintopologicalform}) as $\omega \rightarrow \omega + \mathrm{i} B$.

\subsubsection{Boundary conditions} \label{sec:openboundary}

By requiring BRST invariance of the action coupled to the gauge field (\ref{coupledaction}), one can deduce the possible brane geometries and the corresponding gauge field configurations living on these branes. It appears that for the most interesting case of a Calabi-Yau 3-fold, there are essentially two possibilities: 3-branes \cite{WittenChernString} and 5-branes \cite{KapustinOrlovPaper, KapustinOrlovLectures}.\footnote{See \cite{BigMirrorBook} for a comprehensive review.} For simplicity, we consider the case of a $\mathrm{U} (1)$ gauge field (a single brane). As always in string theory, in the case of a single brane, the worldsheet dynamics, and thus the brane dynamics, is sensitive only to $\mathfrak{F} = B \vert_Y+F$, where $\vert_Y$ means the restriction of the spacetime field $B$ to the worldvolume of the brane $Y$, and $F$ is the field strength of the $\mathrm{U} (1)$ gauge field $A$
\begin{equation}
F = \mathrm{d} A.
\end{equation}
A 3-brane configuration must obey the following conditions:
\begin{align} \label{3branes}
\begin{split}
\sigma =0, \qquad \mathfrak{F} =0,
\end{split}
\end{align}
where we defined $\sigma=\omega \vert_Y$. This means that a 3-brane must wrap a Lagrangian submanifold with a flat connection.
For a 5-brane, one has
\begin{equation} \label{5branes}
\begin{gathered}
\mathfrak{F} \wedge \mathfrak{F} = \sigma \wedge \sigma, \qquad \mathfrak{F} \wedge \sigma = 0.
\end{gathered}
\end{equation}

To preserve a topological symmetry on the worldsheet with boundary, one must also impose boundary conditions on the worldsheet fields $\Phi$, $\chi$ and $\rho$. To make the corresponding formulae shorter, we introduce the notation $\Psi$ defined by
\begin{equation}
\chi^{i} = \Psi^{i}_{+}, \qquad \chi^{\overline{i}} = \Psi^{\overline{i}}_{-}, \qquad \rho^{i}_{\overline{z}} = \Psi^{i}_{-}, \qquad \rho^{\overline{i}}_{z} = \Psi^{\overline{i}}_{+}.
\end{equation}

The boundary conditions can then be written as \cite{OoguriOzYin, KapustinOrlovPaper}
\begin{equation} \label{GeneralRmatrix}
\begin{gathered}
\partial \Phi = R \left(\overline{\partial} \Phi \right), \qquad
\Psi_{+} = R \Psi_{-},
\end{gathered}
\end{equation}
where $R$ is a matrix, whose form depends on the dimensionality of the brane. For a 3-brane, it is
\begin{equation} \label{Lagrangianboundaryconditions}
R_L = -\mathbb{I}_{NY} \oplus \mathbb{I}_{TY},
\end{equation}
and for a 5-brane
\begin{equation} \label{Coisotropicboundaryconditions}
R_C =-\mathbb{I}_{NY} \oplus \left( G - \mathfrak{F} \right)^{-1} \left( G + \mathfrak{F} \right).
\end{equation}
Here $\mathbb{I}_{NY}$ and $\mathbb{I}_{TY}$ are identity matrices in the orthogonal and tangential directions to the brane respectively, and $G=g \vert_{Y}$.

As shown in \cite{KapustinOrlovPaper, KapustinOrlovLectures}, the topological 5-branes in the A-model must be coisotropic manifolds admitting a THF structure, defined below.

A submanifold $Y$ of a K\"{a}hler manifold $X$ is called coisotropic if at any point $p \in Y$, the complement $T_{\perp} Y_p$ of $TY_p$ with respect to the K\"{a}hler form $\omega$ is contained in $TY_p$. There is also an equivalent definition: a submanifold $Y$ is coisotropic if the restriction of $\omega$ to $Y$, which we denoted by $\sigma$, has a constant rank,\footnote{Note that a Lagrangian 3-brane is a particular case of a coisotropic brane with $\mathrm{rank} (\sigma) = 0$.} and its kernel ${\cal L} Y \subset TY$ is an integrable distribution, which means that the commutator of any two vector fields in ${\cal L} Y$ is also in ${\cal L} Y$. The vectors $\mathrm{V} \in {\cal L} Y$ generate the foliation structure: the leaves are given by orbits of $\mathrm{V}$. The dimension of the leaves is equal to the codimension of $Y$ in $X$, giving $\mathrm{dim}({\cal L} Y) = 3$ for a 3-brane, and $\mathrm{dim}({\cal L} Y) = 1$ for a 5-brane. Note that it implies that a 5-brane in a 6-fold is always coisotropic, with a complement $T_{\perp} Y$ coinciding with $\mathrm{V}$. One usually calls ${\cal L} Y$ the tangent bundle of the foliation, and the quotient ${\cal F} Y = TY / {\cal L} Y$ is called the normal bundle of the foliation.

\subsubsection{Transverse holomorphic structure} \label{sec:THF}

A coisotropic brane must also support a non-trivial gauge bundle with field strength $\mathfrak{F}$ obeying (\ref{5branes}), for which one has
\begin{equation} \label{Orthogonality}
\imath_{\mathrm{V}} \mathfrak{F} = \imath_{\mathrm{V}} \sigma = 0.
\end{equation}
The foliation generated by $\mathrm{V}$ must be a THF,\footnote{See \cite{DuchampKalka, GomezMont, GirbauHaefligerSundararaman, BrunellaGhys, Brunella, Ghys} for a mathematical review.} if one wants to wrap a 5-brane on $Y$. This means that $Y$ may be covered by a set of coordinate charts $\left( \tau_i, u_i, \overline{u}_i, v_i, \overline{v}_i \right)$, which transform as
\begin{equation} \label{THFcoordinates}
\tau_j = \tau_j + \vartheta_j \left(  \tau_i, u_i, \overline{u}_i, v_i, \overline{v}_i \right), \qquad u_j = u_j \left( u_i, v_i \right), \qquad v_j = v_j \left( u_i, v_i \right)
\end{equation}
with real functions $\vartheta_j$ as one goes from one chart to the other. The coordinate $\tau$ parameterizes the integral curves of the vector field $\mathrm{V}$.\footnote{This definition can be generalized straightforwardly to manifolds of any odd dimensionality --- for example, in what follows, we will need also three-dimensional manifolds admitting a THF.}
There is a complex structure $J = \left( \sigma \vert_{\mathcal{F} Y} \right)^{-1} \mathfrak{F}$ induced on the 2-complex-dimensional manifold ${\cal F} Y$,\footnote{For a general action generated by $\mathrm{V}$, the space $\mathcal{F} Y$ is singular and is not a manifold.} in which $\hat{\Omega} = \mathfrak{F} + \mathrm{i} \sigma \vert_{\mathcal{F} Y}$ is a $(2,0)$ form. Note that it is different from the complex structure induced on ${\cal F} Y$ by the target space K\"{a}hler form $\omega$ and the metric $g$, which is given by $I = G^{-1} \sigma \vert_{\mathcal{F} Y}$.

The fields $\sigma$ and $\mathfrak{F}$ have the following coordinate expansions in the complex structure $J$:
\begin{align}
\begin{split}
\sigma &= \sigma_{u v} \mathrm{d} u \wedge \mathrm{d} v + \sigma_{\overline{u} \overline{v}} \mathrm{d} \overline{u} \wedge \mathrm{d} \overline{v}, \\
\mathfrak{F} &= \mathfrak{F}_{u v} \mathrm{d} u \wedge \mathrm{d} v + \mathfrak{F}_{\overline{u} \overline{v}} \mathrm{d} \overline{u} \wedge \mathrm{d} \overline{v}.
\end{split}
\end{align}
Substituting these expressions into \eqref{5branes}, one gets
\begin{equation} \label{FintermsofKahlerform}
\mathfrak{F}_{u v} = \mathrm{i} \omega_{u v}, \qquad \mathfrak{F}_{\overline{u} \overline{v}} = - \mathrm{i} \omega_{\overline{u} \overline{v}}.
\end{equation}
Note that when a coisotropic brane touches a Lagrangian cycle, the restriction of the field $\mathfrak{F}$ on the cycle vanishes, which must be the case in order to define the CS theory on the Lagrangian 3-brane.

Generalizing our discussion of the closed A-model in section \ref{sec:closedAmodel}, we come to the conclusion that only holomorphic maps with boundaries on the branes contribute to the open A-model amplitudes. This means that the boundaries of holomorphic maps ending on the coisotropic brane must belong to the orbits of the vector field $\mathrm{V}$, because only in that case the image of the worldsheet is a holomorphic surface. To get a finite contribution, one has to further restrict the boundary to the closed orbits of $\mathrm{V}$ \cite{Herbst, AganagicCostelloMcNamaraVafa}.

The observables of the topological A-model are given by the cohomology of the following operator \cite{OpenStringBRST}:
\begin{equation}
{\cal D} = \mathrm{d}_{\parallel} + \overline{\partial}_{\perp},
\end{equation}
where $\mathrm{d}_{\parallel}$ and $\overline{\partial}_{\perp}$ are the usual de Rham and Dolbeault operators acting in the ${\cal L} Y$ and ${\cal F} Y$ directions, respectively. In the case of a Lagrangian brane, where ${\cal F} Y = \left\{ {\mathrm{pt}} \right\}$, one is left with ${\cal D} = \mathrm{d}$, directly generalizing the closed string observables we discussed.

\subsubsection{Free energy} \label{sec:openfreeenergy}

In topological string theory, we are mostly interested in worldsheet images of two topological types. The first is the topology $\sum_i \mathbb{CP}^1_i \cup \sum_{j} D_j$, where $D_j$ are discs, and $\mathbb{CP}^1_i$ are 2-spheres wrapping the non-trivial 2-cycles in the target space. The discs are attached to the spheres at the points (this attachment is usually denoted by $ \cup $), so the worldsheet is singular. This allows the spheres and the discs to be wrapped a different number of times. The generalization of the free energy (\ref{ClosedFreeEnergy}) to this case has the form
\begin{equation} \label{openfreeenergy1}
\mathrm{F} \left( g_s \right) = \sum\limits_{g, h, \vec{d}, \vec{w}} g_s^{2 g + h - 2} \mathrm{GW}_{g, h}^{\vec{d}, \vec{w}} \prod\limits_{\vec{d}} \mathrm{e}^{ - d_i \mathrm{t}_i} \prod\limits_{\vec{w}} \mathrm{e}^{ - w_j S_j} \mathrm{Tr} ( V_j^{w_{j}} ).
\end{equation}
Here $h$ is the number of disc components of the worldsheet Riemann surface $\Sigma_{g,h}$, whose boundaries must be mapped to the branes' worldvolumes in the target space, $g$ is the total genus of the irreducible domain components contracted to points in the target space $X$ under the embedding $\Phi$, $\mathrm{t}_i$ are the complexified volumes of a basis of the homology group $H_2 \left( X \right)$ to whom the non-contracted 2-spheres get mapped, $V_j$ are the holonomies of the gauge fields along the target space images of the worldsheet boundaries, $S_j$ are the symplectic volumes of the discs in the target space, and $d_i$ and $w_j$ are the corresponding degrees and winding numbers. We introduced the notation $\vec{d} = \left( d_1, ... , d_{b_2 \left( X \right)} \right)$ and $\vec{w} = \left( w_1, ... , w_h \right)$ for brevity.

The second type of worldsheets we need is the sum of annuli $C_i$, for which the free energy is
\begin{equation} \label{openfreeenergy2}
\mathrm{F} (g_s) = \sum\limits_{g, h, \vec{w}} g_s^{2 g + 2 h - 2} \mathrm{GW}_{g, h}^{\vec{w}} \prod\limits_{\vec{w}} \mathrm{e}^{ - w_j S_j} \mathrm{Tr} (V_j^{w_{j}}) \mathrm{Tr} (U_j^{w_{j}}),
\end{equation}
where $h$ is the number of annuli, $g$ is again the total genus of contracted irreducible worldsheet components, $V_j$ and $U_j$ are the holonomies of the gauge fields along the two boundaries of each annulus, and $S_j$ are the annuli symplectic volumes. Here we assume that the worldsheet does not wrap any 2-cycles in the target space.

If some of the branes are compact, and thus dynamical, one must replace the corresponding holonomies in $\left( \ref{openfreeenergy1} \right)$ and $\left( \ref{openfreeenergy2} \right)$ as
\begin{equation}
\mathrm{Tr} (V_j^{w_{j}}) \rightarrow \left< \mathrm{Tr} (V_j^{w_{j}}) \right>,
\end{equation}
where the VEVs are computed in the effective field theories on the branes, governing the dynamics of the gauge fields.

The open GW invariants above are defined similarly to the closed string case $\left( \ref{ClosedGWinvariants} \right)$
\begin{equation}
\mathrm{GW}_{g, h}^{\vec{d}, \vec{w}} = \int\limits_{\overline{\cal M}_{g, h}^{\vec{d}, \vec{w}} \left( X, Y \right)} e \left( {\cal V} \right).
\end{equation}

\section{A review of open Gromov-Witten invariants from equivariant localization} \label{sec:equivariantlocalization}

In this section we review the equivariant localization technique. It was first developed for the case of closed Riemann surfaces in \cite{Kontsevich, GraberPandharipande}, and generalized to open ones in \cite{KatzLiu, LiSong, GraberZaslow}. The method was applied to coisotropic branes in \cite{Saulina}.\footnote{The results obtained there are somewhat different from our result, because of different reality conditions imposed on the worldsheet fermions. We discuss this issue in section \ref{sec:DeformedBoundaryConditions}.} We will closely follow \cite{VafaAlgebraicKnots} in our discussion.

In contrast to the closed GW invariants,\footnote{See \cite{BigMirrorBook} for a detailed discussion.} there is no mathematically rigorous definition of the open GW invariants even for the Lagrangian boundary conditions, and the definition currently available is purely computational. We assume that analogous open string modification works also for coisotropic boundary conditions.

The equivariant localization can be applied when the configurations of the branes, the gauge fields, and the image of the worldsheet in the target space are invariant under the action of some group $G$. Then, the integrals over the moduli space of maps $\overline{\cal M}_{g, h}^{\vec{d}, \vec{w}} \left( X, Y \right)$ can be reduced only to those maps whose images are fixed under $G$. In the problem under consideration, the group is $G = \mathrm{U} (1)$,\footnote{More precisely, $G$ is a non-compact version of $\mathrm{U} (1)$, namely $\mathbb{C}^{\times}$.} so in what follows we consider only this particular case.

The general form of the main formula of the equivariant localization machinery, the Atyiah-Bott fixed point formula, is
\begin{equation}
\int\limits_{\overline{\cal M}_{g, h}^{\vec{d}, \vec{w}} \left( X, Y \right)} \phi = \sum\limits_{i} \int\limits_{F_i} \frac{i^*_{F_i} \phi}{e_G \left( NF_i \right)},
\end{equation}
where the sum runs over the fixed loci $F_i$ of the group action, $\phi$ is any top form on the moduli space, $i_{F_i}$ is the embedding of the fixed locus $F_i$ into the moduli space, and $e_G (NF_i)$ is the $G$-equivariant top Euler class of the bundle normal to $F_i$ in $\overline{\cal M}_{g, h}^{\vec{d}, \vec{w}} \left( X, Y \right)$.

In the resolved conifold geometry, the image of the worldsheet
\begin{equation}
\Sigma = \mathbb{CP}^1_{\Sigma} \cup_{\nu} \Delta
\end{equation}
in the target space $X$ is
\begin{equation}
\Phi \left( \Sigma \right) = \mathbb{CP}^1_X \cup_{\Phi\left( \nu \right)} D.
\end{equation}
Here $\nu$ is a simple node, at which the source disc $\Delta$ is attached to the source 1-cycle $\mathbb{CP}^1_{\Sigma}$, and $D = \Phi \left( \Delta \right)$ is a holomorphic disc. Remember that one can apply the equivariant localization only if all irreducible components of $\Phi \left( \Sigma \right)$ are fixed under the action of $G = \mathbb{C}^{\times}$.

Maps of this type can be represented as decorated graphs consisting of vertices, edges, and legs \cite{Kontsevich, GraberPandharipande, GraberZaslow}. The vertices $v$ correspond to the contracted genus $g_v$ components of the domain $\Sigma$, which must be mapped to the fixed points of the group action. The edges joining the vertices symbolize the $\mathbb{CP}^1_{\Sigma}$'s, which are not contracted, and are labeled by the degree $d_e$ of the map $\mathbb{CP}^1_{\Sigma} \rightarrow \mathbb{CP}^1_X$. Finally, the legs emanating from the vertices stand for the disc components $D = \Phi \left( \Delta \right)$, which are attached to $\mathbb{CP}^1_X$'s at the nodes, and are labeled by the winding $w_l$. In our problem, we are interested in tree-level diagrams, which means that $g = 0$.

The GW invariant of the map corresponding to the graph $\Gamma$ can be computed as
\begin{equation}
\mathrm{GW}_{g, h}^{\vec{d}, \vec{w}} = \sum\limits_{\Gamma} \frac{1}{\vert A_{\Gamma} \vert} \int\limits_{{\cal M}_{\Gamma}} \frac{i^*_{{\cal M}_{\Gamma}} e \left( {\cal V} \right)}{e_G \left( N {\cal M}_{\Gamma} \right)},
\end{equation}
where the sum is over disconnected graph $\Gamma$ components, ${\cal M}_{\Gamma} \subset \overline{\cal M}_{g, h}^{\vec{d}, \vec{w}} \left( X, Y \right)$ corresponds to the graph $\Gamma$, and $\vert A_{\Gamma} \vert$ is the order of the graph automorphism group.

The worldsheet topology we are interested in in the dual deformed conifold geometry is simpler, namely it is a non-singular annulus $C$. There is no room for singularities on this side of the duality, because there are no 2-cycles on the deformed conifold, which could be wrapped by irreducible worldsheet components of $\mathbb{CP}^1$ topology, and be connected to the rest of the worldsheet image at some point.

To proceed, we recall the definition of the tangent-obstruction sequence. For a map $\Phi$ from the Riemann surface $\Sigma$ to the target space $X$, there is an exact sequence of sheaves
\begin{equation}
0 \longrightarrow T \Sigma \longrightarrow \Phi^* T X \longrightarrow N \Sigma \longrightarrow 0,
\end{equation}
where $N \Sigma$ is a bundle normal to the image of $\Sigma$ in $X$ under $\Phi$. This sequence induces a corresponding long exact sequence on the \v{C}ech cohomology
\begin{align} \label{Longsequence}
\begin{split}
0 &\longrightarrow H^0 \left( \Sigma, T \Sigma \right) \longrightarrow H^0 \left( \Sigma, \Phi^* T X \vert_{\partial \Sigma} \right) \longrightarrow H^0 \left( \Sigma, N \Sigma \vert_{\partial \Sigma} \right) \\ & \longrightarrow H^1 \left( \Sigma, T \Sigma \right) \longrightarrow H^1 \left( \Sigma, \Phi^* T X \vert_{\partial \Sigma} \right) \longrightarrow H^1 \left( \Sigma, N \Sigma \vert_{\partial \Sigma} \right) \longrightarrow 0.
\end{split}
\end{align}
All terms in this sequence can be interpreted as infinitesimal automorphisms, deformations and obstructions to deformations for $\Sigma$ and $\Phi$.

The bundle $T \Sigma$ is related to the deformations of $\Sigma$; namely $H^0$ measures infinitesimal automorphisms, $H^1$ measures infinitesimal deformations, and $H^2 \equiv	0$ measures obstructions to deformations (so all deformations are unobstructed).
The bundle $N \Sigma$ is related to the deformations of the map $\Phi$; $H^{-1} \equiv 0$ measures infinitesimal automorphisms, $H^0$ measures deformations, and $H^1$ measures obstructions.
The bundle $\Phi^* T X$ is related to the deformations of the map $\Phi$, where the structure of the source curve $\Sigma$ is held fixed; $H^{-1} \equiv 0$ measures infinitesimal automorphisms, $H^0$ measures deformations, and $H^1$ measures obstructions. This allows us to rewrite $\left( \ref{Longsequence} \right)$ as
\begin{align}
\begin{split}
0 &\longrightarrow \mathrm{Aut} (\Sigma) \longrightarrow \mathrm{Def} (\Phi) \longrightarrow \mathrm{Def} \left( \Sigma, \Phi \right) \\ & \longrightarrow \mathrm{Def} (\Sigma) \longrightarrow \mathrm{Obs} (\Phi) \longrightarrow \mathrm{Obs} \left( \Sigma, \Phi \right) \longrightarrow 0.
\end{split}
\end{align}
It follows from this sequence that in the representation ring of $G$
\begin{equation} \label{defsequence}
\mathrm{Obs} \left( \Sigma, \Phi \right) - \mathrm{Def} \left( \Sigma, \Phi \right) = \mathrm{Aut} (\Sigma) + \mathrm{Obs} (\Phi) - \mathrm{Def} (\Sigma) - \mathrm{Def} (\Phi).
\end{equation}
If there are marked points on the Riemann surface, which correspond to operator insertions, or nodes at which two irreducible parts of the curve are connected, the tangent-obstruction sequence gets naturally generalized to
\begin{align}
\begin{split}
0 &\longrightarrow \mathrm{Aut} \left( \Sigma, p_1, \cdots ,p_n \right) \longrightarrow \mathrm{Def} (\Phi) \longrightarrow \mathrm{Def} \left( \Sigma, p_1, \cdots ,p_n, \Phi \right) \\ & \longrightarrow \mathrm{Def} \left( \Sigma, p_1, \cdots ,p_n \right) \longrightarrow \mathrm{Obs} (\Phi) \longrightarrow \mathrm{Obs} \left( \Sigma, p_1, \cdots ,p_n, \Phi \right) \longrightarrow 0.
\end{split}
\end{align}

As already mentioned, we are most interested in $\Sigma$ and $\Phi \left( \Sigma \right)$ of the topology $\mathrm{Sphere} \cup \mathrm{Disc}$, so the moduli space over which we integrate in order to get the GW invariants is $\overline{\cal M}_{g=0, h=1}^{d, w} \left( X, Y \right)$, where $w$ measures the winding of the map $\Phi \left( \mathrm{Disc} \right)$, and $d$ counts the degree of $\Phi \left( \mathrm{Sphere} \right)$. In order to relate the cohomology groups appearing in $\left( \ref{defsequence} \right)$ to the ones corresponding to irreducible components of the worldsheet, we use the sequence
\begin{equation}
0 \longrightarrow {\cal O}_{\mathbb{CP}^1_{\Sigma} \cup_{\nu} \Delta} \longrightarrow {\cal O}_{\mathbb{CP}^1_{\Sigma}} \oplus {\cal O}_{\Delta} \longrightarrow {\cal O}_{\nu} \longrightarrow 0,
\end{equation}
inducing a long exact sequence on cohomology
\begin{align}
\begin{split}
0 &\longrightarrow \mathrm{Def} (\Phi) \longrightarrow H^0 \left( \Delta , \Phi^* {\cal T}_{\Delta, R} \right) \oplus \mathrm{Def} \left( \mathbb{CP}^1_X \right) \longrightarrow T_{\Phi \left( \nu \right)} X \\ & \longrightarrow \mathrm{Obs} (\Phi) \longrightarrow H^1 \left( \Delta , \Phi^* {\cal T}_{\Delta, R} \right) \oplus \mathrm{Obs} \left( \mathbb{CP}^1_X \right) \longrightarrow 0.
\end{split}
\end{align}
Here ${\cal T}_{\Delta, R}$ is the sheaf of germs of holomorphic sections of the bundle $T^{\left( 1,0 \right)} X \vert_D$ with the boundary condition specified by the matrix $R$, given either by $\left( \ref{Lagrangianboundaryconditions} \right)$ or $\left( \ref{Coisotropicboundaryconditions} \right)$. It gives the two following relations in the representation ring of $G = \mathbb{C}^{\times}$:
\begin{align}
\begin{split}
\mathrm{Obs} \left( \Phi \right)^f - \mathrm{Def} \left( \Sigma \right)^f = &H^1 \left( \Delta , \Phi^* {\cal T}_{\Delta, R} \right)^f - H^0 \left( \Delta , \Phi^* {\cal T}_{\Delta, R} \right)^f \\ &+ \mathrm{Obs} \left( \mathbb{CP}^1_X \right)^f - \mathrm{Def} \left( \mathbb{CP}^1_X \right)^f,
\end{split}
\end{align}
and
\begin{align}
\begin{split}
\mathrm{Obs} \left( \Phi \right)^m - \mathrm{Def} \left( \Sigma \right)^m = &H^1 \left( \Delta , \Phi^* {\cal T}_{\Delta, R} \right)^m - H^0 \left( \Delta , \Phi^* {\cal T}_{\Delta, R} \right)^m \\ &+ \mathrm{Obs} \left( \mathbb{CP}^1_X \right)^m - \mathrm{Def} \left( \mathbb{CP}^1_X \right)^m + T_{\Phi \left( \nu \right)} X,
\end{split}
\end{align}
where the superscripts $f$ and $m$ denote the fixed and moving parts of the corresponding bundles with respect to the action of $G$.

Let us parameterize the source disc $\Delta$ by a complex coordinate $t$, such that $\vert t \vert \leq 1$, and such that the origin connected to $\mathbb{CP}^1_{\Sigma}$ at $\nu$ corresponds to $t=0$. The automorphism group of the disc $\mathrm{Aut} \left( \Delta , 0 \right)$ consists of the infinitesimal automorphisms preserving the
origin, which are generated by the following section $s$:
\begin{equation} \label{AutomorphismsOfTheDisc}
s = t \partial_t.
\end{equation}
Therefore, $\mathrm{Aut} \left( \Delta , 0 \right)^m$ is trivial, and $\mathrm{Aut} \left( \Delta , 0 \right)^f = \mathbb{R}$.\footnote{Note that this is the case also for the annulus.}
Using also
\begin{align}
\begin{split}
& \mathrm{Aut} \left( \Sigma \right)^m = \mathrm{Aut} \left( \mathbb{CP}^1_{\Sigma} , \nu \right)^m + \mathrm{Aut} \left( \Delta , 0 \right)^m, \\ & \mathrm{Aut} \left( \Sigma \right)^f = \mathrm{Aut} \left( \mathbb{CP}^1_{\Sigma} , \nu \right)^f + \mathrm{Aut} \left( \Delta , 0 \right)^f, \\ & \mathrm{Def} \left( \Sigma \right)^f = \mathrm{Def} \left( \mathbb{CP}^1_{\Sigma} , \nu \right)^f, \\ & \mathrm{Def} \left( \Sigma \right)^m = \mathrm{Def} \left( \mathbb{CP}^1_{\Sigma} , \nu \right)^m + T_{\nu} \mathbb{CP}^1_{\Sigma} \otimes T_0 \Delta,
\end{split}
\end{align}
one gets
\begin{align}
\begin{split}
\mathrm{Obs} \left( \Sigma , \Phi \right)^f - \mathrm{Def} \left( \Sigma , \Phi \right)^f = &H^1 \left( \Delta , \Phi^* {\cal T}_{\Delta, R} \right)^f - H^0 \left( \Delta , \Phi^* {\cal T}_{\Delta, R} \right)^f \\ &+ \mathrm{Obs} \left( \mathbb{CP}^1_X \right)^f - \mathrm{Def} \left( \mathbb{CP}^1_X \right)^f \\ &+ \mathrm{Aut} \left( \mathbb{CP}^1_{\Sigma} , \nu \right)^f - \mathrm{Def} \left( \mathbb{CP}^1_{\Sigma} , \nu \right)^f \\ &+ \mathrm{Aut} \left( \Delta , 0 \right)^f,
\end{split}
\end{align}
and
\begin{align}
\begin{split}
\mathrm{Obs} \left( \Sigma , \Phi \right)^m - \mathrm{Def} \left( \Sigma , \Phi \right)^m = &H^1 \left( \Delta , \Phi^* {\cal T}_{\Delta, R} \right)^m - H^0 \left( \Delta , \Phi^* {\cal T}_{\Delta, R} \right)^m \\ &+ \mathrm{Obs} \left( \mathbb{CP}^1_X \right)^m - \mathrm{Def} \left( \mathbb{CP}^1_X \right)^m \\ &+ \mathrm{Aut} \left( \mathbb{CP}^1_{\Sigma} , \nu \right)^m - \mathrm{Def} \left( \mathbb{CP}^1_{\Sigma} , \nu \right)^m \\ &+ \mathrm{Aut} \left( \Delta , 0 \right)^m - T_{\nu} \mathbb{CP}^1_{\Sigma} \otimes T_0 \Delta.
\end{split}
\end{align}

The last equation gives the following relationship between the two normal bundles $e_G \left( N {\cal M}_{\Gamma} \right)$ and $e_G \left( N {\cal M}_{\Gamma^{\prime}} \right)$ (where $\Gamma^{\prime}$ is the graph for the closed curve with the disc `leg' removed, i.e. the map associated to $\mathbb{CP}^1_{\Sigma}$ with one marked point):
\begin{equation}
N {\cal M}_{\Gamma} = N {\cal M}_{\Gamma^{\prime}} - T_{\Phi \left( \nu \right)} X + {\cal R} \mathbb{L}^{-1} - H^1 \left( \Delta , \Phi^* {\cal T}_{\Delta, R} \right)^m - H^0 \left( \Delta , \Phi^* {\cal T}_{\Delta, R} \right)^m,
\end{equation}
where ${\cal R}$ is the representation of $G$ on $T_0 \Delta = \mathbb{C}$, induced by the pullback of the $G$ action on $D$, and $\mathbb{L}$ is the tautological cotangent line bundle on the moduli space, associated to the marked point $\nu$.

Collecting all the above considerations, and taking into account that for the case of $\Sigma = \mathbb{CP}^1_{\Sigma} \cup_{\nu} \Delta$, the graph $\Gamma$ consists of one leg corresponding to $D = \Phi \left( \Delta \right)$, one edge corresponding to $\mathbb{CP}^1_X = \Phi \left( \mathbb{CP}^1_{\Sigma} \right)$, and two vertices corresponding to the two poles of $\mathbb{CP}^1$, at one of which the disc is attached, and that the order of the automorphism group generated by $t \rightarrow \zeta t$, with $\zeta$ being a $w$-th root of unity, is $\vert A_{\Gamma} \vert = w$, one finally gets
\begin{align} \label{GWinvariants}
\begin{split}
\mathrm{GW}_{0, 1}^{d, w} &= \frac{1}{w} \int\limits_{{\cal M}_{\Gamma}} \frac{i^*_{{\cal M}_{\Gamma}} e \left( {\cal V} \right)}{e_G \left( N {\cal M}_{\Gamma} \right)} = \frac{1}{w} \int\limits_{{\cal M}_{\Gamma}} \frac{e_G^m \left( H^1 \left( \Sigma , \Phi^* TX \vert_{\partial \Sigma} \right) \right) e_G^m \left( H^0 \left( \Sigma , T \Sigma \right) \right)}{e_G^m \left( H^0 \left( \Sigma , \Phi^* TX \vert_{\partial \Sigma} \right) \right) e_G^m \left( H^1 \left( \Sigma , T \Sigma \right) \right)} \\ &= \frac{1}{w} \frac{e^m_G \left( H^1 \left( \Delta , \Phi^* {\cal T}_{\Delta, R} \right) \right)}{e^m_G \left( H^1 \left( \Delta , \Phi^* {\cal T}_{\Delta, R} \right) \right)} \int\limits_{{\cal M}_{\Gamma^{\prime}}} \frac{e_G \left( T_{\Phi \left( \nu \right)} X \right)}{e_G \left( N {\cal M}_{\Gamma^{\prime}} \right) e_G \left({\cal R} \mathbb{L}^{-1} \right)}.
\end{split}
\end{align}

The expression for the GW invariants for the annulus amplitude, which we are interested in on the deformed conifold side of the duality, is simpler, because the worldsheet topology is non-singular in this case. This means that if there is a unique $G$-invariant holomorphic cylinder ending on the closed leaf of the THF on the 5-brane, the moduli space reduces just to a point for every value of $w$, and one is left with
\begin{equation} \label{GWinvariantsNonSingular}
\mathrm{GW}^w_{0,2} = \frac{1}{w} \frac{e_G^m \left( H^1 \left( C , \Phi^* TX \vert_{\partial C} \right) \right) e_G^m \left( H^0 \left( C , T C \right) \right)}{e_G^m \left( H^0 \left( C , \Phi^* TX \vert_{\partial C} \right) \right) e_G^m \left( H^1 \left( C , T C \right) \right)}.
\end{equation}
Note that the equivariant Euler classes of the vector bundles are given by products of non-trivial $\mathbb{C}^{\times}$-weights of the basis vectors \cite{Kontsevich,GraberPandharipande}. For closed GW invariants, all the weight dependence cancels, but open GW invariants have some residual weight dependence, which can be mapped under the GV duality to the framing dependence of the Wilson loops in the CS theory \cite{FramedKnotsAtLargeN}.

There is a subtlety in the evaluation of the GW invariants via equivariant localization that we should mention. As one works only with a subset ${\cal M}_{\Gamma} \subset {\cal M}$ fixed under the action of $G$, one loses the information about global properties of the moduli space, in particular an orientation, so the localization computations give an answer for $\mathrm{GW}_{g , h}^{\vec{d}, \vec{w}}$ modulo an overall sign depending on $g$, $h$, $\vec{d}$ and $\vec{w}$. Sometimes, the sign can be deduced from the gauge/gravity duality or from mirror symmetry \cite{AganagicVafa, AganagicKlemmVafa}.

%%%%%%%%%%%%%%%%%%%%%%%%%%%%%
%%%%%%%%%%%%%%%%%%%%%%%%%%%%%
%%%%%%%%%%%%%%%%%%%%%%%%%%%%%
\section{Gopakumar-Vafa duality with Lagrangian branes} \label{sec:GVduality}
%%%%%%%%%%%%%%%%%%%%%%%%%%%%%
%%%%%%%%%%%%%%%%%%%%%%%%%%%%%
\subsection{A review of Chern-Simons gauge theory and its observables} \label{sec:CSreview}

In this subsection we review the gauge theory side of the GV duality, namely the CS theory \cite{WittenChernSimons}. We follow the pedagogical review \cite{MarinoReview}.

The dynamics of a $\mathrm{U} (N)$ CS theory living on a manifold $M$ is governed by the action
\begin{equation}
S_{CS} \left[ A \right] = \mathrm{i} \frac{k}{2 \pi} \int_M \mathrm{Tr} \left[ \frac{1}{2} A \wedge \textrm{d}A + \frac{1}{3} A \wedge A \wedge A \right].
\end{equation}
As the action does not depend on the metric on $M$, there are no local observables in the theory. The only observables are the partition function, defining the topological Reshetikhin-Turaev-Witten invariant of $M$
\begin{equation}
Z_{CS} \left( M \right) = \int D A \mathrm{e}^{-S_{CS} \left[ A \right]},
\end{equation}
and the Wilson loops and their products, corresponding to the knot and link invariants of $M$
\begin{equation}
W_{R_1 \cdots R_L} = \left< W_{R_1}^{{\cal K}_1} \cdots W_{R_L}^{{\cal K}_L} \right> = \frac{1}{Z_{CS} \left( M \right)} \int DA \mathrm{e}^{-S_{CS} \left[ A \right]} W_{R_1}^{{\cal K}_1} \cdots W_{R_L}^{{\cal K}_L},
\end{equation}
where
\begin{equation}
W_R^{{\cal K}} \left( A \right) = \mathrm{Tr}_R \left[ \mathrm{\hat{P}} \mathrm{exp} \left( \oint_{{\cal K}} A \right) \right].
\end{equation}
Here $R$ is a representation of the gauge group $G$, ${\cal K}$ is a knot along which the holonomy is computed, and $\mathrm{\hat{P}}$ is a path-ordering operator. All knots are taken to be oriented.

The simplest invariant of a link ${\cal L}$ consisting of two knots ${\cal K}_1$ and ${\cal K}_2$ is the so-called linking number
\begin{equation}
\mathrm{lk} \left( {\cal K}_1 , {\cal K}_2 \right) = \frac{1}{2} \sum\limits_p \epsilon \left( p \right),
\end{equation}
where the sum is over all crossing points, and $\epsilon \left( p \right) = \pm 1$ is a sign associated to the crossings.
This definition can be easily generalized to a link, consisting of an arbitrary number $L$ of knots ${\cal K}_{\alpha}$ with $\alpha = 1, \cdots , L$, by
\begin{equation}
\mathrm{lk} \left( {\cal L} \right) = \sum\limits_{\alpha < \beta} \mathrm{lk} \left( {\cal K}_{\alpha} , {\cal K}_{\beta} \right).
\end{equation}

One way to do computations in CS theory is to use standard perturbation theory. As usual, one starts with finding the classical solutions to the equations of motion for the gauge field $A = A_{\mu}^a T^a \mathrm{d} x^{\mu}$,
\begin{equation}
\frac{\delta S_{CS}}{\delta A_{\mu}^a} = \mathrm{i} \frac{k}{4 \pi} \epsilon^{\mu \nu \rho} F_{\nu \rho}^a = 0,
\end{equation}
which are simply the flat connections on the manifold $M$. Flat connections are in one-to-one correspondence with group homomorphisms
\begin{equation}
\pi_1 \left( M \right) \rightarrow G.
\end{equation}
We are most interested in the manifold $M = S^3$, which has trivial fundamental group, so there is just one classical solution corresponding to $A^a_{\mu} = 0$. In what follows, we restrict ourselves only to this case.

In order to define the VEV of the Wilson loop at the quantum level, one has to introduce a dependence of $W_R^{\cal K}$ on an integer number, called framing. The easiest way to see where it comes from is to consider the Abelian CS theory, namely to take $G = \mathrm{U} (1)$. Evaluation of the correlation function of the Wilson loops corresponding to different knots ${\cal K}_{\alpha}$ and ${\cal K}_{\beta}$, leads at the leading order in perturbation theory to the following integral:
\begin{equation}
\frac{1}{4 \pi} \oint_{{\cal K}_{\alpha}} \mathrm{d} x^{\mu}  \oint_{{\cal K}_{\beta}} \mathrm{d} y^{\nu} \epsilon_{\mu \nu \rho} \frac{\left( x - y \right)^{\rho}}{\vert x - y \vert^3}.
\end{equation}
The result is precisely equal to $\mathrm{lk} \left( {\cal K}_{\alpha} , {\cal K}_{\beta} \right)$,
which is a topological invariant, i.e. it is invariant under smooth deformations of ${\cal K}_{\alpha}$ and ${\cal K}_{\beta}$. On the other hand, 
%contractions of two holonomies corresponding to the same knot ${\cal K}$ involve the integral
when ${\cal K}_{\alpha} = {\cal K}_{\beta} = {\cal K}$, we obtain the integral
\begin{equation}
\phi \left( {\cal K} \right) = \frac{1}{4 \pi} \oint_{{\cal K}} \mathrm{d} x^{\mu}  \oint_{{\cal K}} \mathrm{d} y^{\nu} \epsilon_{\mu \nu \rho} \frac{\left( x - y \right)^{\rho}}{\vert x - y \vert^3},
\end{equation}
which is finite, but is not invariant under deformations of ${\cal K}$. In order to preserve the topological invariance, one has to modify the definition of the self-linking number. This procedure introduces the framing $p$ by replacing $\mathrm{lk} \left( {\cal K} , {\cal K} \right)$ with $\mathrm{lk} \left( {\cal K} , {\cal K}_p \right)$, where ${\cal K}_p$ is a knot which winds $p$ times around ${\cal K}$.\footnote{In order to work with torus knots, one formally defines a rational framing \cite{TorusKnots}, and even an irrational one may be considered \cite{AganagicCostelloMcNamaraVafa}.} The generalization to non-Abelian gauge groups is straightforward but cumbersome.

The VEV of the link ${\cal L}$ consisting of knots ${\cal K}_1, \cdots , {\cal K}_L$ with framings $p_1 , \cdots , p_L$, differs from the same link with trivially framed knots corresponding to $p_1 = \cdots = p_L = 0$ by an overall factor
\begin{equation}
W_{R_1 \cdots R_L}^{p_1, \cdots , p_L} = \mathrm{e}^{2 \pi \mathrm{i} \sum\limits_{\alpha=1}^L p_{\alpha} \mathrm{h}_{R_{\alpha}}} W_{R_1 \cdots R_L}^{0, \cdots , 0}.
\end{equation}
Here $\mathrm{h}_{R_{\alpha}}$ is a group theoretical factor, which for the most interesting case of $G = \mathrm{U} \left( N \right)$ has the form
\begin{equation}
\mathrm{h}_{R} = \frac{1}{2} \frac{\mathrm{Tr}_R \left( T^a T^a \right)}{k + N}.
\end{equation}

An important quantity in the formalism of the GV duality is the generating function of the Wilson loops in arbitrary representations $R$.\footnote{In what follows, we work with the gauge group $\mathrm{U} \left( N \right)$, but in principle one can also consider $\mathrm{O} \left( N \right)$ and $\mathrm{USp} \left( N \right)$ groups.} In order to construct this, let us start by defining a new basis for the Wilson loop operators, labeled by conjugacy classes of the symmetric group.
%As we work with the gauge group $U \left( N \right)$, we denote the holonomy by $U$ for the sake of brevity. 
Let us denote a specific $\mathrm{U} (N)$ holonomy by $U$. Let $\vec{k} = \left( k_1, k_2, \cdots \right)$ be a vector of infinite entries, almost all of which are zero. We also introduce its norm as
\begin{equation}
\vert \vec{k} \vert = \sum\limits_j k_j.
\end{equation}
This vector defines naturally a conjugacy class $C ( \vec{k} )$ of the symmetric group $S_l$ with
\begin{equation}
l = \sum\limits_j j k_j.
\end{equation}
The class $C ( \vec{k} )$ has $k_j$ cycles of length $j$. 

Next we introduce the operator
\begin{equation}
\Upsilon_{\vec{k}} \left( U \right) = \prod\limits_{j = 1}^{\infty} \left( \mathrm{Tr} (U^j) \right)^{k_j},
\end{equation}
which is the Wilson loop operator in the conjugacy class basis. This operator is a linear combination of the operators
$\mathrm{Tr}_R (U)$ labeled by representations $R$,
\begin{equation} \label{GeneratingOperator}
\Upsilon_{\vec{k}} \left( U \right) = \sum\limits_R \chi_R ( C ( \vec{k} ) ) \mathrm{Tr}_R (U).
\end{equation}
Here $\chi_R ( C ( \vec{k} ) )$ is the character of the representation $R$ of the group $S_l$, evaluated in the conjugacy class $C ( \vec{k} )$. One can also invert this relation,
\begin{equation}
\mathrm{Tr}_R (U) = \sum\limits_{\vec{k}} \frac{\chi_R ( C ( \vec{k} ) )}{z_{\vec{k}}} \Upsilon_{\vec{k}} \left( U \right), \quad z_{\vec{k}} \equiv \prod\limits_j k_j! j^{k_j}.
\end{equation}
We denote the expectation value of $\left( \ref{GeneratingOperator} \right)$ by
\begin{equation}
W_{\vec{k}} = \left< \Upsilon_{\vec{k}} \left( U \right) \right>.
\end{equation}

One can introduce another generating function, called the Ooguri-Vafa operator \cite{KnotInvariantsAndTopologicalStrings}, by introducing a $\mathrm{U} \left( M \right)$ source matrix $V$, and defining
\begin{equation} \label{OoguriVafaCS}
Z \left(U , V\right) = \mathrm{exp} \left[ \sum\limits_{w = 1}^{\infty} \frac{1}{w} \mathrm{Tr} (U^w) \mathrm{Tr} (V^w) \right] = 1 + \sum\limits_{\vec{k}} \frac{1}{z_{\vec{k}}} \Upsilon_{\vec{k}} \left( U \right) \Upsilon_{\vec{k}} \left( V \right),
\end{equation}

\noindent which includes (if we consider all possible values of $M$) the Wilson loops in all possible representations.
The vacuum expectation value $Z_{CS} \left( V \right) = \left< Z \left( U , V \right) \right>$ contains information about the VEVs of the Wilson loop operators, and by taking its logarithm, one can define the connected vacuum expectation values $W_{\vec{k}}^{\left( c \right)}$ as
\begin{equation}
\mathrm{F}_{CS} \left( V \right) = \mathrm{log} (Z_{CS} \left( V \right)) = \sum\limits_{\vec{k}} \frac{1}{z_{\vec{k}} !} W_{\vec{k}}^{\left( c \right)} \Upsilon_{\vec{k}} \left( V \right).
\end{equation}

\subsection{Duality to the topological string theory} \label{sec:gravitydualreview}

In this subsection we review the string theory description of the GV duality and the map between observables, following mostly \cite{VafaAlgebraicKnots} and \cite{MarinoReview}.

We first recall how to represent the CS theory on $S^3$ as the topological A-model on the deformed conifold $X_d = T^{*} S^3$.
The deformed conifold may be defined as the subspace of $\mathbb{C}^4$ parameterized by $\mathrm{X}$, $\mathrm{Y}$, $\mathrm{Z}$ and $\mathrm{W}$\footnote{See \cite{CandelasdelaOssa} for a detailed discussion of $S^2 \times S^3$ conifolds.}
\begin{equation} \label{deformedconifold}
\mathrm{X Y - ZW} = \mu^2, \quad  \mu \in \mathbb{R}.
\end{equation}
The K\"{a}hler form $\omega \vert_{X_d}$ on $T^{*} S^3$ is given by the restriction of
\begin{equation}
\omega_d = \mathrm{i} \left( \mathrm{d} \mathrm{X} \wedge \mathrm{d} \overline{\mathrm{X}} +\mathrm{d} \mathrm{Y} \wedge \mathrm{d} \overline{\mathrm{Y}} + \mathrm{d} \mathrm{Z} \wedge \mathrm{d} \overline{\mathrm{Z}} + \mathrm{d} \mathrm{W} \wedge \mathrm{d} \overline{\mathrm{W}} \right)
\end{equation}
to the conifold via (\ref{deformedconifold}).

The zero section of $T^{*} S^3$,
\begin{equation}\label{zerosectionn}
\mathrm{Y} = \overline{\mathrm{X}}, \qquad \mathrm{W} = - \overline{\mathrm{Z}}, \qquad \mathrm{X} \overline{\mathrm{X}} + \mathrm{Z} \overline{\mathrm{Z}} = \mu^2,
\end{equation}
is a Lagrangian submanifold. Thus, one can put a stack of 3-branes on it. A direct computation in topological string field theory \cite{WittenChernString} shows that only the zero modes $\Phi^I_0$ and $\chi^I_0$ of the worldsheet fields $\Phi^I$ and $\chi^I$ contribute to the dynamics,\footnote{The 1-form field $\rho^I_{z, \overline{z}}$ is a conjugate to $\chi^I$ in the canonical formalism, so does not enter the string field.} and the only relevant term in the expansion of the string field $\Psi$ is
\begin{equation}
\Psi=\chi^I_0 A_{I} \left( \Phi^J_0 \right).
\end{equation}
The full string field theory action
\begin{equation}
S = \frac{1}{g_s} \int \mathrm{Tr} \left[ \frac{1}{2} \Psi \star Q_B \Psi + \frac{1}{3} \Psi \star \Psi \star \Psi \right]
\end{equation}
then reduces to a $\mathrm{U} (N)$ CS action with the dictionary
\begin{equation}
\Psi \rightarrow A, \qquad Q_{B} \rightarrow \mathrm{d}, \qquad \star \rightarrow \wedge, \qquad  \int \rightarrow \int_{S^3}, \qquad g_{s} = - \mathrm{i} \frac{2 \pi}{k+N}.
\end{equation}

This result is valid if there are no finite size instantons in the background. In the case of a stack of branes wrapping the zero section of $T^{*} M$, there are no instantons \cite{WittenChernString}, but for more general configurations they can appear. Let us consider a Calabi-Yau manifold $X$ together with some Lagrangian submanifolds $L_i \subset X$, with $N_i$ 3-branes wrapped over $L_i$. In this case, the spacetime description of topological open strings has two contributions. First of all, we have the contributions of degenerate holomorphic curves. These are captured by $\mathrm{U}(N_i)$ CS theories on the manifolds $L_i$, following the same mechanism that we described for $T^{*} S^3$. However, for a general Calabi-Yau $X$, we may also have open string instantons contributing to the spacetime description, which are embedded holomorphic Riemann surfaces with boundaries ending on the Lagrangian submanifolds $L_i$. An open string instanton $\beta$ intersects the $L_i$ along one-dimensional curves ${\cal K}_i$, which are in general knots inside $L_i$. We know from (\ref{coupledaction}) that the boundary of such an instanton gives the Wilson loop insertion in the spacetime action. 
%We can then take into account the contributions of all instantons by including the corresponding CS theories $S_{CS} \left( A_i \right)$, which account for the degenerate worldsheets coupled in an appropriate way to the holomorphic instantons. 
The spacetime action thus has the form
\begin{equation}
S=\sum\limits_{i} S_{CS} \left( A_i \right)+\sum\limits_{\beta} \mathrm{e}^{-\int_{\beta} \omega} \prod_{i} \mathrm{Tr} \left[ \mathrm{\hat{P}} \mathrm{exp} \left( \oint_{{\cal K}_i} A_i \right) \right],
\end{equation}
where we rescaled $\omega \rightarrow \omega / \mathrm{T}$ (where $\mathrm{T}$ is the string tension). These instanton corrections are crucial for the GV duality to work, and they will play an important role in our generalization.

It is well known \cite{CandelasdelaOssa} that there is a geometric transition preserving the asymptotic form of $T^{*} S^3$. 
As we mentioned in the \nameref{sec:intro}, the manifold $T^{*} S^3$ can be thought as an $S^2 \times S^3$ conifold with a non-vanishing $S^3$ at the tip. The geometric transition in this geometry is done in two steps: shrinking the non-vanishing 3-cycle to zero, and blowing up the vanishing $S^2$. The second step is called `resolution' of the singularity. In this new geometry there are no Lagrangian cycles, and thus there cannot be any 3-branes, while there is now a non-trivial 2-cycle that closed strings can wrap. It is natural to guess, given our understanding of the AdS/CFT correspondence, that these backgrounds with and without branes are equivalent, and this was first conjectured in \cite{GV}. 

In order to prove the GV duality, one has to prove an equality of the partition functions of a $\mathrm{U} (N)$ CS theory on $S^3$ and the topological string A-model on the resolved conifold ${\cal O} \left( - 1 \right) \oplus {\cal O} \left( - 1 \right) \rightarrow \mathbb{CP}^1$, and then map also the Wilson loop observables.
The partition function of the topological A-model on the resolved conifold was first computed in \cite{TopStringsMTheory1, TopStringsMTheory2}, where the embedding of topological strings into eleven-dimensional M-theory was used. It was shown that this partition function is indeed equal to the partition function of a $\mathrm{U} (N)$ CS theory on $S^3$, if one identifies the rank of the gauge group $N$, and the CS level $k$, with the string coupling $g_s$,\footnote{Note that the string coupling stays the same after the geometric transition.} and with the symplectic volume of the resolved 2-cycle $\mathrm{t}_r$, as
\begin{equation} \label{dictionary}
\begin{gathered}
g_s = - \mathrm{i} \frac{2 \pi }{k+N}, \\ \mathrm{t}_r \equiv \int_{S^2} \left( \omega +\mathrm{i} B \right) = - \mathrm{i} \frac{2 \pi N}{k+N}.
\end{gathered}
\end{equation}

Next we review how to include the Wilson loops of the CS theory into the GV duality. We first show how to incorporate them into the A-model on $T^{*} S^3$, following \cite{KnotInvariantsAndTopologicalStrings}.
A knot ${\cal K}$ is given by $q \left( s \right) \in S^3$, $s \in \left[ 0, 2 \pi \right)$. One associates a non-compact Lagrangian manifold $L$ to each knot as follows. At each point $q(s)$ on the loop, we consider the 2-dimensional subspace $T_q^{*} S^3$ of the phase space $T^*S^3$, which is the cotangent space to $S^3$ at the point $q$, spanned by the momenta $p$ orthogonal to $\frac{\mathrm{d} q}{\mathrm{d} s}$. By doing this all along the loop, we can define a 3-cycle with the topology $\mathbb{R}^2 \times S^1$ as the union of these subspaces,
\begin{equation}
L=\left\{ \left(q \left( s \right), p \right) \Bigg| \ p_i \frac{\mathrm{d} q^i}{\mathrm{d} s}=0, \quad 0 \leq s < 2 \pi \right\}.
\end{equation}

Now let us wrap $M$ 3-branes on this cycle. We then have a $\mathrm{U} (M)$ CS theory on $L$, along with a $\mathrm{U} (N)$ CS theory on $S^3$. Let us denote the gauge field living on $S^3$ as $A$, and the one living on $L$ as ${\cal A}$. There is also an additional sector of bi-fundamental strings stretched between the two stacks of branes, localized to their one-dimensional intersection. It has the following action \cite{KnotInvariantsAndTopologicalStrings}:
\begin{equation}
S=\int_{\cal K} \overline{\phi} \left( d + A - {\cal A} \right) \phi.
\end{equation}
The worldsheets of the bi-fundamental strings can be considered as holomorphic instantons with the topology of the annulus, having zero width.
After integrating out the field $\phi$, one is left with the following insertion in the path integral over the gauge field $A$:
\begin{equation} \label{OoguriVafaStrings}
Z \left( U , V \right)= \mathrm{exp} \left[ \sum\limits_{w=1}^{\infty} \frac{1}{w} \mathrm{Tr} (U^w) \mathrm{Tr} (V^w) \right],
\end{equation}
where $U$ and $V$ are the path-ordered holonomies along $q \left( s \right)$ of $A$ and ${\cal A}$, respectively.

Since the branes wrapping $L$ are non-compact, one can consider them as a non-dynamical probe. Their only mission is to `pull' the holonomy $V$ from infinity, where it is fixed as a boundary condition (the CS action on $L$ guarantees that $V$ is equal to the holonomy of $\cal A$ at infinity). Note that the operator $\left( \ref{OoguriVafaStrings} \right)$ exactly coincides with the expression $\left( \ref{OoguriVafaCS} \right)$ that we obtained for the generating function of the Wilson loops in all representations in the CS theory, which completes the dictionary between the CS theory on $S^3$ and the topological A-model on $T^* S^3$.

There is also an alternative derivation of the A-model partition function from the annulus worldsheet perspective, which uses the machinery of the equivariant localization that was discussed in detail in section \ref{sec:equivariantlocalization}.\footnote{See \cite{DiaconescuFloreaGrassi1,DiaconescuFloreaGrassi2} for examples of such computations.}

In order to take the Wilson loops into account on the closed string side, one has to push the non-compact 3-branes through the geometric transition. This procedure may seem non-trivial, because when the $S^3$ at the tip of the conifold vanishes, the stack of non-compact branes $L$ becomes singular, so in order to proceed, we should smoothen the brane singularity somehow. Such a smoothing can be done by lifting the non-compact cycle $L$ off the zero section of $T^* S^3$. Such a lift exists \cite{VafaAlgebraicKnots}, because the dimensionality of the moduli space of $L$ equals the dimensionality of $H_1 \left( L \right)$, which for the cycle of the $\mathbb{R}^2 \times S^1$ topology, is obviously equal to $1$. It turns the bi-fundamental string worldsheet into a finite size cylinder ${\cal C}$, modifying the operator $Z \left( U , V \right)$ to
\begin{equation}
Z \left( U , V \right)= \mathrm{exp} \left[ \sum\limits_{w=1}^{\infty} \frac{\mathrm{e}^{-w S_{\cal C}}}{w} \mathrm{Tr} (U^w) \mathrm{Tr} (V^w) \right],
\end{equation}
where $S_{\cal C}$ is the symplectic volume of ${\cal C}$.

After the lift is done, the image of the cycle $L$ under the geometric transition to the resolved conifold, can be easily found. The dual cycle is given by the same equation as $L$, but now the coordinates $\mathrm{X}$, $\mathrm{Y}$, $\mathrm{Z}$ and $\mathrm{W}$ must be considered as parameterizing not the embedding of the deformed conifold $T^* S^3$ into $\mathbb{C}^4$, but the following embedding of the resolved conifold ${\cal O} \left( - 1 \right) \oplus {\cal O} \left( - 1 \right) \rightarrow \mathbb{CP}^1$ into $\mathbb{C}^4 \times \mathbb{CP}^1$:
\begin{align}
\begin{split}
\mathrm{X} \lambda_1 + \mathrm{Z} \lambda_2 &=0, \\
\mathrm{W} \lambda_1 + \mathrm{Y} \lambda_2 &=0,
\end{split}
\end{align}
where $\mathrm{X}, \mathrm{Y}, \mathrm{Z}, \mathrm{W}$ parameterize $\mathbb{C}^4$, while $\lambda_1$ and $\lambda_2$ are homogeneous coordinates on $\mathbb{CP}^1$.

There is no compact stack of 3-branes on the resolved side of the duality, and thus no holomorphic cylinders, but there are holomorphic discs ending on $L$, which do not exist on the deformed side. The equality of the cylinder amplitudes on the deformed conifold and the disc amplitudes on the resolved conifold is the main statement of the GV duality, and it was checked for a wide class of knots (and corresponding Lagrangian cycles $L$) \cite{VafaAlgebraicKnots}.

\section{Adding topological matter fields and their duality} \label{sec:addingmatter}

In this section we generalize the GV duality by adding 5-branes to the duality of section \ref{sec:GVduality}, giving extra matter fields coupled to the CS theory in a partially topological manner. For simplicity we discuss a single 5-brane (here and in the remainder of this paper) but the generalization to more than one brane is trivial.

\subsection{Open string side} \label{openside}

On the open string side, on which there are $N$ 3-branes wrapping the zero section of the deformed conifold $X_d$, the coupling to the topological matter was described in \cite{AganagicCostelloMcNamaraVafa}.

As we discussed in section \ref{sec:GVduality}, the deformed conifold is the subspace of $\mathbb{C}^4$ parameterized by $\mathrm{X}$, $\mathrm{Y}$, $\mathrm{Z}$ and $\mathrm{W}$
\begin{equation}
\mathrm{X Y - ZW} = \mu^2, \quad  \mu \in \mathbb{R}.
\end{equation}
The K\"{a}hler form $\omega_d \vert_{X_d}$ on $T^{*} S^3$ can be chosen to be the restriction of
\begin{equation}
\omega_d = \mathrm{i} \left( \mathrm{d} \mathrm{X} \wedge \mathrm{d} \overline{\mathrm{X}} +\mathrm{d} \mathrm{Y} \wedge \mathrm{d} \overline{\mathrm{Y}} + \mathrm{d} \mathrm{Z} \wedge \mathrm{d} \overline{\mathrm{Z}} + \mathrm{d} \mathrm{W} \wedge \mathrm{d} \overline{\mathrm{W}} \right)
\end{equation}
to $X_d$,
and we have
%As we will see in section \ref{sec:PTSMReview}, the field theory living on the stack of 
$N$ 3-branes wrapping the base of the conifold
\begin{equation}\label{zerosection}
\mathrm{Y} = \overline{\mathrm{X}}, \qquad \mathrm{W} = - \overline{\mathrm{Z}}, \qquad \mathrm{X} \overline{\mathrm{X}} + \mathrm{Z} \overline{\mathrm{Z}} = \mu^2.
\end{equation}

We would like to add to this background a coisotropic 5-brane, leading to extra matter fields from 3-5 strings. The 5-brane depends on the choice of a THF on its worldvolume. One way to construct such a coisotropic brane is the following \cite{AganagicCostelloMcNamaraVafa}. The THF of the 5-brane induces a THF on the $S^3$ worldvolume of the 3-branes, which the PTCSM theory will depend on (as we will discuss later in section \ref{sec:PTSMReview}). Let us invert the logic, and begin by choosing a THF on the $S^3$, parameterizing the 3-brane worldvolume by $(\tau, u, \overline{u})$ where the vector field in the $\tau$ direction is
%
%depends on a THF structure. We parameterize the vector field which generates it as
\begin{equation}
\mathrm{V}' = \frac{\partial}{\partial \tau} = \mathrm{i} \frac{1}{b} \left( \mathrm{X} \frac{\partial}{\partial \mathrm{X}}-\overline{\mathrm{X}} \frac{\partial}{\partial \overline{\mathrm{X}}} \right)+ \mathrm{i} b \left( \mathrm{Z} \frac{\partial}{\partial \mathrm{Z}}-\overline{\mathrm{Z}} \frac{\partial}{\partial \overline{\mathrm{Z}}} \right).
\end{equation}
Here $b$ is a positive constant, and $b^2$ is irrational. 

We can then describe the deformed conifold $T^* S^3$ using the coordinates $(\tau, u, \overline{u})$ and their conjugate momenta $(p_{\tau}, p_u, p_{\overline{u}})$ , and define the coisotropic brane as $Y_d = T^*_{\mathrm{M}_d, \perp} S^3$ with
\begin{equation}
T^*_{\mathrm{M}_d, \perp} S^3: \ p_{\tau} = \mathrm{M}_d.
\end{equation}
%where $p_{\tau}$ is the momentum conjugate to the coordinate $\tau$. 
%The curvature $\mathfrak{F}_d$ has the form
This is a coisotropic brane when we choose the gauge field strength
\begin{equation} \label{fdeformed}
%\mathfrak{F}_d = \mathrm{d} y \wedge \mathrm{d} p_x - \mathrm{d} x \wedge \mathrm{d} p_y,
\mathfrak{F}_d = \mathrm{i} \left( \mathrm{d} u \wedge \mathrm{d} p_u - \mathrm{d} {\overline{u}} \wedge \mathrm{d} p_{\overline{u}} \right),
\end{equation}
which manifestly satisfies the conditions \eqref{5branes} with the canonical symplectic form
\begin{equation}
\omega = 
%\mathrm{d} x \wedge \mathrm{d} p_x + \mathrm{d} y \wedge \mathrm{d} p_y 
\mathrm{d} u \wedge \mathrm{d} p_u + \mathrm{d} {\overline{u}} \wedge \mathrm{d} p_{\overline{u}}
+ \mathrm{d} \tau \wedge \mathrm{d} p_{\tau}.
\end{equation}
%The coordinates $x$ and $y$ here are defined as $u = x + \mathrm{i} y$, and $\overline{u} = x - \mathrm{i} y$, where $u$ and $\overline{u}$ are the THF-adapted coordinates we discussed in section \ref{sec:THF}. 
Note that we can choose the THF coordinates in many different ways, since the theory does not depend on the $S^3$ metric and there is no condition that the THF is compatible with the metric. Each choice will give a different $\mathfrak{F}_d$.
%Note that these coordinates are not unique if one doesn't require the metric compatibility of the THF \cite{GeometrySUSYPartition}. We do not do this, because our theory is topological, so we are free to choose arbitrary metric leading to the following projectors $(\Pi_{\tau}, \Pi_{u}, \Pi_{\overline{u}})$ to the coordinates $(\tau, u, \overline{u})$:
%\begin{align}
%\begin{split}
%\left( \Pi_{\tau} \right)^{\mu}_{\ \nu} &= \Xi^{\mu} \Xi_{\nu}, \\
%\left( \Pi_{u} \right)^{\mu}_{\ \nu} &= \frac{1}{2} \left( \delta^{\mu}_{\nu} - \mathrm{i} J^{\mu}_{\ \nu} - \Xi^{\mu} \Xi_{\nu} \right), \\ \left( \Pi_{\overline{u}} \right)^{\mu}_{\ \nu} &= \frac{1}{2} \left( \delta^{\mu}_{\nu} + \mathrm{i} J^{\mu}_{\ \nu} - \Xi^{\mu} \Xi_{\nu} \right),
%\end{split}
%\end{align}
%where $\Xi$ is a 1-form dual to the vector $\mathrm{V}'$, defined only up to addition of any 1-form, having no component in $\tau$-direction, and $J$ is a complex structure on the space of leaves of the THF, which is given by $J_{\mu}^{\ \nu} = - \epsilon_{\mu}^{\ \nu \rho} \Xi_{\rho}$, obeying $J_{\mu}^{\ \nu} J_{\nu}^{\ \rho} = -\delta_{\mu}^{\rho} + \Xi_{\mu} \Xi^{\rho}$. The ambiguity in definition of $\Xi$ implies an ambiguity in definition of $\mathfrak{F}_d$, but 
Since the PTCSM theory depends only on the THF, we expect that the partition function of the topological string theory depends only on $b$ and not on the precise value of $\mathfrak{F}_d$. That is exactly what we will find in section \ref{sec:annulus}.

Changing coordinates from $\left(u , \overline{u} , \tau \ ; \ p_u , p_{\overline{u}} , p_{\tau} \right)$ to $\left( \mathrm{X}, \mathrm{Y}, \mathrm{Z}, \mathrm{W} \right)$, we can write the 5-brane embedding as
\begin{equation} \label{5brane}
Y_d: \ \mathrm{H}_d \left( \mathrm{X} , \mathrm{Y} , \mathrm{Z} , \mathrm{W} \right) = \frac{1}{b} \left( \mathrm{X} \overline{\mathrm{X}} - \mathrm{Y} \overline{\mathrm{Y}} \right)  +b \left( \mathrm{Z} \overline{\mathrm{Z}} - \mathrm{W} \overline{\mathrm{W}} \right) = \mathrm{M}_d.
\end{equation}
The parameter $\mathrm{M}_d$ is real and positive, and governs the masses of 3-5 and 5-3 strings. This follows from the fact that the 5-brane \eqref{5brane} touches the base of the conifold, where the 3-branes sit, only if $\mathrm{M}_d$ vanishes.
The function $\mathrm{H}_d$ \eqref{5brane} gives rise to a vector field $\mathrm{V}_d=\frac{\partial}{\partial \tau}$ on the 5-brane, generating the THF such that
\begin{equation}
\mathrm{d} \mathrm{H}_d = - \imath_{\mathrm{V}_d} \omega_d.
\end{equation}
For $\mathrm{H}_d$ of \eqref{5brane} this field has the form
\begin{equation} \label{Vectorfield}
\mathrm{V}_d = \mathrm{i} \frac{1}{b} \left( \mathrm{X} \frac{\partial}{\partial \mathrm{X}} - \mathrm{Y} \frac{\partial}{\partial \mathrm{Y}} \right)+\mathrm{i} b \left(\mathrm{Z} \frac{\partial}{\partial \mathrm{Z}} - \mathrm{W} \frac{\partial}{\partial \mathrm{W}} \right) + \mathrm{c.c.}
\end{equation}

We ignore the 5-5 strings in our discussion, because the 5-brane is non-compact, and thus we expect these strings to be non-dynamical. For a general geometry of the 5-brane, the action of the effective field theory on the coisotropic brane $Y$ proposed in \cite{AganagicCostelloMcNamaraVafa} is
\begin{equation}
S = \frac{1}{g_s} \int_Y \mathrm{CS}_5 \left( {\cal A} + \mathrm{i} \mathfrak{a} \right),
\end{equation}
where $\mathrm{CS}_5$ is the Chern-Simons 5-form, and $\mathfrak{a}$ is defined locally as
\begin{equation}
\omega = \mathrm{d} \mathfrak{a}.
\end{equation}
Variation of this action gives precisely the string worldsheet boundary condition \eqref{5branes}, with the solution \eqref{fdeformed} for $\mathfrak{F} = \mathrm{d} {\cal A}$.
%(\refeq{5branes})

The 3-5 strings add matter fields coupled to the CS theory, and it was argued in \cite{AganagicCostelloMcNamaraVafa} that they are coupled in a specific partially topological manner, depending on the THF, that we will review in section \ref{sec:fieldtheory}. In any case, the string field theory of the deformed conifold with $N$ 3-branes and a 5-brane \eqref{5brane} implicitly defines a PTCSM theory.

\subsection{Closed string side} \label{closedside}

To find the closed string dual of the aforementioned configuration, we need to follow the coisotropic 5-brane through the Gopakumar-Vafa duality.

As mentioned in section \ref{sec:GVduality}, the resolved conifold $X_r = {\cal O} \left( - 1 \right) \oplus {\cal O} \left( - 1 \right) \rightarrow \mathbb{CP}^1$ is given by the following subset of $\mathbb{C}^4 \times \mathbb{CP}^1$ \cite{CandelasdelaOssa}:
\begin{align} \label{ResolvedConifold}
\begin{split}
\mathrm{X} \lambda_1 + \mathrm{Z} \lambda_2 &=0, \\
\mathrm{W} \lambda_1 + \mathrm{Y} \lambda_2 &=0,
\end{split}
\end{align}
where $\mathrm{X} , \mathrm{Y} , \mathrm{Z} , \mathrm{W}$ parameterize $\mathbb{C}^4$, while $\lambda_1$ and $\lambda_2$ are homogeneous coordinates on $\mathbb{CP}^1$. Note that when not all of the coordinates on $\mathbb{C}^4$ are simultaneously zero, the coordinates on $\mathbb{CP}^1$ are fixed uniquely, but if all of $\mathrm{X} , \mathrm{Y} , \mathrm{Z} , \mathrm{W}$ vanish, the coordinates $\lambda_1$ and $\lambda_2$ are unconstrained, parameterizing an entire $\mathbb{CP}^1$. We use the two following coordinate charts on $X_r$:
\begin{align}
\begin{split}
& H_{+}: \ \left( \mathrm{X} , \mathrm{W} , \xi \right), \quad \xi = \frac{\lambda_1}{\lambda_2}, \quad \lambda_2 \neq 0, \\
& H_{-}: \ ( \mathrm{Y} , \mathrm{Z} , \eta ), \quad \eta = \frac{\lambda_2}{\lambda_1}, \quad \lambda_1 \neq 0.
\end{split}
\end{align}
The K\"{a}hler form on the resolved conifold $\omega \vert_{X_r}$ is obtained by the restriction of the 2-form
\begin{equation}
\omega_r = \mathrm{i} \left( \mathrm{d} \mathrm{X} \wedge \mathrm{d} \overline{\mathrm{X}} +\mathrm{d} \mathrm{Y} \wedge \mathrm{d} \overline{\mathrm{Y}} + \mathrm{d} \mathrm{Z} \wedge \mathrm{d} \overline{\mathrm{Z}} + \mathrm{d} \mathrm{W} \wedge \mathrm{d} \overline{\mathrm{W}} \right) + \mathrm{t}_r \omega_{\mathbb{CP}^1}
\end{equation}
from $\mathbb{C}^4 \times \mathbb{CP}^1$ to $X_r$. Here $\mathrm{t}_r$ corresponds to the volume of the $\mathbb{CP}^1$ at the apex of $X_r$, and $\omega_{\mathbb{CP}^1}$ is the 2-sphere K\"{a}hler form
\begin{equation}
\omega_{\mathbb{CP}^1} = \mathrm{i} \frac{1}{2 \pi} \frac{\mathrm{d} \xi \wedge \mathrm{d} \overline{\xi}}{\left( 1 + \xi \overline{\xi} \right)^2} = \mathrm{i} \frac{1}{2 \pi} \frac{\mathrm{d} \eta \wedge \mathrm{d} \overline{\eta}}{\left( 1 + \eta \overline{\eta} \right)^2}.
\end{equation}

We choose the 5-brane geometry in the resolved conifold in such a way that the vector field $\mathrm{V}_r$ generating the THF on the 5-brane, being continued to the whole $X_r$, generates a symmetry of the target space \eqref{ResolvedConifold}, and such that it coincides with $\mathrm{V}_d$ at infinity. It takes the form
\begin{equation}
\mathrm{V}_r = \mathrm{i} \frac{1}{b} \left( \mathrm{X} \frac{\partial}{\partial \mathrm{X}} - \mathrm{Y} \frac{\partial}{\partial \mathrm{Y}} \right)+\mathrm{i} b \left(\mathrm{Z} \frac{\partial}{\partial \mathrm{Z}} - \mathrm{W} \frac{\partial}{\partial \mathrm{W}} \right) + \mathrm{i} q \xi \frac{\partial}{\partial \xi} + \mathrm{c.c.},
\end{equation}
where we introduced the notation
\begin{equation}
q = b - \frac{1}{b}.
\end{equation}
Now, using the equation
\begin{equation}
\mathrm{d} \mathrm{H}_r = - \imath_{\mathrm{V}_r} \omega_r,
\end{equation}
and requiring the result to be symmetric under $b \rightarrow 1/b$, we find that the equation for the position of the 5-brane $\mathrm{H}_r$ is determined uniquely, and in the $H_{+}$ chart it has the form
\begin{equation} \label{n5brane}
Y_r: \ \mathrm{H}_r = \frac{1}{b} \left( \mathrm{X} \overline{\mathrm{X}} - \mathrm{Y} \overline{\mathrm{Y}} \right) + b \left( \mathrm{Z} \overline{\mathrm{Z}} - \mathrm{W} \overline{\mathrm{W}} \right) - \frac{q}{2 \pi} \mathrm{t}_r \left[ \frac{1}{1 + \xi \overline{\xi}} - \frac{1}{2} \right] = \mathrm{M}_r.
\end{equation}
Note that this transforms nicely as we go from $H_+$ to $H_-$, and asymptotically it is the same as \eqref{5brane} (we expect the geometric transition not to affect the behavior at infinity).
We have not been able to find an explicit solution $\mathfrak{F}_r$ for the 5-brane field, such that \eqref{n5brane} is a coisotropic 5-brane, but we expect that it should be possible to do this.
We will see in section \ref{sec:disc} that we do not need to know the precise value of $\mathfrak{F}_r$ in order to compute the partition function, 
%the partition function does not depend on the value of the Chan-Paton curvature on the closed leaves of the THF, 
so we will consider there the most general ansatz for it allowed by the symmetries of the problem.

Since we started with a 5-brane, disjoint from the zero section of $X_d$ that becomes singular in the transition, we expect to have a brane lifted off from the zero section of $X_r$ after the transition as well, at least for small $\mathrm{t}_r$. 
%It means that we should take $\mathrm{M}_r$ sufficiently large in order the equation
The 5-brane \eqref{n5brane} intersects the $\mathbb{CP}^1$ at the tip when $\mathrm{M}_r \leq |\frac{q \mathrm{t}_r}{4\pi}|$, and we will focus on the case of large enough values of $\mathrm{M}_r$, so that this does not arise.
%\begin{equation}
%\mathrm{H}_r \left(0, 0, 0, 0, l \right) = - \frac{q}{2 \pi} \mathrm{t}_r \left[ \frac{1}{1 + \vert l \vert^2} - \frac{1}{2} %right] = \mathrm{M}_r,
%\end{equation}
%and its counterpart in the $H_{-}$ chart
%\begin{equation}
%\mathrm{H}_r (0, 0, 0, 0, \tilde{l}) = \frac{q}{2 \pi} \mathrm{t}_r \left[ \frac{1}{1 + \vert \tilde{l} \vert^2} - \frac{1}{2} \right] = \mathrm{M}_r,
%\end{equation}
%do not have solutions for $l$ and $\tilde{l}$ respectively.

Our main conjecture is that the PTCSM theory is dual to the topological string on the resolved conifold, with the same mapping of parameters as in the Gopakumar-Vafa duality, and with an extra 5-brane \eqref{n5brane}. We will provide evidence for this in the next sections, by comparing the subleading order partition functions between the two topological string theories described above. This will also enable us to match the parameters $\mathrm{M}_d$ and $\mathrm{M}_r$.

\section{The open string side of the duality} \label{sec:annulus}

In this section, we compute the annulus topological A-model amplitude on the deformed conifold with $N$ Lagrangian 3-branes and a coisotropic 5-brane.

\subsection{The worldsheet geometry} \label{sec:DeformedGeometricSetUp}

We described the form of the deformed conifold $X_d$ and of the coisotropic 5-brane $Y_d$ \eqref{5brane} in section \ref{openside}.

In order to be able to apply the equivariant localization, we demand that the brane embeddings, the worldsheet annuli, and the field configurations are invariant under the action of a symmetry group, acting on the manifold $X_d$. We take this group $T_d$ to be $\mathbb{C}^{\times}$ generated by the action of $\mathrm{V}_d$, continued from the brane $Y_d$ to the whole space $X_d$. It acts on the coordinates as
\begin{equation} \label{vaction}
( \mathrm{X} , \mathrm{Y} , \mathrm{Z} , \mathrm{W} ) \rightarrow ( \mathrm{e}^{\mathrm{i} \frac{1}{b} \tau} \mathrm{X}, \mathrm{e}^{- \mathrm{i} \frac{1}{b} \tau} \mathrm{Y}, \mathrm{e}^{ \mathrm{i} b \tau} \mathrm{Z}, \mathrm{e}^{ - \mathrm{i} b \tau} \mathrm{W} ),
\end{equation}
so the coordinates have the following weights under the action generated by $T_d$:
\begin{equation} \label{Tweights}
\lambda_{\mathrm{X}} = \frac{1}{b}, \qquad \lambda_{\mathrm{Y}} = - \frac{1}{b}, \qquad \lambda_{\mathrm{Z}} = b, \qquad \lambda_{\mathrm{W}} = - b.
\end{equation}

As described in section \ref{sec:equivariantlocalization}, the partition function is the sum over specific holomorphic maps (worldsheet instantons). On the deformed conifold, the instantons we are interested in are the holomorphic annuli with one boundary lying on the base of the conifold, and the other one lying on the 5-brane, which are closed leaves of the THF. The bulk of these annuli must be invariant under the action of $T_d$. If $b^2$ is irrational, there are only two compact orbits of the THF-generating field $\mathrm{V}_d$ on the 5-brane. They are given by
\begin{equation} \label{xcycle5}
{\cal B}_5^{\mathrm X}: \ \mathrm{Z} = \mathrm{W} = 0, \quad \mathrm{Y} = a \overline{\mathrm{X}}, \quad a = \frac{b \mathrm{M}_d}{2 \mu^2} \left[ \sqrt{1+ \left( \frac{2 \mu^2}{b \mathrm{M}_d} \right)^2}-1 \right],
\end{equation}
and
\begin{equation} \label{zcycle5}
{\cal B}_5^{\mathrm{Z}}: \ \mathrm{X} = \mathrm{Y} = 0, \quad \mathrm{W} = c \overline{Z}, \quad c = - \frac{\mathrm{M}_d}{2 b \mu^2} \left[ \sqrt{1+ \left(\frac{2 b \mu^2}{\mathrm{M}_d} \right)^2 }-1 \right].
\end{equation}

There are two corresponding holomorphic $\mathrm{V}_d$-invariant cylinders:
\begin{align}
\begin{split}
& {\cal C}^{\mathrm{X}}: \ \mathrm{X Y} = \mu^2, \quad \mathrm{Z} = \mathrm{W} = 0, \\
& {\cal C}^{\mathrm{Z}}: \ - \mathrm{Z W} = \mu^2, \quad \mathrm{X} = \mathrm{Y} = 0,
\end{split}
\end{align}
which intesect the 5-brane along ${\cal B}_5^{\mathrm X}$ and ${\cal B}_5^{\mathrm{Z}}$, respectively, and which intersect the zero section \eqref{zerosection} along the two following cycles:
\begin{equation} \label{xcycle3}
{\cal B}_3^{\mathrm{X}}: \ {\mathrm{Z}} = {\mathrm{W}} = 0, \quad {\mathrm{X}} = \overline{{\mathrm{Y}}},
\end{equation}
and
\begin{equation} \label{zcycle3}
{\cal B}_3^{\mathrm{Z}}: \ {\mathrm{X}} = {\mathrm{Y}} = 0, \quad {\mathrm{W}} = - \overline{{\mathrm{Z}}}.
\end{equation}
Note that these two cycles are precisely the closed leaves of the THF on $S^3$, generated by the field $\mathrm{V}^{\prime}_d$, obtained by the restriction of $\mathrm{V}_d$ to the base of the conifold, and given by
\begin{equation}
\mathrm{V}^{\prime}_d = \mathrm{i} \frac{1}{b} \left( \mathrm{X} \frac{\partial}{\partial \mathrm{X}}-\overline{\mathrm{X}} \frac{\partial}{\partial \overline{\mathrm{X}}} \right)+ \mathrm{i} b \left( \mathrm{Z} \frac{\partial}{\partial \mathrm{Z}}-\overline{\mathrm{Z}} \frac{\partial}{\partial \overline{\mathrm{Z}}} \right).
\end{equation}
That is how the THF of the 5-brane is induced on the base of $T^*S^3$ by the instantons.

In the vicinity of ${\cal B}_5^{\mathrm X}$ given by $\left( \ref{xcycle5} \right)$, the coordinate $\mathrm{X}$ is non-zero, so we can express $\mathrm{Y}$ as
\begin{equation}
\mathrm{Y} = \frac{\mu^2 + \mathrm{Z W}}{\mathrm{X}},
\end{equation}
and work with coordinates $\left( \mathrm{X}, \mathrm{Z}, \mathrm{W} \right)$. Similarly, in the neighborhood of ${\cal B}_5^{\mathrm Z}$, one can express the coordinate $\mathrm{W}$ in terms of $\left( \mathrm{X}, \mathrm{Y}, \mathrm{Z} \right)$. Using these two sets of coordinates, it is easy to compute the actions of instantons.
The symplectic volumes of the holomorphic cylinders ${\cal C}^{\mathrm{X}}$ and ${\cal C}^{\mathrm{Z}}$ are
\begin{align} \label{volumes}
\begin{split}
& V_{{\cal C}^{\mathrm{X}}} = \int_{{\cal C}^{\mathrm{X}}} \omega_d = 2 \pi b \mathrm{M}_d, \\
& V_{{\cal C}^{\mathrm{Z}}} = \int_{{\cal C}^{\mathrm{Z}}} \omega_d = 2 \pi \frac{\mathrm{M}_d}{b}.
\end{split}
\end{align}
In principle, there are also $T_d$-invariant holomorphic discs, ending on the closed leaves of the THF on the 5-brane in the deformed conifold geometry. They are given by
\begin{align}
\begin{split}
& {\cal D}^{\mathrm{X}}_d: \ \mathrm{X \overline{X}} \leq b \mathrm{M}_d, \quad \mathrm{Y} = \mathrm{Z} = \mathrm{W} = 0, \\
& {\cal D}^{\mathrm{Z}}_d: \ \mathrm{Z \overline{Z}} \leq \frac{\mathrm{M}_d}{b}, \quad \mathrm{X} = \mathrm{Y} = \mathrm{W} = 0.
\end{split}
\end{align}
However, their volumes blow up at their centers
\begin{align}
\begin{split}
& V_{{\cal D}^{\mathrm{X}}_d} = \mathrm{i} \int\limits_{\mathrm{X \overline{X}} \leq b \mathrm{M}_d} \mathrm{ d X \wedge d \overline{X}} \left[ 1 + \frac{\mu^4}{\left( \mathrm{X \overline{X}} \right)^2} \right] \rightarrow \infty, \\ & V_{{\cal D}^{\mathrm{Z}}_d} = \mathrm{i} \int\limits_{\mathrm{Z \overline{Z}} \leq \frac{\mathrm{M}_d}{b}} \mathrm{ d Z \wedge d \overline{Z}} \left[ 1 + \frac{\mu^4}{\left( \mathrm{Z \overline{Z}} \right)^2} \right] \rightarrow \infty,
\end{split}
\end{align}
so they do not contribute. Thus, the leading corrections to the partition function of the CS theory in this configuration come from the annulus worldsheets described above.
%There are known to be no non-trivial disc instantons ending on the zero section due to the ``vanishing theorem" %\cite{WittenChernString}.
Note that the volumes \eqref{volumes} of the holomorphic annuli are independent of the size of the zero section of $T^{*}S^3$ given by $\mu$, as they must be in a topological string theory.

The A-model amplitudes for these two annuli are very similar, and the amplitude related to ${\cal C}^{\mathrm{Z}}$ may be obtained from the one related to ${\cal C}^{\mathrm{X}}$ by a substitution $b \rightarrow 1/b$, so in what follows, we concentrate only on ${\cal C}^{\mathrm{X}}$.
As we discussed in section \ref{sec:openAmodel}, there must be a non-zero 2-form field $\mathfrak{F}_d$ on the 5-brane obeying $\left( \ref{5branes} \right)$. Evaluating the K\"{a}hler form $\omega_d$ on the cycle ${\cal B}_5^{\mathrm X}$ gives
\begin{equation}
\sigma_d = \mathrm{i} \left( \mathrm{d} \mathrm{Z} \wedge \mathrm{d} \overline{\mathrm{Z}} + \mathrm{d} \mathrm{W} \wedge \mathrm{d} \overline{\mathrm{W}} \right).
\end{equation}

As we discussed in section \ref{openside}, the expression for $\mathfrak{F}_d$ is not unique for a given $b$, but the results should not depend on the precise choice, so we consider the most general real $T_d$-invariant 2-form that we can write down at the location of the cycle ${\cal B}_5^{\mathrm X}$. Taking into account $\left( \ref{Orthogonality} \right)$, the resulting expression for $\mathfrak{F}_d \vert_{{\cal B}^{\mathrm{X}}_{5}}$ is
\begin{equation}
\mathfrak{F}_d \vert_{{\cal B}^{\mathrm{X}}_{5}} = A_d \mathrm{d Z} \wedge \mathrm{d W} + \overline{A}_d \mathrm{d \overline{Z}} \wedge \mathrm{d \overline{W}} + C_d \mathrm{d Z} \wedge \mathrm{d \overline{Z}} + D_d \mathrm{d W} \wedge \mathrm{d \overline{W}},
\end{equation}
and after imposing the condition $\left( \ref{5branes} \right)$, one is left with
\begin{equation} \label{Bfield}
\mathfrak{F}_d \vert_{{\cal B}^{\mathrm{X}}_{5}} = A_d \mathrm{d Z} \wedge \mathrm{d W} + \overline{A}_d \mathrm{d \overline{Z}} \wedge \mathrm{d \overline{W}} \pm \sqrt{1-A_d \overline{A}_d} \left( \mathrm{d Z} \wedge \mathrm{d \overline{Z}} - \mathrm{d W} \wedge \mathrm{d \overline{W}} \right),
\end{equation}
where one must have $\vert A_d \vert \geq 1$ in order to get a real form. Since the 5-brane has the topology $\mathbb{R}^2 \times S^3$, its fundamental group is trivial, so one can globally define an ${\cal A}_d$ such that
\begin{equation}
\mathfrak{F}_d = \mathrm{d} {\cal A}_d.
\end{equation}
An example of an ${\cal A}_d$ whose derivative gives $\mathfrak{F}_d$ of \eqref{Bfield} above on the cycle ${\cal B}^{\mathrm X}_5$
is given by
\begin{equation}
{\cal A}_d = A_d \mathrm{Z} \mathrm{d W} + \overline{A}_d \mathrm{\overline{Z}} \mathrm{d \overline{W}} \pm \sqrt{1-A_d \overline{A}_d} \left( \mathrm{Z} \mathrm{d \overline{Z}} - \mathrm{W} \mathrm{d \overline{W}} \right).
\end{equation}
The contribution of its holonomy to the annulus instanton action is
\begin{equation}
\oint_{{\cal B}^{\mathrm{X}}_{5}} {\cal A}_d =\oint_{{\cal B}^{\mathrm{X}}_{5}} \left[ A_d \mathrm{Z} \mathrm{d W} + \overline{A}_d \mathrm{\overline{Z}} \mathrm{d \overline{W}} \pm \sqrt{1-A_d \overline{A}_d} \left( \mathrm{Z} \mathrm{d \overline{Z}} - \mathrm{W} \mathrm{d \overline{W}} \right) \right] = 0.
\end{equation}

The fluctuations of the fields on the coisotropic brane are frozen due to the infinite volume of this brane, which means that the VEV of the holonomy $V$ of ${\cal A}_d$ along the cycle ${\cal B}^{\mathrm{X}}_{5}$ is
\begin{equation}
\left< V \right>_{{\cal B}^{\mathrm{X}}_{5}} = \left< \mathrm{exp} \left[\mathrm{i} \oint_{{\cal B}^{\mathrm{X}}_{5}} {\cal A}_d \right] \right> = 1.
\end{equation}
\subsection{Boundary conditions} \label{sec:DeformedBoundaryConditions}

Substituting the expression $\left( \ref{Bfield} \right)$ for the field $\mathfrak{F}_d \vert_{{\cal B}^{\mathrm{X}}_{5}}$ into the boundary conditions $\left( \ref{GeneralRmatrix} \right)$, using $\left( \ref{Lagrangianboundaryconditions} \right)$ and $\left( \ref{Coisotropicboundaryconditions} \right)$, we get for the Lagrangian branes
\begin{equation} \label{LagragianBoundary}
\left[\begin{array}{c}
\chi^{\mathrm{X}} \\
\rho^{\overline{\mathrm{X}}}_z \\ \chi^{\mathrm{Z}} \\
\rho^{\overline{\mathrm{Z}}}_z \\ \chi^{\mathrm{W}} \\
\rho^{\overline{\mathrm{W}}}_z \end{array} \right] = \left[\begin{array}{cccccc}
0 & -\mathrm{e}^{2 \mathrm{i} \theta} & 0 & 0 & 0 & 0 \\
-\mathrm{e}^{- 2 \mathrm{i} \theta}  & 0 & 0 & 0 & 0 & 0 \\ 0 & 0 & 0 & 0 & 0 &  -1 \\ 0 & 0 & 0 & 0 & -1 & 0 \\ 0 & 0 & 0 & -1 & 0 & 0 \\ 0 & 0 & -1 & 0 & 0 & 0 \end{array} \right] \left[\begin{array}{c}
\rho^{\mathrm{X}}_{\overline{z}} \\
\chi^{\overline{\mathrm{X}}} \\ \rho^{\mathrm{Z}}_{\overline{z}} \\ \chi^{\overline{\mathrm{Z}}} \\ \rho^{\mathrm{W}}_{\overline{z}} \\ \chi^{\overline{\mathrm{W}}} \end{array} \right],
\end{equation}
and for the coisotropic brane
\begin{equation} \label{CoisotropicBoundary}
\left[\begin{array}{c}
\chi^{\mathrm{X}} \\
\rho^{\overline{\mathrm{X}}}_z \\ \chi^{\mathrm{Z}} \\
\rho^{\overline{\mathrm{Z}}}_z \\ \chi^{\mathrm{W}} \\
\rho^{\overline{\mathrm{W}}}_z \end{array} \right] = \left[\begin{array}{cccccc}
0 & -\mathrm{e}^{2 \mathrm{i} \theta} & 0 & 0 & 0 & 0 \\
-\mathrm{e}^{- 2 \mathrm{i} \theta}  & 0 & 0 & 0 & 0 & 0 \\ 0 & 0 & 0 & 0 & 0 &  R^{\mp}_d \\ 0 & 0 & 0 & 0 & \widetilde{R}^{\pm}_d & 0 \\ 0 & 0 & 0 & - R^{\pm}_d & 0 & 0 \\ 0 & 0 & - \widetilde{R}^{\mp}_d & 0 & 0 & 0 \end{array} \right] \left[\begin{array}{c}
\rho^{\mathrm{X}}_{\overline{z}} \\
\chi^{\overline{\mathrm{X}}} \\ \rho^{\mathrm{Z}}_{\overline{z}} \\ \chi^{\overline{\mathrm{Z}}} \\ \rho^{\mathrm{W}}_{\overline{z}} \\ \chi^{\overline{\mathrm{W}}} \end{array} \right],
\end{equation}
where we defined $\mathrm{X} = \vert \mathrm{X} \vert \mathrm{e}^{\mathrm{i} \theta} = \sqrt{b \mathrm{M}_d} \mathrm{e}^{\mathrm{i} \theta}$, and
\begin{equation}
R^{\mp}_d = \frac{ 1 \mp \sqrt{1 - A_d \overline{A}_d}}{A_d}, \qquad \widetilde{R}^{\pm}_d = \frac{ 1 \pm \sqrt{1 - A_d \overline{A}_d}}{\overline{A}_d},
\end{equation}
where the choice of sign is correlated to the choice in \eqref{Bfield}.

There is a subtlety in solving equations $\left( \ref{LagragianBoundary} \right)$ and $\left( \ref{CoisotropicBoundary} \right)$, because in order to solve them, one has to choose a way in which the $\Psi^{\overline{i}}$ components are related to the $\overline{\Psi^i}$ components. In physical string theory, they are related as
\begin{equation}
\overline{\Psi^I_{\alpha}} = \epsilon_{\alpha \beta} I^I_K \Psi^K_{\beta},
\end{equation}
where $I^I_K$ is a complex structure in the target space. In topological string theory, we cannot use this condition, because it would relate the fields $\chi^I$ and $\rho^I_{z, \overline{z}}$, which have different Lorentz structures. So, the condition we use is that the conjugation of all the components of $\chi^I$ and $\rho^I_{z, \overline{z}}$ is
\begin{equation}
\overline{\chi^i} = \chi^{\overline{i}}, \qquad \overline{\rho^i_{\overline{z}}} = \rho^{\overline{i}}_z.
\end{equation}

\subsection{Gromov-Witten invariants and the free energy} \label{sec:OpenGWInvariants}

Let us evaluate the annulus amplitude on the deformed conifold. Using the formula $\left( \ref{GWinvariants} \right)$ for the GW invariants, and taking into account that the holomorphic $T_d$-invariant cylinders are unique, giving ${\cal M}_{\Gamma} = \left\{ \mathrm{pt} \right\}$, the general expression $\left( \ref{openfreeenergy2} \right)$ reduces to
\begin{equation}
\mathrm{F}_{g = 0, h = 2}^d = \mathrm{f} \left( b \right) + \mathrm{f} \left( 1/b \right),
\end{equation}
with
\begin{align} \label{Freeenergydeformed}
\begin{split}
\mathrm{f} \left( b \right) &= \sum\limits_{w=1}^{\infty} \mathrm{GW}^w_{0,2} \left< \mathrm{Tr} (U_0^{w}) \right>_{b^2} \left< \mathrm{Tr} (V^w) \right>_{{\cal B}^{\mathrm{X}}_{5}} \mathrm{e}^{ - w V_{{\cal C}^{\mathrm{X}}}} \\ &= \sum\limits_{w=1}^{\infty} \frac{1}{w} \frac{e_{\mathbb{C}^{\times}}^m \left( H^1 \left( C , \Phi^* TX_d \vert_{\partial C} \right) \right) e_{\mathbb{C}^{\times}}^m \left( H^0 \left( C , T C \right) \right)}{e_{\mathbb{C}^{\times}}^m \left( H^0 \left( C , \Phi^* TX_d \vert_{\partial C} \right) \right) e_{\mathbb{C}^{\times}}^m \left( H^1 \left( C , T C \right) \right)} \left< \mathrm{Tr} (U_0^{w}) \right>_{b^2} \mathrm{e}^{ - 2 \pi b \mathrm{M}_d w}.
\end{split}
\end{align}
Here $C$ is the annulus worldsheet mapping to ${\cal C}^{\mathrm{X}}$, and $\left< \mathrm{Tr} (U_0^{w}) \right>_{b^2}$ is the CS VEV of the unknot with irrational framing $b^2$, which was argued in \cite{AganagicCostelloMcNamaraVafa} to be the correct framing in this setup.

Let us compute the cohomology groups $H^0 \left( C , \Phi^* TX_d \vert_{\partial C} \right)$ and $H^1 \left( C , \Phi^* TX_d \vert_{\partial C} \right)$, which depend on the boundary conditions in $X_d$.
The map $\Phi$ from the worldsheet annulus $C$ to the target space annulus ${\cal C}^{\mathrm{X}}$ for a $w$-fold covering is given by
\begin{equation}
\mathrm{X} \left( t \right) = t^w, \qquad \mathrm{Z} \left( t \right) = 0, \qquad \mathrm{W} \left( t \right) = 0.
\end{equation}
This means that the $T_d$-weight of $t$ is
\begin{equation}
\lambda_t = \frac{\lambda_{\mathrm{X}}}{w}.
\end{equation}
The range of $t$ here is given by $\vert \mathrm{X} \vert_{\mathrm{min}}^{1/w} \leq \vert t \vert < \vert \mathrm{X} \vert_{\mathrm{max}}^{1/w}$, where $\vert \mathrm{X} \vert_{\mathrm{min}}$ and $\vert \mathrm{X} \vert_{\mathrm{max}}$ correspond to the Lagrangian and coisotropic boundaries respectively, and are given by
\begin{equation}
\vert \mathrm{X} \vert_{\mathrm{min}}^2 = \mu^2, \qquad \vert \mathrm{X} \vert_{\mathrm{max}}^2 = \mu^2 \left[  \sqrt{1 + \left( \frac{b \mathrm{M}_d}{2 \mu^2} \right)^2} + \frac{b \mathrm{M}_d}{2 \mu^2} \right].
\end{equation}
We choose the following covering of the domain annulus $C$ for each winding $w$ ,
\begin{equation}
\mathrm{U}_d = \left\{\vert \mathrm{X} \vert_{\mathrm{min}}^{1/w} \leq \vert t \vert < \vert \mathrm{X} \vert_{\mathrm{max}}^{1/w} \right\}, \qquad \mathrm{V}_d = \left\{\vert \mathrm{X} \vert_{\mathrm{min}}^{1/w} < \vert t \vert \leq \vert \mathrm{X} \vert_{\mathrm{max}}^{1/w} \right\}.
\end{equation}

The group $H^0 \left( C , \Phi^* TX_d \vert_{\partial C} \right)$ is generated by the global sections of the form
\begin{equation}
s^{d, \chi} = \sum\limits_{k = - \infty}^{\infty} \left[ \alpha_{d,k} t^k \partial_{\mathrm{X}} + \beta_{d,k} t^k \partial_{\mathrm{Z}} + \gamma_{d,k} t^k \partial_{\mathrm{W}} \right],
\end{equation}
obeying the boundary conditions $\left( \ref{LagragianBoundary} \right)$ and $\left( \ref{CoisotropicBoundary} \right)$.
The coefficients $\beta_{d,k}$ and $\gamma_{d,k}$ must be zero for any value of $A_d$ in order to satisfy the conditions (\ref{CoisotropicBoundary}), so
\begin{equation} \label{sdc}
s^{d,\chi} = \sum\limits_{k = -\infty}^{\infty}  \alpha_{d,k} t^k \partial_{\mathrm{X}}.
\end{equation}

Now, applying the identical boundary conditions for the $\mathrm{X}$-component of the section on two boundaries, which follows from $\left( \ref{LagragianBoundary} \right)$ and $\left( \ref{CoisotropicBoundary} \right)$, one is left with
\begin{equation}
\alpha_{d,k} = - \overline{\alpha}_{d,2w-k} \vert {\mathrm{X}} \vert_{\mathrm{min}}^{2 \left( 1-k/w \right)} = - \overline{\alpha}_{d,2w-k} \vert {\mathrm{X}} \vert_{\mathrm{max}}^{2 \left( 1-k/w \right)}.
\end{equation}
Since $\vert \mathrm{X} \vert_{\mathrm{min}}$ and $\vert \mathrm{X} \vert_{\mathrm{max}}$ are different, a non-zero solution is possible only for $k=w$, but this term in the expansion \eqref{sdc} has a trivial weight under the action of $T_d$, so, as we discussed in section \ref{sec:equivariantlocalization}, it does not contribute to $e^m_{\mathbb{C}^{\times}} \left( H^0 \left( C , \Phi^* TX_d \vert_{\partial C} \right) \right)$, and we have
\begin{equation}
e^m_{\mathbb{C}^{\times}} \left( H^0 \left( C , \Phi^* TX_d \vert_{\partial C} \right) \right) = 1.
\end{equation}

Now let us turn to $e^m_{\mathbb{C}^{\times}} \left( H^1 \left( C , \Phi^* TX_d \vert_{\partial C} \right) \right)$, which is given by those sections $\rho^I_{z , \overline{z}}$ on $\mathrm{U}_d \cap \mathrm{V}_d$, that cannot be represented as a difference of two sections defined on $\mathrm{U}_d$ and $\mathrm{V}_d$. Let us denote these sections as
\begin{align}
\begin{split}
s^{d,\rho}_{\mathrm{U}_d \cap \mathrm{V}_d} &= \sum\limits_{k = - \infty}^{\infty} \left[ A_{d,k} t^k \partial_{\mathrm{X}} + B_{d,k} t^k \partial_{\mathrm{Z}} + C_{d,k} t^k \partial_{\mathrm{W}} \right], \\ s^{d,\rho}_{\mathrm{V}_d} &= \sum\limits_{k = - \infty}^{\infty} \left[ \alpha_{d,k} t^k \partial_{\mathrm{X}} + \beta_{d,k} t^k \partial_{\mathrm{Z}} + \gamma_{d,k} t^k \partial_{\mathrm{W}} \right], \\ s^{d,\rho}_{\mathrm{U}_d} &= \sum\limits_{k = - \infty}^{\infty} \left[ a_{d,k} t^k \partial_{\mathrm{X}} + b_{d,k} t^k \partial_{\mathrm{Z}} + c_{d,k} t^k \partial_{\mathrm{W}} \right].
\end{split}
\end{align}
Using the boundary conditions (\ref{LagragianBoundary}) and (\ref{CoisotropicBoundary}), one gets
\begin{align}
\beta_{d,k} = \gamma_{d,k} = 0, \qquad c_{d,k} = - \overline{b}_{d,-k} \vert {\mathrm{X}} \vert_{\mathrm{min}}^{- 2 k/w}.
\end{align}

Let us choose $b_{d,k>0}$ and $c_{d,k \geq 0}$ as independent coefficients. It means that the $\mathrm{Z}$ and $\mathrm{W}$ parts of the section $s^{d,\rho}_{\mathrm{U}_d \cap \mathrm{V}_d}$, which cannot be represented as a difference of $s^{d,\rho}_{\mathrm{V}_d}$ and $s^{d,\rho}_{\mathrm{U}_d}$, are
\begin{equation} \label{Annulussection}
\sum\limits_{k = - \infty}^{0} B_{d,k} t^k \partial_{\mathrm{Z}} + \sum\limits_{k = - \infty}^{-1} C_{d,k} t^k \partial_{\mathrm{W}}.
\end{equation}
The equivariant Euler class $e^m_{\mathbb{C}^{\times}} \left( H^1 \left( C , \Phi^* TX_d \vert_{\partial C} \right) \right)$ is given by the product of non-trivial $T_d$-weights of basis vectors of $\left( \ref{Annulussection} \right)$, which is
\begin{equation}
\prod\limits_{k=-\infty}^{0} \left( \lambda_{\mathrm{X}} \frac{k}{w} - \lambda_{\mathrm{Z}} \right) \prod\limits_{k=-\infty}^{-1} \left( \lambda_{\mathrm{X}} \frac{k}{w} - \lambda_{\mathrm{W}} \right).
\end{equation}
Now, recalling the $T_d$-weights of the coordinates $\left( \ref{Tweights} \right)$, using the $\zeta$-function regularization
\begin{equation}
\prod\limits_{k=-\infty}^{-1} \left( \frac{k}{b w} \right)^2 \longrightarrow 2 \pi b w,
\end{equation}
and the product representation of the sine function
\begin{equation}
\prod\limits_{k=-\infty}^{-1} \left( 1 - \frac{b^4 w^2}{k^2} \right) = \frac{\mathrm{sin} \left( \pi b^2 w \right)}{\pi b^2 w},
\end{equation}
we are left with
\begin{equation}
e^m_{\mathbb{C}^{\times}} \left( H^1 \left( C , \Phi^* TX_d \vert_{\partial C} \right) \right) = - 2 \mathrm{sin} \left( \pi b^2 w \right).
\end{equation}

The $\mathrm{X}$ part of the cohomology is trivial. This is very easy to see by putting $a_{d,k} = 0$ for $k<0$. Then one can always solve
\begin{equation}
A_{d,k} = \alpha_{d,k} - a_{d,k}
\end{equation}
as
\begin{align}
\begin{split}
\alpha_{d,k} = A_{d,k}, \quad &k < 0 \\
a_{d,k} = - \overline{\alpha}_{d,2w-k} \vert \mathrm{X} \vert_{\mathrm{max}}^{2 \left( 1-k/w \right)} - A_{d,k}, \quad &k \geq 0.
\end{split}
\end{align}

Now we compute the \v{C}ech cohomology groups $H^0 \left( C , T C \right)$ and $H^1 \left( C , T C \right)$, which do not depend on the target space boundary conditions.
The group $H^0 \left( C , T C \right)$ consists of automorphisms preserving two boundaries of the source cylinder $C$, which is just a $\mathrm{U} (1)$ group generated by $t \partial_t$. The $T_d$-weight of this section is $0$, so
\begin{equation}
e_{\mathbb{C}^{\times}}^m \left( H^0 \left( C , T C \right) \right) = 1.
\end{equation}
The deformation space $ H^1 \left( C , T C \right)$ is also trivial, giving
\begin{equation}
e_{\mathbb{C}^{\times}}^m \left( H^1 \left( C , T C \right) \right) = 1.
\end{equation}
Collecting all the Euler classes, we get
\begin{equation}
\mathrm{GW}^w_{0,2} = - 2 \mathrm{sin} \left( \pi b^2 w \right) ,
\end{equation}
which leads us to 
\begin{equation}
\mathrm{F}_{0,2}^d 
= \sum_{w=1}^\infty \frac{1}{w} (-2\sin{(\pi b^2 w)}) 
\left< {\rm Tr}(U_0^w) \right>_{b^2} \mathrm{e}^{-2\pi b \mathrm{M}_d w} + \left( b \rightarrow 1/b \right) .
\label{eq:annulus_decomposition}
\end{equation}

The final ingredient that we need is the value of the Wilson loop $\left< \mathrm{Tr} (U_0^{w}) \right>_{b^2}$ at the leading order in the $1/N$ expansion.
%, corresponding to the subleading order computation on the closed string side of the GV duality. 
In order to get it, we generalize the result for the Wilson loop with arbitrary rational framing $p$ to the irrational one $b^2$. The Wilson loop with rational framing $p$ has the form \cite{TorusKnots}
\begin{equation} \label{WLrational}
\left< \mathrm{Tr} (U_0^{w}) \right>_p = \frac{N}{\mathrm{t}_d} \sum\limits_{d=0}^w (-1)^{w+d} \frac{1}{w!} {w\choose d} \prod\limits_{k=1}^{w-1} \left( p w + d - k \right) \mathrm{e}^{ \frac{1}{2} \left( p - 1 \right) w \mathrm{t}_d} \mathrm{e}^{ d \mathrm{t}_d},
\end{equation}
where we define
\begin{equation}
\mathrm{t}_d = \mathrm{i} \frac{2 \pi N}{k + N}.
\end{equation}
%Changing the framing in the last equation to $p \rightarrow b^2$, substituting everything into $\left( \ref{Freeenergydeformed} \right)$, and ignoring an overall sign, to which the equivariant localization is not sensitive, 
Plugging this into \eqref{eq:annulus_decomposition}, and ignoring the signs, which cannot be determined from the localization computation, we finally have
\begin{equation} \label{FinalOpenFreeEnergy}
\mathrm{F}_{0, 2}^d = \frac{2}{g_s} \sum\limits_{w=1}^{\infty} \sum\limits_{d=0}^{w} \frac{ \mathrm{sin} \left(\pi b^2 w \right)}{w w!} {w \choose d} \prod\limits_{k=1}^{w-1} \left( b^2 w + d - k \right) \mathrm{e}^{ - 2 \pi b \left( \mathrm{M}_d - \frac{q}{4 \pi} \mathrm{t}_d \right) w } \mathrm{e}^{d \mathrm{t}_d} + \left( b \rightarrow 1/b \right),
\end{equation}
where
\begin{equation}
g_s = - \mathrm{i} \frac{2 \pi}{k + N}.
\end{equation}

\section{The closed string side of the duality} \label{sec:disc}

Now we compute the disc amplitude on the resolved conifold $X_r$, in order to compare it with the annulus amplitude on the deformed conifold $X_d$, and to test the generalization of the GV duality with a coisotropic brane. The computation in this section is very similar to the previous one, so we will not delve into details.

\subsection{The worldsheet geometry} \label{sec:ResolvedGeometricSetUp}

We discussed the resolved conifold $X_r$ and the embedding of the 5-brane $Y_r$ into it in section \ref{closedside}.
We again take the action of the group $T_r = \mathbb{C}^{\times}$ to be analogous to that of $T_d$ above, namely the action of $\mathrm{V}_r$ continued to the target space $X_r$. The weights of the coordinates under its action are
\begin{equation}
\lambda_{\mathrm{X}} = \frac{1}{b}, \qquad \lambda_{\mathrm{Y}} = - \frac{1}{b}, \qquad \lambda_{\mathrm{Z}} = b, \qquad \lambda_{\mathrm{W}} = - b, \qquad \lambda_{\xi} = q, \qquad \lambda_{\eta} = - q.
\end{equation}
The disc amplitude on the resolved conifold with the 5-brane, corresponds to $T_r$-invariant holomorphic discs, ending on the closed leaves of the THF on the coisotropic brane, with possible multi-coverings of the non-trivial $\mathbb{CP}^1$ at the tip.

There are two closed leaves of the THF on the 5-brane, given by
%\begin{equation} \label{xcycle5after}
\begin{align}
\begin{split}
& {\cal A}^{\mathrm{X}}_5: \ \mathrm{Y} = \mathrm{Z} = \mathrm{W} = \xi = 0, \quad \mathrm{X \overline{X}} = b \left( \mathrm{M}_r + \frac{q}{4 \pi} \mathrm{t}_r \right),\\
%\end{equation}
%\begin{equation} \label{zcycle5after}
& {\cal A}^{\mathrm{Z}}_5: \ \mathrm{X} = \mathrm{Y} = \mathrm{W} = \eta = 0, \quad \mathrm{Z \overline{Z}} = \frac{1}{b} \left( \mathrm{M}_r - \frac{q}{4 \pi} \mathrm{t}_r \right),
\end{split}
\end{align}
%\end{equation}
and two corresponding $T_r$-invariant holomorphic discs
\begin{align}
\begin{split}
& {\cal D}^{\mathrm{X}}_r: \ \mathrm{Y} = \mathrm{Z} = \mathrm{W} = \xi = 0, \quad \mathrm{X \overline{X}} \leq b \left( \mathrm{M}_r + \frac{q}{4 \pi} \mathrm{t}_r \right), \\
& {\cal D}^{\mathrm{Z}}_r: \ \mathrm{X} = \mathrm{Y} = \mathrm{W} = \eta = 0, \quad \mathrm{Z \overline{Z}} \leq \frac{1}{b} \left( \mathrm{M}_r - \frac{q}{4 \pi} \mathrm{t}_r \right).
\end{split}
\end{align}
We see that the discs lie on the fibers of $X_r = {\cal O} \left( - 1 \right) \oplus {\cal O} \left( - 1 \right) \rightarrow \mathbb{CP}^1$, touching the base $\mathbb{CP}^1$ at the two poles $\xi = 0$ and $\eta = 0$.
The symplectic volumes of the holomorphic discs ${\cal D}^{\mathrm{X}}_r$ and ${\cal D}^{\mathrm{Z}}_r$ are
\begin{align}
\begin{split}
& V_{{\cal D}^{\mathrm{X}}_r} = \int_{{\cal D}^{\mathrm{X}}_r} \omega_r = 2 \pi b \left( \mathrm{M}_r + \frac{q}{4 \pi} \mathrm{t}_r \right), \\
& V_{{\cal D}^{\mathrm{Z}}_r} = \int_{{\cal D}^{\mathrm{Z}}_r} \omega_r = 2 \pi \frac{1}{b} \left( \mathrm{M}_r - \frac{q}{4 \pi} \mathrm{t}_r \right).
\end{split}
\end{align}
In complete analogy to the deformed conifold side of the duality, the ${\cal D}^{\mathrm{Z}}_r$ amplitude differs from the one of ${\cal D}^{\mathrm{X}}_r$ by $b \rightarrow 1/b$, so we consider only the latter. The disc ${\cal D}^{\mathrm{X}}_r$ can be covered entirely by the $H_{+}$ chart, so in what follows we will often use the coordinates $\mathrm{X}, \mathrm{W}$ and $\xi$ in related computations.

Next we consider the holonomy of the gauge field along the cycle ${\cal A}^{\mathrm{X}}_5$ on the 5-brane. The field strength must be non-zero in order to obey the conditions $\left( \ref{5branes} \right)$. The value of the K\"{a}hler form on ${\cal A}^{\mathrm{X}}_5$, expressed in the coordinate chart $H_{+}$ is
\begin{equation}
\sigma_r = \mathrm{i} \mathrm{d} \mathrm{W} \wedge \mathrm{d} \overline{\mathrm{W}} + \mathrm{i} b \left( \mathrm{M}_r + \frac{Q}{4 \pi} \mathrm{t}_r \right) \mathrm{d} \xi \wedge \mathrm{d} \overline{\xi},
\end{equation}
where 
\begin{equation}
Q = b + \frac{1}{b}.
\end{equation}
In order to simplify the notation, we rescale $\xi$ on the cycle ${\cal A}^{\mathrm{X}}_5$ as $\Lambda = \xi \sqrt{b \left( \mathrm{M}_r + \frac{Q}{4 \pi} \mathrm{t}_r \right)}$, leaving us with
\begin{equation}
\sigma_r = \mathrm{i} \left( \mathrm{d} \mathrm{W} \wedge \mathrm{d} \overline{\mathrm{W}} + \mathrm{d} \Lambda \wedge \mathrm{d} \overline{\Lambda} \right).
\end{equation}
The most general $T_r$-symmetric real 2-form, evaluated on ${\cal A}^{\mathrm{X}}_5$, is
\begin{equation}
\mathfrak{F}_r \vert_{{\cal A}^{\mathrm{X}}_{5}} = B_r \mathrm{X} \mathrm{d W} \wedge \mathrm{d} \Lambda + \overline{B}_r \overline{\mathrm{X}} \mathrm{d \overline{\mathrm{W}}} \wedge \mathrm{d} \overline{\Lambda} + C_r \mathrm{d W} \wedge \mathrm{d \overline{W}} + D_r \mathrm{d} \Lambda \wedge \mathrm{d} \overline{\Lambda}.
\end{equation}
Plugging this into $\left( \ref{5branes} \right)$, the general solution is
\begin{equation} \label{Bfieldresolved}
\mathfrak{F}_r \vert_{{\cal A}^{\mathrm{X}}_{5}} = A_r \mathrm{e}^{\mathrm{i} \theta} \mathrm{d W} \wedge \mathrm{d} \Lambda + \overline{A}_r \mathrm{e}^{ - \mathrm{i} \theta} \mathrm{d \overline{\mathrm{W}}} \wedge \mathrm{d} \overline{\Lambda} \pm \sqrt{1 - A_r \overline{A}_r} \left( \mathrm{d W} \wedge \mathrm{d \overline{W}} - \mathrm{d} \Lambda \wedge \mathrm{d} \overline{\Lambda} \right),
\end{equation}
where $\mathrm{X} = \vert \mathrm{X} \vert \mathrm{e}^{\mathrm{i} \theta} = \sqrt{b \left( \mathrm{M}_r + \frac{q}{4 \pi} \mathrm{t}_r \right)} \mathrm{e}^{\mathrm{i} \theta}$, $A_r = B_r \sqrt{b \left( \mathrm{M}_r + \frac{q}{4 \pi} \mathrm{t}_r \right)}$, and $\vert A_r \vert \geq 1$ in order to maintain the reality of $\mathfrak{F}_r \vert_{{\cal A}^{\mathrm{X}}_{5}}$.

As on the deformed conifold side of the duality, we compute the holonomy of the gauge field with field strength $\mathfrak{F}_r \vert_{{\cal A}^{\mathrm{X}}_{5}}$ along the boundary of the holomorphic disc. We again choose some ${\cal A}_r$, such that
\begin{equation}
\mathfrak{F}_r = \mathrm{d} {\cal A}_r
\end{equation}
on ${\cal A}^{\mathrm{X}}_5$, and find the contribution of the holonomy $V$ of ${\cal A}_r$ to the disc instanton action
\begin{equation}
\oint_{{\cal A}^{\mathrm{X}}_{5}} {\cal A}_r =\oint_{{\cal A}^{\mathrm{X}}_{5}} \left[ A_r \mathrm{X} \mathrm{W} \mathrm{d} \Lambda + \overline{A}_r \overline{\mathrm{X}} \overline{\mathrm{W}} \mathrm{d} \overline{\Lambda} \pm \sqrt{1-A_r \overline{A}_r} \left( \mathrm{W} \mathrm{d \overline{W}} - \mathrm{\Lambda} \mathrm{d} \overline{\Lambda} \right) \right] = 0.
\end{equation}
The fluctuations of the fields on the coisotropic brane are frozen, so one has
\begin{equation}
\left< V \right>_{{\cal A}^{\mathrm{X}}_{5}} = \left< \mathrm{exp} \left[\mathrm{i} \oint_{{\cal A}^{\mathrm{X}}_{5}} {\cal A}_r \right] \right> = 1.
\end{equation}

\subsection{Boundary conditions} \label{sec:ResolvedBoundaryConditions}

Now, we substitute the expression $\left( \ref{Bfieldresolved} \right)$ for the field $\mathfrak{F}_r \vert_{{\cal A}^{\mathrm{X}}_{5}}$ into the fermion boundary conditions $\left( \ref{GeneralRmatrix} \right)$, using $\left( \ref{Coisotropicboundaryconditions} \right)$, and get
\begin{equation} \label{CoisotropicBoundaryResolved}
\left[\begin{array}{c}
\chi^{\mathrm{X}} \\
\rho^{\overline{\mathrm{X}}}_z \\ \chi^{\mathrm{W}} \\
\rho^{\overline{\mathrm{W}}}_z \\ \chi^{\Lambda} \\
\rho^{\overline{\Lambda}}_z \end{array} \right] = \left[\begin{array}{cccccc}
0 & -\mathrm{e}^{2 \mathrm{i} \theta} & 0 & 0 & 0 & 0 \\
-\mathrm{e}^{- 2 \mathrm{i} \theta}  & 0 & 0 & 0 & 0 & 0 \\ 0 & 0 & 0 & 0 & 0 & \mathrm{e}^{- \mathrm{i} \theta} R^{\mp}_r \\ 0 & 0 & 0 & 0 & \mathrm{e}^{\mathrm{i} \theta} \widetilde{R}^{\pm}_r & 0 \\ 0 & 0 & 0 & - \mathrm{e}^{ - \mathrm{i} \theta} R^{\pm}_r & 0 & 0 \\ 0 & 0 & - \mathrm{e}^{ \mathrm{i} \theta} \widetilde{R}^{\mp}_r & 0 & 0 & 0 \end{array} \right] \left[\begin{array}{c}
\rho^{\mathrm{X}}_{\overline{z}} \\
\chi^{\overline{\mathrm{X}}} \\ \rho^{\mathrm{W}}_{\overline{z}} \\ \chi^{\overline{\mathrm{W}}} \\ \rho^{\Lambda}_{\overline{z}} \\ \chi^{\overline{\Lambda}} \end{array} \right],
\end{equation}
where we defined
\begin{equation}
R^{\mp}_r = \frac{ 1 \mp \sqrt{1 - A_r \overline{A}_r}}{A_r}, \qquad \widetilde{R}^{\pm}_r = \frac{ 1 \pm \sqrt{1 - A_r \overline{A}_r}}{\overline{A}_r},
\end{equation}
and the choice of sign is correlated to the choice in \eqref{Bfieldresolved}.

\subsection{Gromov-Witten invariants and the free energy}

The general form of the topological A-model amplitude on the resolved conifold is a particular case of the general formula $\left( \ref{openfreeenergy1} \right)$:
\begin{align}
\begin{split}
\mathrm{F}_{g=0, h=1}^r &= \frac{1}{g_s} \sum\limits_{w=1}^{\infty} \sum\limits_{d=0}^{\infty} \mathrm{GW}^{d,w}_{0,1} \left< \mathrm{Tr} (V^w) \right>_{{\cal A}^{\mathrm{X}}_{5}} \mathrm{e}^{ - w V_{{\cal D}^{\mathrm{X}}_r}} \mathrm{e}^{ - d \mathrm{t}_r} + \left( b \rightarrow 1/b \right) \\ &= \frac{1}{g_s} \sum\limits_{w=1}^{\infty} \sum\limits_{d=0}^{\infty} \mathrm{GW}^{d,w}_{0,1} \mathrm{e}^{ - 2 \pi b \left( \mathrm{M}_r + \frac{q}{4 \pi} \mathrm{t}_r \right) w} \mathrm{e}^{ - d \mathrm{t}_r} + \left( b \rightarrow 1/b \right).
\end{split}
\end{align}

We begin the computation of the GW invariants $\mathrm{GW}^{d,w}_{0,1}$ with the special case of $d = 0$, because the contribution of non-zero $d$ is independent of the conditions on the boundary of the disc, and can be found in the literature. In this case, 
like for the annulus, we have a unique invariant disc ending on each closed leaf, so we have
%the GW invariants are computed exactly as for the annulus, that was considered in section \ref{sec:OpenGWInvariants}, %so we start with the amplitude
\begin{equation}
\tilde{\mathrm{F}}_{0,1}^r = \sum\limits_{w=1}^{\infty} \frac{1}{w} \frac{e_{\mathbb{C}^{\times}}^m \left( H^1 \left( \Delta , \Phi^* TX_r \vert_{\partial \Delta} \right) \right) e_{\mathbb{C}^{\times}}^m \left( H^0 \left( \Delta , T \Delta \right) \right)}{e_{\mathbb{C}^{\times}}^m \left( H^0 \left( \Delta , \Phi^* TX_r \vert_{\partial \Delta} \right) \right) e_{\mathbb{C}^{\times}}^m \left( H^1 \left( \Delta , T \Delta \right) \right)} \mathrm{e}^{ - 2 \pi b \left( \mathrm{M}_r + \frac{q}{4 \pi} \mathrm{t}_r \right) w} + \left( b \rightarrow 1/b \right),
\end{equation}
where $\Delta$ is the domain disc, and the map $\Phi$ from the worldsheet $\Delta$ to the target space disc ${\cal D}^{\mathrm{X}}_r$ is given by
\begin{equation}
\mathrm{X} \left( t \right) = t^w, \qquad \mathrm{W} \left( t \right) = 0, \qquad \Lambda \left( t \right) = 0.
\end{equation}
The range of $t$ is now $0 \leq \vert t \vert < t_{\mathrm{max}}$, where $t_{\mathrm{max}} = \left[b \left( \mathrm{M}_r + \frac{q}{4 \pi} \mathrm{t}_r \right) \right]^{\frac{1}{2w}}$.

Next we compute the cohomologies $H^0 \left( \Delta , \Phi^* TX_r \vert_{\partial \Delta} \right)$ and $H^1 \left( \Delta , \Phi^* TX_r \vert_{\partial \Delta}  \right)$, which depend on the boundary condition in the target space. Let us choose the following covering of the domain disc $\Delta$ for each winding $w$
\begin{equation}
\mathrm{U}_r = \left\{0 \leq \vert t \vert < t_{\mathrm{max}} \right\}, \qquad \mathrm{V}_r = \left\{0 < \vert t \vert \leq t_{\mathrm{max}} \right\}.
\end{equation}
The group $H^0 \left( \Delta , \Phi^* TX_r \vert_{\partial \Delta} \right)$ is generated by the global sections of the form
\begin{equation}
s^{r,\chi}=\sum\limits_{k=0}^{\infty} \left[ \alpha_{r,k} t^k \partial_{\mathrm{X}} + \beta_{r,k} t^k \partial_{\mathrm{W}} + \gamma_{r,k} t^k \partial_{\Lambda} \right]
\end{equation}
obeying the boundary condition $\left( \ref{CoisotropicBoundaryResolved} \right)$.
The coefficients $\beta_{r,k}$ and $\gamma_{r,k}$ must vanish in order to satisfy $\left( \ref{CoisotropicBoundaryResolved} \right)$, so
\begin{equation}
s^{r,\chi} = \sum\limits_{k=0}^{\infty}  \alpha_{r,k} t^k \partial_{\mathrm{X}}
\end{equation}
with
\begin{equation}
\alpha_{r,k} = - t_{\mathrm{max}}^{ 2 w - 2 k } \overline{\alpha}_{r,2w-k}.
\end{equation}
This means that $H^0 \left( \Delta , \Phi^* TX_r \vert_{\partial \Delta} \right)$ is generated by
\begin{equation} \label{H0resolved}
s^{r,\chi}= \sum\limits_{k=0}^{2w}  \alpha_{r,k} t^k \partial_{\mathrm{X}}.
\end{equation}
%Various components of $\left (\ref{H0resolved} \right)$
%transform according to the following real representations of $T_r = \mathbb{C}^{\times}$
%\begin{equation}
%(0) \oplus \sum\limits_{k=0}^{w-1} (k),
%\end{equation}
%which means 
Multiplying the non-zero weights of the $\alpha_{r,k}$ under $T_r = \mathbb{C}^{\times}$, we find
that the equivariant Euler class $e^m_{\mathbb{C}^{\times}} \left( H^0 \left( \Delta , \Phi^* TX_r \vert_{\partial \Delta} \right) \right)$ is given by
\begin{equation}
e^m_{\mathbb{C}^{\times}} \left( H^0 \left( \Delta , \Phi^* TX_r \vert_{\partial \Delta} \right) \right) = \prod\limits_{k=0}^{w-1} \lambda_{\mathrm{X}} \left(\frac{k}{w} -1 \right) = \prod\limits_{k=0}^{w-1} \frac{1}{b} \left(\frac{k}{w} -1 \right).
\end{equation}

The cohomology group $H^1 \left( \Delta , \Phi^* TX_r \vert_{\partial \Delta} \right)$ is generated by the sections $\rho_{\mathrm{U}_r \cap \mathrm{V}_r}$, which cannot be written as the difference of two sections $\rho_{\mathrm{V}_r} - \rho_{\mathrm{U}_r}$. We denote them as
\begin{align}
\begin{split}
s^{r,\rho}_{\mathrm{U}_r \cap \mathrm{V}_r} &= \sum\limits_{k = - \infty}^{\infty} \left[ A_{r,k} t^k \partial_{\mathrm{X}} + B_{r,k} t^k \partial_{\mathrm{W}} + C_{r,k} t^k \partial_{\Lambda} \right], \\ s^{r,\rho}_{\mathrm{V}_r} &= \sum\limits_{k = - \infty}^{\infty} \left[ \alpha_{r,k} t^k \partial_{\mathrm{X}} + \beta_{r,k} t^k \partial_{\mathrm{W}} + \gamma_{r,k} t^k \partial_{\Lambda} \right], \\ s^{r,\rho}_{\mathrm{U}_r} &= \sum\limits_{k = 0}^{\infty} \left[ a_{r,k} t^k \partial_{\mathrm{X}} + b_{r,k} t^k \partial_{\mathrm{W}} + c_{r,k} t^k \partial_{\Lambda} \right].
\end{split}
\end{align}

Let us start with the $\mathrm{X}$ and $\overline{\mathrm{X}}$ components of $s^{\rho}$. The section of $\rho$ which can be written as the difference of two sections $\rho_{\mathrm{V}_r} - \rho_{\mathrm{U}_r}$ is
\begin{equation} \label{Sectiondifference}
\sum\limits_{k=-\infty}^{\infty} \alpha_{r,k} t^k \partial_{\mathrm{X}} = \sum\limits_{k=-\infty}^{\infty} A_{r,k} t^k \partial_{\mathrm{X}} - \sum\limits_{k=0}^{\infty} a_{r,k} t^k \partial_{\mathrm{X}},
\end{equation}
where $A_{r,k}$ must obey the boundary condition
\begin{equation}
A_{r,k} = - t_{\mathrm{max}}^{2 w - 2 k} \overline{A}_{r,2w-k}.
\end{equation}
Since $a_{r,k}$ are arbitrary, one can tune them in such a way that any section on $\mathrm{U}_r \cap \mathrm{V}_r$ can be written as in (\ref{Sectiondifference}), so the $\mathrm{X}$ part of the cohomology is trivial.

Now let us turn to the $\mathrm{W}$ and $\Lambda$ components. As $\beta_{r,k}$ and $\gamma_{r,k}$ are zero, it follows that all the sections on $\mathrm{V}_r$ are trivial, and only the $k < 0$ part of the section $s^{r,\rho}_{\mathrm{U}_r \cap \mathrm{V}_r}$ contributes to $H^1 \left( \Delta , \Phi^* TX_r \vert_{\partial \Delta} \right)$, because the coefficients $b_{r,k}$ and $c_{r,k}$ are arbitrary, and can always be chosen such that
\begin{equation}
b_{r,k \geq 0} = B_{r,k \geq 0}, \qquad c_{r,k \geq 0} = C_{r,k \geq 0}.
\end{equation}
So, the cohomology $H^1 \left( \Delta , \Phi^* TX_r \vert_{\partial \Delta} \right)$ is given by the sections of the form
\begin{equation}
s^{r,\rho}_{\mathrm{U}_r \cap \mathrm{V}_r} = \sum\limits_{k=-\infty}^{-1} \left[ B_{r,k} t^k \partial_{\mathrm{W}} + C_{r,k} t^k \partial_{\Lambda} \right].
\end{equation}
As always, the equivariant Euler class $e^m_{\mathbb{C}^{\times}} \left( H^1 \left( \Delta , \Phi^* TX_r \vert_{\partial \Delta} \right) \right)$ is given by the product of non-trivial $T_r$-weights of basis vectors of $s^{r,\rho}_{\mathrm{U}_r \cap \mathrm{V}_r}$, which is
\begin{equation}
e^m_{\mathbb{C}^{\times}} \left( H^1 \left( \Delta , \Phi^* TX_r \vert_{\partial \Delta} \right) \right) = \prod\limits_{k=-\infty}^{-1} \left( \lambda_{\mathrm{X}} \frac{k}{w} - \lambda_{\mathrm{W}} \right) \left( \lambda_{\mathrm{X}} \frac{k}{w} - \lambda_{\Lambda} \right).
\end{equation}

Let us compute the cohomology groups $H^0 \left( \Delta , T \Delta \right)$ and $H^1 \left( \Delta , T \Delta \right)$, which do not depend on the target space boundary conditions. Equation $\left( \ref{AutomorphismsOfTheDisc} \right)$ states that the group $H^0 \left( \Delta , T \Delta \right)$ is generated by $t \partial_t$ with vanishing $T_r$-weight, but in the case of $d=0$, there is no special point at the center of the disc, at which the 2-sphere is attached, and thus no requirement that it must be fixed by the action of the automorphism group is imposed. This means that the group gets enhanced from $\mathrm{U} (1)$ to $\mathrm{SL} \left( 2 , \mathbb{R} \right)$, generated by
\begin{equation}
s=a_0 \partial_t + a_1 t \partial_t + a_2 t^2 \partial_t
\end{equation}
with
\begin{equation}
a_0=-\overline{a}_2, \qquad a_1 = - \overline{a}_1.
\end{equation}
It follows that there is only one independent basis vector with non-trivial $T_r$-weight, which may be either $\partial_t$ or $t^2 \partial_t$,\footnote{These two choices are related by a sign change of the $T_r$-weight, but as we discussed in section \ref{sec:equivariantlocalization}, the equivariant localization in any case does not allow to determine an overall sign.} so
\begin{equation}
e^m_{\mathbb{C}^{\times}} \left( H^0 \left( \Delta, T \Delta \right) \right) = \frac{\lambda_{\mathrm{X}}}{w} = \frac{1}{b} \frac{1}{w}.
\end{equation}
The deformation space $H^1 \left( \Delta, T \Delta \right)$ is again trivial, resulting in
\begin{equation}
e_{\mathbb{C}^{\times}}^m \left( H^1 \left( \Delta , T \Delta \right) \right) = 1.
\end{equation}

Collecting all the Euler classes, one obtains (up to an overall indeterminable sign)
\begin{align}
\begin{split}
\mathrm{GW}^{0,w}_{0,1} &=\frac{1}{w}  \frac{\lambda_{\mathrm{X}}}{w} \frac{\prod\limits_{k=-\infty}^{-1} \left( \lambda_{\mathrm{X}} \frac{k}{w} - \lambda_{\mathrm{W}} \right) \left( \lambda_{\mathrm{X}} \frac{k}{w} - \lambda_{\Lambda} \right)}{\prod\limits_{k=0}^{w-1} \lambda_{\mathrm{X}} \left(\frac{k}{w} -1 \right)} \\ &= \frac{1}{w} \frac{1}{b w} \frac{\prod\limits_{k=-\infty}^{-1} \left( \frac{k}{b w} \right)^2 \prod\limits_{k=-\infty}^{-1} \left( 1 - \frac{b^4 w^2}{k^2} \right) \prod\limits_{k=0}^{w-1} \left( b^2 w - k \right)}{\prod\limits_{k=0}^{w-1} \left( w - k \right)}.
\end{split}
\end{align}
Now, using
\begin{align}
\begin{split}
& \prod\limits_{k=-\infty}^{-1} \left( \frac{k}{b w} \right)^2 \longrightarrow 2 \pi b w, \qquad
\prod\limits_{k=-\infty}^{-1} \left( 1 - \frac{b^4 w^2}{k^2} \right) = \frac{\mathrm{sin} \left( \pi b^2 w \right)}{\pi b^2 w}, \\ 
& \prod\limits_{k=0}^{w-1} \left( b^2 w - k \right) = b^2 w \prod\limits_{k=1}^{w-1} \left( b^2 w - k \right), \qquad 
\prod\limits_{k=0}^{w-1} \left( w- k \right) = w!,
\end{split}
\end{align}
we are left with
\begin{equation}
\mathrm{GW}^{0,w}_{0,1} = 2 \mathrm{sin} \left(\pi b^2 w \right) \frac{\prod\limits_{k=1}^{w-1} \left( b^2 w - k \right)}{w w!}. 
\end{equation}

Next, we need to add the possibility of a non-zero wrapping of degree $d$ of the disc over the minimal $\mathbb{CP}^1$ at the tip. For this one must compute the integral in formula $\left( \ref{GWinvariants} \right)$. The result is independent of what happens on the boundary of the worldsheet, so the computation is the same as the one deduced via mirror symmetry in \cite{FramedKnotsAtLargeN}, and obtained by direct computation using the graph method (briefly discussed in section \ref{sec:equivariantlocalization}) in \cite{FangLiu}. The result amounts to the following substitution in $\mathrm{GW}^{0,w}_{0,1}$:
\begin{equation}
\prod\limits_{k=1}^{w-1} \left( b^2 w - k \right) \rightarrow \sum\limits_{d = 0}^{w} {w \choose d} \prod\limits_{k=1}^{w-1} \left( b^2 w + d - k \right).
\end{equation}
So finally we get the disc amplitude
\begin{equation} \label{FinalClosedFreeEnergy}
\mathrm{F}_{0, 1}^r = \frac{2}{g_s} \sum\limits_{w=1}^{\infty} \sum\limits_{d=0}^{w} \frac{ \mathrm{sin} \left(\pi b^2 w \right)}{w w!} {w \choose d} \prod\limits_{k=1}^{w-1} \left( b^2 w + d - k \right) \mathrm{e}^{ - 2 \pi b \left( \mathrm{M}_r + \frac{q}{4 \pi} \mathrm{t}_r \right) w} \mathrm{e}^{ - d \mathrm{t}_r} + \left( b \rightarrow 1/b \right).
\end{equation}

Comparing the free energies on the open $\left( \ref{FinalOpenFreeEnergy} \right)$ and closed $\left( \ref{FinalClosedFreeEnergy} \right)$ string sides of the duality, we see that they are identical if we identify the symplectic volume $\mathrm{t}_r$ of the non-trivial $\mathbb{CP}^1$ of the resolved conifold with the 't Hooft parameter $\mathrm{t}_d$ of the CS theory living on the zero section of the deformed conifold,
\begin{equation}
\mathrm{t}_d = - \mathrm{t}_r,
\end{equation}
which requires an analytic continuation of $\mathrm{t}_r$ to purely imaginary value, as in the standard Gopakumar-Vafa duality, and we also simply identify
\begin{equation}
\mathrm{M}_r =  \mathrm{M}_d.
\end{equation}

\section{Higher order corrections}\label{sec:higher}

In principle, one can go beyond the $1/g_s$ order, and compute the higher order terms in the $g_s$ expansion. Let's consider the next order, namely $g_s^0$. Recalling that the worldsheet amplitude of genus $g$ with $h$ boundaries is weighted by $g_s^{2g + h - 2}$, we see that there are two options for the resolved conifold:
\begin{itemize}
	\item $g = 1$, $h = 0$;
	\item $g = 0$, $h = 2$.
\end{itemize}

The first option corresponds to the worldsheet image of the topology $S^2 \cup T$, where $T$ is a torus, which is mapped to one of the poles of the zero section, fixed under the action of $G = T_r$. Under the GV duality, this amplitude is mapped to the $N^0$ order of the expansion of the pure CS theory partition function, and is not related to coisotropic brane dynamics.

The second option is more interesting. The topology of the corresponding worldsheet is $D_1 \cup S^2 \cup D_2$, where $D_1$ and $D_2$ are mapped to the holomorphic discs $\mathcal{D}^{\mathrm{X}}_r$ and $\mathcal{D}^{\mathrm{Z}}_r$, which we discussed in section \ref{sec:disc}. The general form of the amplitude is
\begin{equation}
\mathrm{F}^r_{0,2} = g_s^0 \sum\limits_{d, w_{\mathrm{X}}, w_{\mathrm{Z}}} \mathrm{GW}^{d,w_{\mathrm{X}}, w_{\mathrm{Z}}}_{0,2} \mathrm{e}^{ - w_{\mathrm{X}} V_{{\cal D}^{\mathrm{X}}_r}} \mathrm{e}^{ - w_{\mathrm{Z}} V_{{\cal D}^{\mathrm{Z}}_r}} \mathrm{e}^{ - d \mathrm{t}_r}.
\end{equation}

Under the generalization of the GV duality that we propose, this amplitude must be equated to the deformed conifold amplitude, corresponding to two holomorphic cylinders, and is of the form
\begin{equation}
\mathrm{F}^d_{0,4} = g_s^2 \sum\limits_{w_{\mathrm{X}}, w_{\mathrm{Z}}} \mathrm{GW}^{w_{\mathrm{X}}, w_{\mathrm{Z}}}_{0,4} \left< \mathrm{Tr} (U_{0,b^2}^{w_{\mathrm{X}}}) \mathrm{Tr} (U_{0,1/b^2}^{w_{\mathrm{Z}}}) \right>_{N^2} \mathrm{e}^{ - w_{\mathrm{X}} V_{{\cal C}^{\mathrm{X}}}} \mathrm{e}^{ - w_{\mathrm{Z}} V_{{\cal C}^{\mathrm{Z}}}},
\end{equation}
where $\left< \mathrm{Tr} (U_{0,b^2}^{w_{\mathrm{X}}}) \mathrm{Tr} (U_{0,1/b^2}^{w_{\mathrm{Z}}}) \right>_{N^2}$ is the leading order (in $1/N$) VEV of the Hopf link, and the framings of the two knots are $b^2$ and $1/b^2$ as before.

One can calculate the higher loop corrections using the so-called Faber algorithm \cite{Faber}, which allows to express the integrals over the moduli space ${\cal M}_{\Gamma}$ at arbitrary genus $g$ in terms of integrals over the moduli spaces of lower genera. We leave this for future work.

\section{The large N limit of the partially topological Chern-Simons-matter partition function} \label{sec:fieldtheory}

\subsection{Partially topological Chern-Simons-matter theory} \label{sec:PTSMReview}

\subsubsection{Classical theory}

In this subsection we give a brief review of the PTCSM theory, proposed in \cite{AganagicCostelloMcNamaraVafa} as a field theory description of the topological A-model on the deformed conifold with an additional coisotropic 5-brane of topology $\mathbb{R}^2 \times S^3$.

One can couple the CS theory reviewed in section \ref{sec:CSreview}, to partially topological matter, if the 3-fold on which the CS theory lives, admits a THF. The discussion of a THF structure on a 3-fold is completely analogous to the 5-fold case, which was reviewed in section \ref{sec:openAmodel} --- a 3-fold admits a THF if it may be covered by a set of coordinate charts $\left( \tau_i, u_i, \overline{u}_i \right)$, which transform holomorphically as
\begin{equation} \label{THFcoordinates3}
\tau_j = \tau_j + \vartheta_j \left( \tau_i, u_i, \overline{u}_i \right), \qquad u_j = u_j \left( u_i \right),
\end{equation}
\noindent with real functions $\vartheta_j$ as one goes from one chart to the other. Locally, there is one leaf of the foliation parameterized by $\tau$, for each point with coordinates $\left(u, \overline{u} \right)$.

Compact 3-manifolds with THF are all classified \cite{BrunellaGhys, Brunella, Ghys}, and admit a finite number of deformation parameters each (see also \cite{GeometrySUSYPartition} for more details). They are all either Seifert manifolds, i.e. circle fibrations over a Riemann surface, or $T^2$ fibrations over $S^1$. In what follows, we consider only the case of $S^3$, which admits a THF parameterized by a single complex number $b$ (a detailed discussion of THFs on $S^3$ can be found in \cite{THFSphere}). For simplicity, we take it to be a real positive number, with $b^2$ irrational, in all calculations.

The form (\ref{THFcoordinates3}) of the coordinate changes means, in particular, that the space of holomorphic 1-forms $\Omega^{1,0}$ on a manifold with THF is well-defined. Similarly, we denote by $\Omega^{0,1}$ the space of anti-holomorphic 1-forms. Note that the $\mathrm{d} {\tau}$ direction is not well-defined, because of the mixing with $\left( u, \overline{u} \right)$ as we go from one patch to another. The quotient spaces $\Omega / \Omega^{1,0}$ and $\Omega / \Omega^{0,1}$ are well-defined, where $\Omega$ is the space of all 1-forms, as is the projection $p: \Omega \rightarrow \Omega / \Omega^{1,0}$, which in local coordinates simply means
\begin{equation}
\begin{gathered}
p: f \mathrm{d} \tau + g \mathrm{d} u + h \mathrm{d} \overline{u} \rightarrow f \mathrm{d} \tau + h \mathrm{d} \overline{u}, \\ p \cdot \mathrm{d} = \mathrm{d} \tau \frac{\partial}{\partial \tau} + \mathrm{d} \overline{u} \frac{\partial}{\partial \overline{u}}.
\end{gathered}
\end{equation}

The matter fields of the PTCSM theory of \cite{AganagicCostelloMcNamaraVafa} are $\phi \in \Omega^{1,0} \otimes R$, and $\psi \in \Omega / \Omega^{1,0} \otimes \overline{R}$, where $R$ is the fundamental representation of $\mathrm{U} (N)$, and $\overline{R}$ is its conjugate. The action of the theory has the form
\begin{equation} \label{chernsimonsmatteraction}
S=\mathrm{i} \frac{k}{2 \pi} \int \mathrm{Tr} \left[ \frac{1}{2} A \wedge \textrm{d}A+ \frac{1}{3} A \wedge A \wedge A \right]+\int \mathrm{Tr} \left[ \psi \wedge \mathrm{D}_A \phi \right].
\end{equation}
Note that this theory has an additional gauge symmetry, given by
\begin{equation} \label{Symmetry}
\begin{gathered}
\psi \rightarrow \psi + \hat{\mathrm{D}}_A \epsilon, \\ A \rightarrow A + \frac{4 \pi \mathrm{i}}{k} \mu \left( \phi, \epsilon \right),
\end{gathered}
\end{equation}

\noindent where $\epsilon \in \overline{R}$ is an arbitrary function of coordinates, $\hat{\mathrm{D}}_A$ is equal to $p\cdot (\mathrm{d} - \mathrm{i} A)$, and $\mu: R \otimes \overline{R} \rightarrow \mathfrak{g}$ is the moment map.

The matter system can be made massive in a way that preserves the partial topological invariance, and gives the $\left( \phi , \psi \right)$ system a real mass $\mathrm{M}$. This is done by introducing a background complex connection $A_b$ to the matter action (\ref{chernsimonsmatteraction}), which locally looks like
\begin{equation}
A \rightarrow A-A_b=A-\mathrm{i M d} \tau.
\end{equation}

In order to compare the field theory results with the string theory ones, it is convenient to view the sphere $S^3$, on which the PTCSM theory lives, as being embedded in $\mathbb{C}^2$ via
\begin{equation}
\mathrm{X} \overline{\mathrm{X}} + \mathrm{Z} \overline{\mathrm{Z}} = \mu^2, \quad \mu \in \mathbb{R}.
\end{equation}
In these coordinates, the THF that we are interested in is generated by the vector field
\begin{equation}
\mathrm{V}^{\prime}=\mathrm{i} \frac{1}{b} \left( \mathrm{X} \frac{\partial}{\partial \mathrm{X}}-\overline{\mathrm{X}} \frac{\partial}{\partial \overline{\mathrm{X}}} \right)+ \mathrm{i} b \left( \mathrm{Z} \frac{\partial}{\partial \mathrm{Z}}-\overline{\mathrm{Z}} \frac{\partial}{\partial \overline{\mathrm{Z}}} \right).
\end{equation}
For irrational $b^2$ this THF has two closed leaves going to themselves under the THF action, corresponding to $\mathrm{X} = 0$ or $\mathrm{Z} = 0$.

\subsubsection{Quantization}

One can quantize the PTCSM theory using the Batalin–Vilkovisky (BV) formalism \cite{AganagicCostelloMcNamaraVafa}. In order to do this, we introduce the ghost field $\varkappa \in \overline{R}$ for the symmetry (\refeq{Symmetry}), and anti-fields
\begin{align}
\begin{split}
\phi^* &= \phi^*_{\overline{u} \tau} \mathrm{d} \overline{u} \wedge \mathrm{d} \tau \in \Omega^2 / \left( \Omega^{1,0} \wedge \Omega^1 \right) \otimes \overline{R}, \\ \psi^* &= \psi^*_{u \tau} \mathrm{d} u \wedge \mathrm{d} \tau + \psi^*_{u \overline{u}} \mathrm{d} u \wedge \mathrm{d} \overline{u} \in \left( \Omega^{1,0} \wedge \Omega^1 \right) \otimes R, \\ \varkappa^* &= \varkappa^*_{u \overline{u} \tau} \mathrm{d} u \wedge \mathrm{d} \overline{u} \wedge \mathrm{d} \tau \in \Omega^3 \otimes R.
\end{split}
\end{align}
After packing these fields into
\begin{equation}
\varphi = \phi + \psi^* + \varkappa^*, \qquad \varrho = \varkappa + \psi + \phi^*,
\end{equation}
the BV action becomes
\begin{equation}
S_{BV} = \mathrm{CS}_3 \left( \Delta \right) + \int \varrho \wedge \mathrm{\hat{D}}_{\Delta} \varphi,
\end{equation}
where the field $\Delta$ includes the fields, ghosts and anti-fields of usual BV quantization of the CS theory.

The only fields in the BRST cohomology are $\phi_u$ and $\varkappa$, so the space of local operators of the theory is spanned by the products of $\partial_u^n \phi_u$ and $\partial_u^m \varkappa$.

%It was argued in \cite{AganagicCostelloMcNamaraVafa} that the partition function of this theory is equal to that of a $3d$ ${\cal N}=2$ $\mathrm{U} (N)$ CS-matter theory with a single chiral multiplet, and has the form
%\begin{equation}
%Z_{CSM} = \frac{1}{N!} \int\limits_{-\infty}^{\infty} \mathrm{d}^N x \prod\limits_{i<j} \mathrm{sinh} \left( \pi b x_{ij} \right) \mathrm{sinh} \left( \pi b^{-1} x_{ij} \right) \prod\limits_{j = 1}^{N} \mathrm{e}^{\mathrm i k \pi x_{ij}^2} s_b \left( \mathrm{i} \frac{Q}{2} + \mathrm{M} - x_j \right),
%\end{equation}

%\noindent where $Q = b + b^{-1}$, $x_{ij} = x_i - x_j$, and $s_b \left( z \right)$ is a double sine function, defined by
%\begin{equation}
%s_b \left( z \right) = \prod\limits_{m,n \geq 0} \frac{m b + n b^{-1} + Q/2 - \mathrm{i} z}{m b + n b^{-1} + Q/2 + \mathrm{i} z}.
%\end{equation} 

%%%%%%%%%%%%%%%%%%%%%%%
%%%%%%%%%%%%%%%%%%%%%%%
%%%%%%%%%%%%%%%%%%%%%%%
\subsection{The large N partition function of the Chern-Simons-matter theory on \boldmath $S_b^3$}
%%%%%%%%%%%%%%%%%%%%%%%
%%%%%%%%%%%%%%%%%%%%%%%
%%%%%%%%%%%%%%%%%%%%%%%

In this section we compute the partition function of the PTCSM theory \eqref{chernsimonsmatteraction} in the large $N$ limit. 
It was argued in \cite{AganagicCostelloMcNamaraVafa} that this is the same as the partition function of a 3d $\mathcal{N}=2$ supersymmetric $\mathrm{U} (N)_k$ CS theory coupled to a single fundamental chiral multiplet with $\mathrm{U} (1)_R$-charge $0$ and real mass $\mathrm{M}$, with the same THF structure. 
And it was shown in \cite{GeometrySUSYPartition} that this is the same as the partition function of this theory  on a squashed sphere $S_b^3$. Thus we can use the known results for these partition functions.

Applying the localization method \cite{Pestun,KapustinWillettYaakov},
the partition function is given by the following $N$-dimensional integral \cite{HamaHosomichiLee}:\footnote{An expression for generic $R$-charge $\Delta_R$ for the chiral multiplet can be obtained by replacing $\mathrm{M} \rightarrow \mathrm{M} - \frac{\mathrm{i}Q}{2}\Delta_R$.}
\be
Z= \frac{1}{N!}\int\limits_{-\infty}^\infty \mathrm{d}^N x 
\prod_{i<j} \sinh{(\pi b x_{ij})} \sinh{ \left( \pi b^{-1} x_{ij} \right)}
\prod_{j=1}^N \mathrm{e}^{\mathrm{i} k \pi x_j^2} s_b \left( \frac{\mathrm{i} Q}{2} + \mathrm{M}  -x_j \right) ,
\ee
where $Q = b + \frac{1}{b}$ as before, $x_{ij} = x_i - x_j$, and $s_b (z)$ is the double sine function defined by
\be
s_b (z)
=\prod_{m,n\geq 0} \frac{mb+nb^{-1}+\frac{Q}{2}- \mathrm{i} z}{mb+nb^{-1}+\frac{Q}{2}+\mathrm{i} z}.
\ee
We are interested in the planar limit
\be
N\rightarrow \infty, \qquad \frac{N}{k'}=\lambda ={\rm fixed}, \qquad
k' =k-\frac{1}{2},
\ee
where the factor $(-1/2)$ in $k'$ comes from the CS level shift.\footnote{A more appropriate definition of $k'$ would be $k' =k-\frac{1}{2}{\rm sgn} \left( \mathrm{M} \right)$. Note that $k'$ and not $k$ will appear in the modified duality relations, but this does not affect the leading order answers.}

Since the matter has only $\mathcal{O}(N)$ degrees of freedom,
the leading large $N$ behavior of $Z$ is captured by the $\mathcal{N}=2$ SUSY pure CS theory without the matter:
\be
Z_{\rm CS} (k,N)
= \frac{1}{N!}\int\limits_{-\infty}^\infty \mathrm{d}^N x 
\prod_{i<j} \sinh{(\pi b x_{ij})} \sinh{(\pi b^{-1} x_{ij})} 
\prod_{j=1}^N \mathrm{e}^{\mathrm{i} k \pi x_j^2}  ,
\ee
which can be exactly computed for any $N$\footnote{See e.g. equation (2.9) of \cite{NishiokaYonekura}.}
\be \label{purecs}
Z_{\rm CS} (k,N)
= \frac{\mathrm{e}^{\frac{\mathrm{i} \pi}{12k}N(N^2 -1)(b^2 +b^{-2})}}{k^{N/2}} \prod_{j=1}^{N} \left( 2\sin{\frac{\pi j}{k}} \right)^{N-j} .
%= \frac{e^{-\frac{i\pi}{6k}N(N^2 -1)(1-\frac{Q^2}{2})}}{k^{N/2}} \prod_{j=1}^{N} \left( 2\sin{\frac{\pi j}{k}} \right)^{N-j} .
\ee
In the language of the free energy 
\be
\mathrm{F} (N) = \log({Z(N)}),
\ee
the large $N$ expansion of \eqref{purecs} starts at $\mathcal{O}(N^2 )$, and it has an expansion in powers of $1/N^2$, corresponding by the GV duality to perturabtive corrections on the closed string side of the duality. In this paper we focus on the $\mathcal{O}(N)$ part of the CSM free energy, which is the first order affected by the fundamental matter.

%%%%%%%%%%%%%%%%%%%%%%%
%%%%%%%%%%%%%%%%%%%%%%%
\subsection{The \boldmath $\mathcal{O}(N)$ free energy for \boldmath ${\rm Im}(b^2) >0$}
%%%%%%%%%%%%%%%%%%%%%%%
%%%%%%%%%%%%%%%%%%%%%%%
For ${\rm Im}(b^2) >0$, the double sine function has the following representation:
\be
s_b (z)
= \mathrm{e}^{-\frac{\mathrm{i} \pi z^2}{2} -\frac{\mathrm{i} \pi}{24} \left( b^2 + \frac{1}{b^2} \right)}
\frac{(-p_+^{1/2} \mathrm{e}^{2\pi b z};p_+ )_\infty}
{(-p_-^{-1/2} \mathrm{e}^{2\pi b^{-1} z};p_-^{-1} )_\infty} ,
\label{eq:sb_upper}
\ee
where
\be 
p_\pm \equiv \mathrm{e}^{2\pi \mathrm{i} b^{\pm 2}}, \qquad  (a; \mathrm{q})_\infty \equiv \prod_{n=0}^{\infty} (1-a \mathrm{q}^n ) .
\ee
In terms of this representation, we can rewrite the partition function as
\begin{align} \label{eq:tildeZ}
\begin{split}
&Z = \mathrm{e}^{-\frac{\mathrm{i} \pi N}{2} \left( \frac{\mathrm{i} Q}{2} + \mathrm{M} \right)^2 - \frac{\mathrm{i} \pi \lambda (\mathrm{i} Q/2 + \mathrm{M})^2}{4} - \frac{\mathrm{i} \pi N}{24}( b^2 + b^{-2} )} \tilde{Z}, \\
&\tilde{Z}
= \frac{1}{N!}
\int_{-\infty -\frac{\mathrm{i} Q}{4k'}}^{\infty -\frac{\mathrm{i} Q}{4k'}} \mathrm{d}^N x 
\prod_{i<j} \sinh{(\pi b x_{ij})} \sinh{(\pi b^{-1} x_{ij})} 
\prod_{j=1}^N \mathrm{e}^{\mathrm{i} k' \pi x_j^2}
\prod_{n=0}^\infty 
\frac{1 - p_+^{n+1} \mathrm{e}^{2\pi b (\tilde{\mathrm{M}}-x_j ) }}
{1 - p_-^{-n} \mathrm{e}^{2\pi b^{-1}(\tilde{\mathrm{M}}-x_j ) }},
\end{split}
\end{align}
where we have changed the integral variable as $x \rightarrow x + \frac{\frac{\mathrm{i} Q}{2} + \mathrm{M}}{2k'}$, and defined
\be
\tilde{\mathrm{M}} = \mathrm{M} -\frac{\mathrm{i} Q/2 + \mathrm{M}}{2k'} .
\label{eq:tildem}
\ee

Now we take the large $N$ limit and focus on the $\mathcal{O}(N)$ free energy, dropping terms of order $N^0$:
\begin{align}
\begin{split}
&\mathrm{F} \vert_{\mathcal{O}(N)}
= \left. \log({Z}) \right|_{\mathcal{O}(N)}
= - \frac{ \mathrm{i} \pi N}{2} \left[  \left( \frac{\mathrm{i} Q}{2} + \mathrm{M} \right)^2 + \frac{1}{12} \left( b^2 + \frac{1}{b^2} \right) \right]
+ \mathrm{\tilde{F}} \vert_{\mathcal{O}(N)} , \\
&\mathrm{\tilde{F}} = \log({\tilde{Z}})  .
\end{split}
\end{align}
In the large $N$ limit, we can make two simplifications. First, we can replace $\tilde{\mathrm{M}}$ by $\mathrm{M}$ and ignore the imaginary shift of the integral contour in \eqref{eq:tildeZ}
\bea
\tilde{Z}
\simeq \frac{1}{N!}
\int\limits_{-\infty }^{\infty} \mathrm{d}^N x 
\prod_{i<j} \sinh{(\pi b x_{ij})} \sinh{(\pi b^{-1} x_{ij})} 
\prod_{j=1}^N \mathrm{e}^{\mathrm{i} k' \pi x_j^2}
\prod_{n=0}^\infty 
\frac{1 - p_+^{n+1} \mathrm{e}^{2\pi b (\mathrm{M}-x_j )}}
{1 - p_-^{-n} \mathrm{e}^{2\pi b^{-1}(\mathrm{M}-x_j )}}.
\eea
Second, after redefining the integration variable as $x \rightarrow - x$, the large $N$ factorization implies\footnote{The trace is taken in the fundamental representation i.e.~${\rm Tr} \left( f(x) \right) = \sum_{j=1}^N f(x_j )$.}
\be
\tilde{\mathrm{F}} |_{\mathcal{O}(N)}
= \sum_{n=0}^\infty 
\left< {\rm Tr} \left[\log  \left( 1 - p_+^{n+1} \mathrm{e}^{2\pi b \mathrm{M}} \mathrm{e}^{2\pi b x} \right) \right] \right>_{{\rm CS}_b}
+ \left( b \rightarrow 1/b \right) ,
\ee
where $\left< \cdots \right>_{{\rm CS}_b}$ denotes the planar limit of a normalized VEV in the following matrix model:
\be
\int\limits_{-\infty}^\infty \mathrm{d}^N x \ \mathrm{e}^{\mathrm{i} k' \pi \sum_j x_j^2}  
\prod_{i<j} \sinh{(\pi b x_{ij})} \sinh{(\pi b^{-1} x_{ij})}   .
\ee

Now we expand the first term as
\begin{align}
\begin{split}
\sum_{n=0}^\infty 
\left< {\rm Tr} \left[ \log  \left( 1 -p_+^{n+1} \mathrm{e}^{2\pi b \mathrm{M}} \mathrm{e}^{2 \pi b x} \right) \right] \right>_{{\rm CS}_b}
&= -\sum_{n=1}^\infty \sum_{w=1}^\infty \frac{1}{w}  
\left( p_+^n \mathrm{e}^{2\pi b \mathrm{M}}  \right)^w
\left< {\rm Tr} \left[ \mathrm{e}^{2\pi b x w} \right]  \right>_{{\rm CS}_b} \\
&= - \sum_{w=1}^\infty \frac{1}{w} \frac{p_+^w}{1-p_+^w} 
\mathrm{e}^{2\pi b \mathrm{M} w} 
\left< {\rm Tr} \left[ \mathrm{e}^{2\pi b x w} \right] \right>_{{\rm CS}_b}.
\end{split}
\end{align}
Repeating this for the second term, we find
\bea \label{eq:beforeVEV}
\tilde{\mathrm{F}} |_{\mathcal{O}(N)}
= \sum_{w=1}^\infty \frac{1}{w} \frac{p_+^w}{1-p_+^w} 
\mathrm{e}^{2\pi b \mathrm{M} w} 
\left< {\rm Tr} \left[ \mathrm{e}^{2\pi b x w} \right] \right>_{{\rm CS}_b}\ 
+ \left( b \rightarrow 1/b \right).
\eea

This expression shows that our problem is now reduced to computing some quantities in the pure CS theory on $S_b^3$. 
The quantities are interpreted as supersymmetric Wilson loops in the fundamental representation, defined by \cite{Tanaka}
\be
W ={\rm Tr} \left\{ \mathrm{\hat{P}} \exp{ \left[
	\oint_{\mathrm{C}} \mathrm{d} \tau (\mathrm{i}A_\mu \dot{x}^\mu +\sigma_s |\dot{x}|) \right]} \right\} ,
\ee
where $\sigma_s$ is the scalar field in the 3d ${\cal N}=2$ vector multiplet.

The difference between the first and second terms in \eqref{eq:beforeVEV} comes from different contours $\mathrm{C}$.
Let us choose $S_b^3$ to be the following ellipsoid:
\be
b^2 |z_1 |^2 + \frac{1}{b^2} |z_2 |^2 =1 ,
\ee
and take the torus fibration coordinates
\be
z_1 = \frac{1}{b} \mathrm{e}^{\mathrm{i} \phi_1} \cos \theta, \qquad
z_2 = b \mathrm{e}^{\mathrm{i} \phi_2} \sin \theta.
\ee
There are (at least) three following types of SUSY-preserving contours on the squashed sphere, where two of them exist for any $b$, while the other exists only for rational $b^2$:
\begin{enumerate}
	\item $\theta =0$, $\phi_1 = b\tau +{\rm const.}$\\
	This winds the $\phi_1$ cycle and is an unknot.
	When this loop winds this cycle $w$ times, the
	localization formula for this loop is
	\be
	\left< W_{\theta =0}^{(w)} \right>_{{\rm CS}_b} 
	=\left< {\rm Tr} \left[ \mathrm{e}^{2\pi b^{-1} x w} \right] \right>_{{\rm CS}_b}  .
	\ee
	It is known that this is formally the same as the unknot Wilson loop with framing $b^{-2}$ in a pure CS theory without a level shift \cite{TorusKnots}.
	
	\item $\theta = \pi /2$, $\phi_2 = b^{-1}\tau +{\rm const.}$\\
	This winds the $\phi_2$ cycle and  is an unknot. When this loop winds this cycle $w$ times, the localization formula for this loop is
	\be
	\left< W_{\theta =\pi /2}^{(w)} \right>_{{\rm CS}_b} 
	=\left< {\rm Tr} \left[ \mathrm{e}^{2\pi b x w} \right] \right>_{{\rm CS}_b}  ,
	\ee
	which is known to be the same as an unknot Wilson loop with framing $b^2$ in a pure CS theory without a level shift \cite{TorusKnots}.
	
	\item $\theta \neq \{0,\pi/2\}$, $\phi_1 = b\tau +{\rm const.}$,
	$\phi_2 = b^{-1}\tau +{\rm const.}$\\
	When $b^2 =b_1 /b_2$ with coprime $b_1 ,b_2 \in\mathbb{Z}$,
	this winds the $\phi_1$ cycle $b_1$ times and the $\phi_2$ cycle $b_2$ times. Therefore, this is a $(b_1 ,b_2 )$ torus knot.
	Note that this contour is not closed for irrational $b^2$.
	When the loop winds this cycle $w$ times, the localization formula for this loop is
	\be
	\left< W_{(b_1 ,b_2 )}^{(w)}  \right>_{{\rm CS}_{b=\sqrt{b_1 /b_2}}} 
	= \left< {\rm Tr} \left[ \mathrm{e}^{2\pi \sqrt{b_1 b_2} x w} \right]  \right>_{{\rm CS}_{b=\sqrt{b_1 /b_2}}}  .
	\ee
	Note that this loop cannot be distinguished from the other two loops at the level of the localization formula, because
	\be
	\left< W_{(b_1 ,b_2 )}^{(w)}  \right>_{{\rm CS}_b} 
	=\left< W_{\theta =0}^{(b_1 w)} \right>_{{\rm CS}_b} 
	=\left< W_{\theta =\pi /2}^{(b_2 w)} \right>_{{\rm CS}_b} 
	=\left< {\rm Tr} \left[ \mathrm{e}^{2\pi \sqrt{b_1 b_2} x w} \right] \right>_{{\rm CS}_b} 
	\ee
	for $b=\sqrt{\frac{b_1}{b_2}}$.
	
\end{enumerate}
The bottom line is that the $\mathcal{O}(N)$ free energy \eqref{eq:beforeVEV} for generic values of $b$ can be written as
\be
\tilde{\mathrm{F}} |_{\mathcal{O}(N)}
= \sum_{w=1}^\infty \frac{1}{w} \frac{p_+^w}{1-p_+^w} 
\mathrm{e}^{2\pi b \mathrm{M} w} \left< W_{\theta =\pi /2}^{(w)} \right>_{{\rm CS}_b} + \left( b \rightarrow 1/b \right).
\ee

There are known exact results on the Wilson loops even for finite $N$ \cite{TorusKnots}:
\bea
\left< {\rm Tr} \left[ \mathrm{e}^{2\pi b w x} \right] \right>_{{\rm CS}_b} 
= \mathrm{e}^{\frac{b^2 -1}{2} w \mathrm{t} } \sum_{d = 0}^w  (-1)^{w+d}  \mathrm{e}^{d \mathrm{t}}
\frac{ \prod\limits_{j=1}^{w-1} \left[ \mathrm{e}^{\frac{\pi \mathrm{i}(b^2 w - w + d + j)}{k'}} - \mathrm{e}^{-\frac{\pi \mathrm{i} (b^2 w - w + d + j)}{k'}} \right]}
{\prod\limits_{j=1}^d \left[ \mathrm{e}^{\frac{\pi \mathrm{i}}{k'}j} - \mathrm{e}^{-\frac{\pi \mathrm{i}}{k'}j} \right]
	\prod\limits_{j=1}^{w-d} \left[ \mathrm{e}^{\frac{\pi \mathrm{i} }{k'}j} - \mathrm{e}^{-\frac{\pi \mathrm{i}}{k'}j} \right]} ,
\eea
where
\be
\mathrm{t}=\frac{2 \pi \mathrm{i} N}{k'} .
%,\quad q=e^{\frac{2\pi i}{k'}} .
\ee
In the planar limit, we can use the approximation $ \left[ \mathrm{e}^{\frac{\pi \mathrm{i}}{k'}j} - \mathrm{e}^{-\frac{\pi \mathrm{i}}{k'}j} \right] \simeq  \frac{2 \pi \mathrm{i}}{k'} j$,
% and $[n]! \simeq \left( \frac{2\pi i}{k}\right)^n n!$,
and find
\be
\frac{1}{N} \left< {\rm Tr} \left[ \mathrm{e}^{2 \pi b x w} \right] \right>_{{\rm CS}_b}  = \frac{w \mathrm{e}^{\frac{b^2 - 1}{2}w \mathrm{t}}}{\mathrm{t}}
\sum_{d = 0}^w  W_{w,d}(b^2) \mathrm{e}^{d \mathrm{t}} ,
\label{eq:W_planar}
\ee
where \footnote{Here we defined $(n)_d \equiv n(n+1)\cdots(n+d-1)$. 
	There is also a representation in terms of a hypergeometric function:
	\be
	\frac{1}{N} \left< {\rm Tr} \left[ \mathrm{e}^{2\pi b x w} \right] \right>_{{\rm CS}_b} 
	=  \frac{(-1)^w}{\mathrm{t}} \frac{\Gamma (b^2 w)}{w! \Gamma (b^2 w -w+1)}
	\mathrm{e}^{\frac{b^2 - 1}{2} w \mathrm{t}}\ 
	_2 F_1 \left( -w, b^2 w ;b^2 w -w+1 ; \mathrm{e^t} \right) .
	\ee
}
\be
W_{w,d} (f)
%= \frac{(-1)^{w+\ell}}{w w!}\begin{pmatrix} w \cr \ell \end{pmatrix} \prod_{j=-\ell +1}^{w-\ell-1} (wf -j) 
= \frac{(-1)^{w+d}}{w w!} {w \choose d}
\prod_{k=1}^{w-1} (fw+d-k) 
= \frac{(-1)^{w} }{w w!} 
\frac{(-w)_d (fw)_d (fw-w+1)_{w-1}}{d ! (fw -w+1)_d} .
\label{eq:W_wl}
\ee
Thus, the $\mathcal{O}(N)$ free energy is given by
\be
\tilde{\mathrm{F}}|_{\mathcal{O}(N)} = - \frac{N}{\mathrm{t}} \sum_{w=1}^\infty \sum_{d =0}^w
\frac{p_+^w \mathrm{e}^{2\pi b \mathrm{M} w}}{1-p_+^w}   \mathrm{e}^{\frac{(b^2 -1)w \mathrm{t}}{2}} \mathrm{e}^{d \mathrm{t}}
W_{w,d}(b^2 ) + \left( b \rightarrow 1/b \right).
\label{eq:F_upper}  
\ee
Note that for ${\rm Im}(b^2) >0$, the small $p_+$ and large $p_-$ expansions in this expressions are convergent, while the  large $p_+$ and small $p_-$ expansions are non-convergent. We can also write this expression as
\be \label{finalfield}
\tilde{\mathrm{F}}|_{\mathcal{O}(N)}  
= \frac{N}{\mathrm{t}} \sum_{w=1}^\infty \sum_{d=0}^w
\frac{p_+^w \mathrm{e}^{2 \pi b \mathrm{M}^{\prime} w}}{1-p_+^w}    \mathrm{e}^{-d \mathrm{t}} W_{w,d }(-b^2 ) + \left( b \rightarrow 1/b \right),
\ee
where we have used the identity $W_{w,d}(f) =-W_{w,w-d}(-f)$, and $\mathrm{M}^{\prime}$ is defined by
\be
\mathrm{M}^{\prime} = \mathrm{M} +\frac{Q \mathrm{t}}{4\pi}.
\ee

%%%%%%%%%%%%%%%%%%%%%%%
%%%%%%%%%%%%%%%%%%%%%%%
\subsection{The \boldmath $\mathcal{O}(N)$ free energy for \boldmath ${\rm Im}(b^2) < 0$}
%%%%%%%%%%%%%%%%%%%%%%%
%%%%%%%%%%%%%%%%%%%%%%%
For ${\rm Im}(b^2) <0$, the double sine function has the representation
\be
s_b (z)
= \mathrm{e}^{-\frac{\mathrm{i} \pi z^2}{2} -\frac{\mathrm{i} \pi}{24}(b^2 +b^{-2})}
\frac{(-p_-^{1/2} \mathrm{e}^{2\pi b^{-1} x};p_- )_\infty}
{(-p_+^{-1/2} \mathrm{e}^{2\pi bx};p_+^{-1} )_\infty} ,
\ee
which can be formally obtained by taking $b\rightarrow b^{-1}$ 
in the representation \eqref{eq:sb_upper} for ${\rm Im}(b^2) >0$.
Then the $\mathcal{O}(N)$ free energy for generic $b$ is given by
\begin{align}
\begin{split}
\tilde{\mathrm{F}} |_{\mathcal{O}(N)}
&= \sum_{w=1}^\infty \frac{1}{w} \frac{p_+^w}{1-p_+^w} 
\mathrm{e}^{2 \pi b \mathrm{M} w} 
\left< {\rm Tr} \left[ \mathrm{e}^{2\pi b x w} \right] \right>_{{\rm CS}_b}\ + \left( b \rightarrow 1/b \right) \\
&= - \frac{N}{\mathrm{t}} \sum_{w=1}^\infty \sum_{d =0}^w
\frac{p_+^w \mathrm{e}^{2 \pi b \mathrm{M} w}}{1-p_+^w}   \mathrm{e}^{\frac{(b^2 -1)w \mathrm{t}}{2}} \mathrm{e}^{d \mathrm{t}}
W_{w,d }(b^2 ) + \left( b \rightarrow 1/b \right),
\end{split}
\end{align}
which is formally the same as the result \eqref{eq:F_upper} for ${\rm Im}(b^2 )>0$. Therefore, this expression is valid both for\footnote{
	\cite{AganagicCostelloMcNamaraVafa} used the following expression for the double sine function:
	\[
	s_b (z)
	= \mathrm{e}^{-\frac{ \mathrm{i} \pi z^2}{2} -\frac{\mathrm{i} \pi}{24}(b^2 +b^{-2})}
	(-p_+^{1/2} \mathrm{e}^{2 \pi bx};p_+^{1} )_\infty (-p_-^{1/2} \mathrm{e}^{2\pi b^{-1} x};p_- )_\infty .
	\]
	Although this expression is divergent for all values of $b$,
	the $\mathcal{O}(N)$ free energy formally computed in a similar way, leads us to the same result \eqref{eq:F_upper}.
} ${\rm Im}(b^2 )>0$ and ${\rm Im}(b^2 )<0$.

%%%%%%%%%%%%%%%%%%%%%%%
%%%%%%%%%%%%%%%%%%%%%%%
\subsection{The \boldmath $\mathcal{O}(N)$ free energy for real and irrational \boldmath $b^2$}
%%%%%%%%%%%%%%%%%%%%%%%
%%%%%%%%%%%%%%%%%%%%%%%
We have seen that the expression \eqref{eq:F_upper} is valid for all ${\rm Im}(b^2 ) \neq 0$.
It is natural to wonder if \eqref{eq:F_upper} still makes sense for real $b^2$. 
To see this, we take a limit of \eqref{eq:F_upper} from complex $b^2$ to real $b^2$.
We can easily see that the limit to irrational $b^2$ is perfectly smooth and finite.
%while it is apparently divergent for rational $b^2$ (we will discuss this issue in appendix \ref{rational}).
This implies that
the expression \eqref{eq:F_upper} is still valid for real and irrational $b^2$. The arguments reviewed above then suggest that it should be equal to our topological string partition function computations.

For the comparison to the open string side on the deformed conifold, 
it is useful to rewrite the free energy \eqref{eq:beforeVEV} as
\begin{equation}
\tilde{\mathrm{F}}|_{\mathcal{O}(N)}  
= \sum_{w=1}^\infty \frac{1}{w}
\left( \frac{1}{2 \mathrm{i}} \frac{\mathrm{e}^{\mathrm{i} \pi b^2 w}}{\sin{(\pi b^2 w)}} \right) \left< {\rm Tr}(U_0^w) \right>_{b^2} \mathrm{e}^{2\pi b \mathrm{M} w}
+(b \rightarrow 1/b) .
\label{eq:QFT_decomposition} 
\end{equation}
%it is useful to rewrite our result \eqref{finalfield} in the form
%\begin{equation} \label{FinalClosedFreeEnergy2}
% \tilde{F}|_{\mathcal{O}(N)}   = \frac{N}{2it} \sum\limits_{w=1}^{\infty} \sum\limits_{d=0}^{w} \frac{ (-1)^{d+1} e^{\pi i b^2 w}}{w w! \ \mathrm{sin} \left(\pi b^2 w \right)} {w \choose d} \prod\limits_{k=1}^{w-1} \left( b^2 w + d - k \right) \mathrm{e}^{ 2 \pi b m' w} \mathrm{e}^{ - d t} + \left( b \rightarrow 1/b \right),
%\end{equation}
%where we plugged in the form of $W_{(w,\ell)}$ and renamed $\ell \to d$, $j+\ell \to k$.
Comparing to \eqref{eq:annulus_decomposition},
%\eqref{FinalClosedFreeEnergy}, 
we see that the expressions are very similar if we identify
\begin{equation}
\mathrm{M} = - \mathrm{M}_d, \qquad \mathrm{t} = \mathrm{t}_d
\end{equation}
to get
\begin{equation}
\tilde{\mathrm{F}}|_{\mathcal{O}(N)}  
= \sum_{w=1}^\infty \frac{1}{w}
\left( \frac{1}{2 \mathrm{i}} \frac{\mathrm{e}^{\mathrm{i} \pi b^2 w}}{\sin{(\pi b^2 w)}} \right)    \left< {\rm Tr}(U_0^w) \right>_{b^2} \mathrm{e}^{-2\pi b \mathrm{M}_d w}
+(b\rightarrow 1/b) .
\end{equation}
%$m'+i{b\over 2}$ with $-(\mathrm{M}_r + \frac{q}{4 \pi} \mathrm{t}_r)$.
The main difference is that the factor of $\mathrm{sin} (\pi b^2 w)$ is now in the denominator instead of the numerator. 

Similarly, since we have seen the agreement between the open and closed string sides
under the identifications $\mathrm{M}_d = \mathrm{M}_r$ and $\mathrm{t}_d =- \mathrm{t}_r$, the QFT free energy is also similar to the one of the closed string on the resolved conifold if we identify
\begin{equation}
\mathrm{M} = - \mathrm{M}_r, \qquad \mathrm{t} = - \mathrm{t}_r.
\end{equation}
Again the main difference is whether the factor $\mathrm{sin} (\pi b^2 w)$ is in the denominator or the numerator.

%%%%%%%%%%%%%%%%%%%%%%%
%%%%%%%%%%%%%%%%%%%%%%%
\subsection{The \boldmath $\mathcal{O}(N)$ free energy for rational \boldmath $b^2$} \label{sec:rational}
%%%%%%%%%%%%%%%%%%%%%%%
%%%%%%%%%%%%%%%%%%%%%%%
Finally, let us consider the rational $b^2$ case, where we take $b^2 =b_1 /b_2$ with coprime integers $(b_1 ,b_2 )$.
The limit of \eqref{eq:F_upper} to rational $b^2$ is subtle 
since the prefactors $p_\pm^w /(1-p_\pm^w )$ become divergent for some $w$ with $p_\pm^w =1$. Nevertheless, we will show in appendix~\ref{rational} that the apparent divergences are finally canceled and \eqref{eq:F_upper} has a finite limit, which is the same as the result of a direct computation of the partition function at $b^2 =b_1 /b_2$. The final result is
\begin{align} \label{eq:result_rational}
\begin{split}
\tilde{\mathrm{F}} |_{\mathcal{O}(N)} = &- \sum_{w_1 = 0}^{\infty}   \sum_{w_2 = 1}^{b_2 - 1} \left. \frac{\mathrm{e}^{2\pi \sqrt{\frac{b_1}{b_2}}\mathrm{M} w }}{w}
\frac{\mathrm{e}^{2\pi \mathrm{i} \frac{b_1}{b_2} w_2} }{1-\mathrm{e}^{2\pi \mathrm{i} \frac{b_1}{b_2} w_2}}
\left< {\rm Tr} \left[ \mathrm{e}^{2 \pi \sqrt{\frac{b_1}{b_2}} x w} \right] \right> \right|_{w=b_2 w_1 +w_2}
+(b_1 \leftrightarrow b_2 ) \\
& + \sum_{w_1 =1}^\infty \frac{\mathrm{e}^{2\pi\sqrt{b_1 b_2} \mathrm{M} w_1}}{w_1}  
\left< {\rm Tr} \left[ \left( \frac{\mathrm{i}}{2\pi b_1 b_2 w_1 } - \frac{\mathrm{i}( \mathrm{i} Q/2 + \mathrm{M} +x)}{\sqrt{b_1 b_2}} \right) \mathrm{e}^{2\pi \sqrt{b_1 b_2 } x w_1} \right] \right>.
\end{split}
\end{align}
Here $\left< \cdots \right>$ actually means $\left< \cdots \right>_{{\rm CS}_{\sqrt{b_1 /b_2}}}$.

While most of the terms appearing here are the Wilson loops \eqref{eq:W_planar} in the pure CS theory, we also have the quantity
\be
\frac{1}{N} \left< {\rm Tr} \left[ x \mathrm{e}^{2\pi \sqrt{b_1 b_2} x w} \right] \right>_{{\rm CS}_{\sqrt{b_1 /b_2}}} ,
\ee
whose physical interpretation is unclear\footnote{
	For $b=1$, this quantity is proportional to Bremsstrahlung function \cite{LewkowyczMaldacena}.
}.
One can compute this quantity by considering the derivative with respect to $w$:\footnote{
	Differentiation by $w$ is subtle, 
	since we have assumed $w\in\mathbb{Z}$ in the derivation.
	Yet most parts of $\left< {\rm Tr} \left[ \mathrm{e}^{2\pi \sqrt{b_1 b_2} x w} \right] \right>_{{\rm CS}_b} $ with $w\in\mathbb{Z}$
	seem to have a natural extension to complex $w$. 
	The only exception is the factor $(-1)^w$ in $W_{w,d}(f)$ \eqref{eq:W_wl}
	whose extension is quite ambiguous.
	Here we have chosen $(-1)^w =\cos{(\pi w)}$ which leads us to a vanishing derivative at $w\in\mathbb{Z}$,
	while other choices give an imaginary part.
	Heuristically this seems to be the appropriate choice, as this choice has the expected behavior for $b=1$.
}
\begin{align}
\begin{split}
&\frac{1}{N} \left< {\rm Tr} \left[ x \mathrm{e}^{2 \pi \sqrt{b_1 b_2} x w} \right] \right>_{{\rm CS}_b} 
= \frac{1}{2\pi N \sqrt{b_1 b_2}}\frac{\del}{\del w} 
\left< {\rm Tr} \left[ \mathrm{e}^{2\pi \sqrt{b_1 b_2} x w} \right] \right>_{{\rm CS}_b} \\
&= \frac{(b_1 -b_2 ) \mathrm{t}}  {4 \pi N \sqrt{b_1 b_2}}  
\left< {\rm Tr} \left[ \mathrm{e}^{2\pi \sqrt{b_1 b_2} x w} \right] \right>_{{\rm CS}_b}
+\frac{\mathrm{e}^{\frac{b_1 - b_2}{2}w \mathrm{t}}}{2\pi \mathrm{t} \sqrt{b_1 b_2}} 
\sum_{d = 0}^{b_2 w} \mathrm{e}^{d \mathrm{t}}  \frac{\del}{\del w} 
\left( b_2  w W_{b_2 w, d }\left( \frac{b_1}{b_2} \right) \right)  , 
\end{split}
\end{align}
where
\begin{align}
\begin{split}
&\frac{\del}{\del w} \left[ w W_{b_2 w, d } \left( \frac{b_1}{b_2} \right) \right] =  w W_{b_2 w, d } \left( \frac{b_1}{b_2} \right) \Biggl[ b_1 \psi_0 (b_1 w) -(b_1 - b_2 ) \psi_0 (b_1 w -b_2 w +1) \\ &- b_2 \psi_0^d (-b_2 w) 
- b_2 \psi_0 (b_2 w +1) + b_1 \psi_0^d (b_1 w )  -(b_1 -b_2 ) \psi_0^d (b_1 w -b_2 w +1 ) \Biggr]. 
\end{split}
\end{align}

%%%%%%%%%%%%%%%%%%%%%%%
%%%%%%%%%%%%%%%%%%%%%%%
%%%%%%%%%%%%%%%%%%%%%%%
\section{Conclusions and future directions} \label{sec:conclusions}
%%%%%%%%%%%%%%%%%%%%%%%
%%%%%%%%%%%%%%%%%%%%%%%
%%%%%%%%%%%%%%%%%%%%%%%
In this paper, we considered adding matter to the CS theory on $S^3$, in its dual description as a topological string on the deformed conifold $X_d = T^* S^3$. As suggested in \cite{AganagicCostelloMcNamaraVafa}, a specific type of matter may be described by adding coisotropic 5-branes to the conifold with $N$ Lagrangian 3-branes wrapping its zero section. A generalization of the GV duality suggests an equivalent description of this in terms of coisotropic 5-branes on the resolved conifold $X_r = {\cal O} \left( - 1 \right) \oplus {\cal O} \left( - 1 \right) \rightarrow \mathbb{CP}^1$, and we conjectured a duality of this form. Using the equivariant localization, we computed the leading order topological A-model vacuum amplitudes in the deformed and resolved conifold backgrounds with coisotropic branes, and found an agreement between the two, testing the conjectured duality. Surprisingly, our result differs slightly from the large $N$ limit of the partition function of the specific PTCSM theory, which was proposed in \cite{AganagicCostelloMcNamaraVafa} as a field theory description of the topological A-model on the deformed conifold with coisotropic branes. There are several possible resolutions of the discrepancy.

One is that our topological string computations are not precise, for instance because the equivariant localization method does not work properly for topological strings with coisotropic boundary conditions. As we discussed in section \ref{sec:GVduality}, even in the case of Lagrangian boundary conditions, one has to tune the toric weights appearing in the A-model amplitude appropriately in order to match the Wilson loop framing dependence in the CS theory. It seems possible that such a procedure fails for more general coisotropic boundary conditions. It is also possible that equivariant localization for coisotropic branes leads to extra quantum mechanical degrees of freedom that need to separately be taken into account, as in \cite{WittenNewLook}. However, if the topological string computations are not correct, it is somewhat surprising that the two separate computations agree with each other.

Another possibility is that our field theory computation of the 3d ${\cal N}=2$ partition function at large $N$ is not correct. 
One might think that since our computation used the Wilson loops in the pure CS theory with irrational framing rather than the standard integer one,
there might be subtleties.
%The computation uses a Chern-Simons theory with irrational framing rather than the standard integer one, and there may be subtleties there.
However, this point seems irrelevant for the mismatch
since the mismatch between the open string and field theory sides
appears at the level of the decompositions in terms of the Wilson loops,
as seen from \eqref{eq:annulus_decomposition} and \eqref{eq:QFT_decomposition}.
Therefore we are not currently aware of any subtleties relevant for the mismatch in the field theory computation.

It is also possible that one of the two claims from \cite{AganagicCostelloMcNamaraVafa} that we used is not correct. One claim is that the topological string with 3-branes and 5-branes gives rise to the PTCSM theory \eqref{chernsimonsmatteraction}. This was not derived directly on the conifold but rather in flat space, so there may be subtleties in the derivation. Perhaps, as discussed in \cite{AganagicCostelloMcNamaraVafa}, the PTCSM theory may be generalized to include a non-trivial potential, and this may solve the problem. The other claim we used is that the partition function of \eqref{chernsimonsmatteraction} is the same as that of the 3d ${\cal N}=2$ partition function with the same THF structure, and again there may be various subtleties here.

It is important to understand the mismatch in order to properly incorporate coisotropic branes into the GV duality, and thus into the beautiful web of dualities between the CS theory and topological string theories.

There are many future directions to consider.

Additional computations may shed light on the discrepancy, and also be interesting in their own right. In this paper, on the topological string side we only considered the case of irrational $b^2$, while on the field theory side we computed in section \ref{sec:rational} the result also for rational $b^2$. For rational $b^2$ there are extra surfaces contributing on the topological string side, and in particular continuous families of surfaces contribute, modifying the computation. It would be nice to repeat the computation in this case, and to see to what extent it agrees with the field theory result \eqref{eq:result_rational}. It may also be possible to go to higher orders in $g_s$, as we briefly discussed in section \ref{sec:higher}.

An important aspect of any holographic duality is the map between observables. In this paper we considered the partition functions on both sides of the duality, but as discussed in section \ref{sec:PTSMReview}, there are also local observables in the PTCSM. Gauge-invariant combinations of these observables should map to operators made from the 5-5 strings on the closed string side. Due to topologicity of the theory, there is no room for any non-trivial coordinate dependence of the correlation functions, but it is still very interesting to understand their physical significance, and their images on the closed string side of the duality.

It would be interesting to generalize our discussion to other manifolds. For most spaces the topological string dual of CS theory is not known, but for Lens spaces the GV duality is discussed in \cite{LensSpaces1, LensSpaces2}, and the 3d ${\cal N}=2$ partition function is also known, so it should be possible to repeat all of our computations.

One of the most interesting dualities involving Lagrangian branes in the A-model is mirror symmetry. The original construction \cite{HoriVafa} of this duality was generalized to include Lagrangian branes in \cite{AganagicVafa, AganagicKlemmVafa}, and to the simplest configuration of coisotropic branes in \cite{CoisotropicNonCommutative}, but an understanding of the duality for more general 5-brane setups is lacking.

It would also be great to include 5-branes in the Topological Vertex formalism \cite{TheTopologicalVertex}, which allows to find the A-model solutions for very complicated toric backgrounds by appropriately `gluing' the topological string amplitudes in simpler backgrounds.

Sometimes, the topological string dualities can be embedded into physical string theories \cite{VafaSuperstrings}, and into M-theory \cite{TopStringsMTheory1, TopStringsMTheory2, Acharya, AtiyahMaldacenaVafa, DijkgraafVafaVerlinde, AganagicOoguriVafaYamazaki, DedushWitten}. Therefore, inclusion of coisotropic branes in the GV duality may shed light also on some aspects of the physical branes dynamics.

Another problem which better understanding of topological 5-brane dynamics could help to solve is the relation between the higher-spin Vasiliev theory \cite{FradkinVasiliev1, FradkinVasiliev2} and string theory. It is known (see \cite{KP} for the original proposal on holography for vector models, and \cite{GiombiReview} for a recent review) that the Vasiliev theory on $AdS_4$ is holographically dual to the CSM theory on $S^3$,\footnote{There are also some generalizations of this duality with supersymmetries and with non-Abelian gauge symmetries on both sides \cite{ABJTriality, ABJQuadrality}.} so we hope that the PTCSM/A-model duality can be a step towards inclusion of the Vasiliev theory into string theory. In particular, the topological string background with coisotropic branes that we found should be equivalent to some twist of the version of the Vasiliev higher-spin gravity on $AdS_4$ that is dual to the 3d ${\cal N}=2$ CSM theory \cite{ABJTriality}. It would be interesting to understand this better.

There should also be a relation between the GV duality with coisotropic branes and the topological holography program, whose aim is to construct the holographic duals to topologically (or holomorphically) twisted gauge theories \cite{Yangian, TwistedSupergravityAndItsQuantization, CostelloMTheoryOmegaBackground, HolographyAndKoszulDuality,CostelloGaiotto}, because the PTCSM theory is related to the holomorphic twist of supersymmetric Yang-Mills theory via a dimensional reduction.

All this suggests that there may exist a web of dualities between the PTCSM theories, topological string theories on conifolds and their generalizations (whose solutions are constructed using the Topological Vertex, as we mentioned), string theories on $AdS_4 \times M$, Vasiliev theories on $AdS_4$, and certain generalizations \cite{TwistedSupergravityAndItsQuantization} of BCOV theory \cite{BCOVLong}.

\section*{Acknowledgements}
We would like to thank  E. Balzin, M. Isachenkov, Z. Komargodski, J. McNamara, and P. Putrov for useful discussions, and to thank especially M. Aganagic for useful comments on a draft of this manuscript. We also thank C. Chen for pointing out misprints in the first version of this paper. AF would like to thank Novosibirsk State University for hospitality. This work was supported in part  by an Israel Science Foundation center for excellence grant (grant number 1989/14) and by the Minerva foundation with funding from the Federal German Ministry for Education and Research. OA is the Samuel Sebba Professorial Chair of Pure and Applied Physics. 
The work of MH has been partially supported by STFC consolidated grant ST/P000681/1.

%%%%%%%%%%%%%%%%%%%%%%%
%%%%%%%%%%%%%%%%%%%%%%%
%%%%%%%%%%%%%%%%%%%%%%%
\appendix
\section{Derivations of the \boldmath $\mathcal{O}(N)$ free energy for rational \boldmath $b^2$} 
\label{rational}

In this appendix we present two derivations of the result \eqref{eq:result_rational} for rational $b^2$.
%%%%%%%%%%%%%%%%%%%%%%%
%%%%%%%%%%%%%%%%%%%%%%%
\subsection{The limit to rational \boldmath $b^2$ from complex \boldmath $b^2$}
%%%%%%%%%%%%%%%%%%%%%%%
%%%%%%%%%%%%%%%%%%%%%%%
%In this appendix we consider the rational $b^2$ case,
%where we take $b^2 =b_1 /b_2$ with coprime integers $(b_1 ,b_2 )$.
The limit of \eqref{eq:F_upper} to rational $b^2$ is subtle 
since the prefactors $p_\pm^w /(1-p_\pm^w )$ become divergent for some $w$ with $p_\pm^w =1$.
To see this explicitly, let us rewrite $\tilde{\mathrm{F}}|_{\mathcal{O}(N)}$ as
\be \label{fonapp}
\tilde{\mathrm{F}}|_{\mathcal{O}(N)}  
= - \frac{N}{\mathrm{t}} \sum_{w=1}^\infty \left[ \frac{\mathrm{e}^{2\pi \mathrm{i} b^2 w} }{1-\mathrm{e}^{2\pi \mathrm{i} b^2 w}} f_w (b^2) + \left( b \rightarrow 1/b \right) \right],
\ee
where
\be
f_w (b^2 ) 
= \mathrm{e}^{2\pi b \mathrm{M} w}   \sum_{d=0}^w  \mathrm{e}^{\frac{(b^2 -1)w \mathrm{t}}{2}}  W_{w,d }(b^2 ) \mathrm{e}^{d \mathrm{t}}  
=\frac{\mathrm{t}}{N} \frac{\mathrm{e}^{2\pi b \mathrm{M} w}}{w} \left< {\rm Tr} \left[ \mathrm{e}^{2\pi b x w} \right] \right>_{{\rm CS}_b} .
\ee
We can easily see in the limit $b\rightarrow \sqrt{\frac{b_1}{b_2}}$ that 
the first and second terms in \eqref{fonapp} are divergent for $w=b_2 \mathbb{Z}$ and $w=b_1 \mathbb{Z}$ respectively.
Therefore, it is convenient to decompose $\tilde{\mathrm{F}}|_{\mathcal{O}(N)}$  into a harmless part and an apparently divergent part
\be
\tilde{\mathrm{F}}|_{\mathcal{O}(N)}  
=  \mathrm{F}_{\rm safe} + \mathrm{F}_{\rm danger} ,
\ee
where 
\begin{align} \label{safedanger}
\begin{split}
&\mathrm{F}_{\rm safe} = - \frac{N}{\mathrm{t}} \sum_{w_1 =0}^\infty \left[
\sum_{w_2 =1}^{b_2 -1}  \left.  \frac{\mathrm{e}^{2\pi \mathrm{i} b^2 w} }{1- \mathrm{e}^{2 \pi \mathrm{i} b^2 w}} f_w (b^2)  \right|_{w=b_2 w_1 +w_2} 
+ \left( b_1 \leftrightarrow b_2 \right) \right], \\
&\mathrm{F}_{\rm danger} 
= - \frac{N}{\mathrm{\mathrm{t}}} \sum_{w_1 = 1}^\infty \left[  
\left. \frac{\mathrm{e}^{2\pi \mathrm{i} b^2 w} }{1- \mathrm{e}^{2\pi \mathrm{i} b^2 w}} f_w (b^2)  \right|_{w=b_2 w_1}
+ \left( b_1 \leftrightarrow b_2 \right)  \right].
\end{split}
\end{align}

$\mathrm{F}_{\rm safe}$ does not have any divergence when $b \rightarrow \sqrt{\frac{b_1}{b_2}}$, and we can easily take the limit
\be
\lim_{b\rightarrow \sqrt{b_1 /b_2}} \mathrm{F}_{\rm safe} 
= - \frac{N}{\mathrm{t}}  \sum_{w_1 =0}^\infty \left[
\sum_{w_2 =1}^{b_2 -1} \frac{\mathrm{e}^{2\pi \mathrm{i} \frac{b_1}{b_2} w_2} }{1-\mathrm{e}^{2\pi \mathrm{i} \frac{b_1}{b_2} w_2}} f_{b_2 w_1 +w_2} \left( \frac{b_1}{b_2} \right) 
+ \left( b_1 \leftrightarrow b_2 \right) \right]  .
\ee

We have to be careful with $\mathrm{F}_{\rm danger}$, since it is apparently divergent. However, we will show that the apparent divergence is canceled and $\mathrm{F}_{\rm danger}$ has a finite limit.  To see this, first note that as we approach a rational value
\begin{align}
\begin{split}
\left. \frac{\mathrm{e}^{ 2\pi \mathrm{i} b^2 w}}{1- \mathrm{e}^{ 2\pi \mathrm{i} b^2 w}} \right|_{w=b_2 w_1} 
&=-\frac{1}{2\pi \mathrm{i} b_2 w_1} \frac{1}{\delta b^2 }  -\frac{1}{2}  +\mathcal{O}(\delta b^2 ), \\
\left. \frac{\mathrm{e}^{ 2 \pi \mathrm{i} b^{-2} w}}{1- \mathrm{e}^{ 2\pi \mathrm{i} b^{-2} w}} \right|_{w=b_1 w_1}
&= +\frac{1}{2\pi \mathrm{i} b_1 w_1} \left( \frac{b_1}{b_2} \right)^2 \frac{1}{\delta b^2 }  -\frac{1}{2} -\frac{\mathrm{i}}{2\pi b_2 w_1} +\mathcal{O}(\delta b^2 ) ,
\end{split}
\end{align}
where $\delta b^2 =b^2 -b_1 /b_2$.
Then, the divergent part is
\bea
\frac{1}{2\pi \mathrm{i} \delta b^2} \frac{N}{\mathrm{t}} 
\sum_{w_1 =1}^\infty  \left[  
-\frac{1}{b_2 w_1} f_{b_2 w_1} \left( \frac{b_1}{b_2} \right)  
+ \frac{1}{b_1 w_1} \left( \frac{b_1}{b_2} \right)^2 f_{b_1 w_1} \left( \frac{b_2}{b_1} \right) \right].
\eea
We immediately see that this is canceled, because
\be
f_{b_2 w_1} \left( \frac{b_1}{b_2} \right) 
=\frac{\mathrm{t}}{N} \frac{\mathrm{e}^{2\pi\sqrt{b_1 b_2} \mathrm{M} w_1}}{b_2 w_1}
\left< {\rm Tr} \left[ \mathrm{e}^{2\pi \sqrt{b_1 b_2} x w_1} \right]
\right>_{{\rm CS}_{\sqrt{b_1 /b_2}}} 
=\frac{b_1}{b_2} f_{b_1 w_1} \left( \frac{b_2}{b_1} \right)  .
\label{eq:id_f}
\ee
Thus, expanding in powers of $\delta b^2$, the expression \eqref{safedanger} has the following finite limit:
\begin{align}
\begin{split}
\lim_{b\rightarrow \sqrt{b_1 /b_2}} \mathrm{F}_{\rm danger} 
= &- \frac{N}{\mathrm{t}} \sum_{w_1 =1}^\infty \Biggl[ 
-\frac{1}{2} f_{b_2 w_1}\left( \frac{b_1}{b_2} \right)
-\left( \frac{1}{2} +\frac{\mathrm{i}}{2\pi b_2 w_1 } \right) f_{b_1 w_1} \left( \frac{b_2}{b_1} \right) \\
& -\frac{1}{2\pi \mathrm{i} b_2 w_1 }  f_{b_2 w_1}^{\prime} \left( \frac{b_1}{b_2} \right) - \frac{1}{2\pi \mathrm{i} b_1 w_1 }  f_{b_1 w_1}^{\prime} \left( \frac{b_2}{b_1} \right) \Biggr].
\end{split}
\end{align}

Combining this with $\mathrm{F}_{\rm safe}$, we obtain
\begin{align} \label{eq:F_limit}
\begin{split}
\lim_{b\rightarrow \sqrt{b_1 /b_2}} \tilde{\mathrm{F}} |_{\mathcal{O}(N)}  
= &- \sum_{w_1 =0}^\infty \sum_{w_2 = 1}^{b_2 - 1} \left. \frac{\mathrm{e}^{2\pi \sqrt{\frac{b_1}{b_2}} \mathrm{M} w }}{w}
\frac{\mathrm{e}^{2\pi \mathrm{i} \frac{b_1}{b_2} w_2} }{1- \mathrm{e}^{2\pi \mathrm{i} \frac{b_1}{b_2} w_2}}
\left< {\rm Tr} \left[ \mathrm{e}^{2\pi \sqrt{\frac{b_1}{b_2}} x w} \right] \right> \right|_{w=b_2 w_1 +w_2}
+(b_1 \leftrightarrow b_2 ) \\
&+ \sum_{w_1 =1}^\infty \frac{\mathrm{e}^{2\pi\sqrt{b_1 b_2} \mathrm{M} w_1}}{w_1}  
\left< {\rm Tr} \left\{ \left[ \frac{\mathrm{i}}{2\pi b_1 b_2 w_1 } - \frac{\mathrm{i}(\mathrm{i} Q/2 + \mathrm{M} +x)}{\sqrt{b_1 b_2}} \right]
\mathrm{e}^{2\pi \sqrt{b_1 b_2 } x w_1} \right\} \right>,
\end{split}
\end{align}
where the VEVs are $\left< \cdots \right> = \left< \cdots \right>_{{\rm CS}_{\sqrt{b_1 /b_2}}}$, and we have used\footnote{
	This identity is correct only for large $N$.
	For finite $N$, we also have corrections from the derivative of 
	the factor $\prod\limits_{i<j} \sinh{(\pi b x_{ij})} \sinh{(\pi b^{-1} x_{ij})}$ with respect to $b$.
}
\be
-\frac{1}{2\pi \mathrm{i}  w_1 } \left[ \frac{1}{b_2}  f_{b_2 w_1}^{\prime} \left( \frac{b_1}{b_2} \right) 
+\frac{1}{ b_1}  f_{b_1 w_1}^{\prime} \left( \frac{b_2}{b_1} \right) \right] 
=\frac{\mathrm{i}}{w_1 \sqrt{b_1 b_2}} 
\left< {\rm Tr} \left[ ( \mathrm{M} + x) \mathrm{e}^{2\pi \sqrt{b_1 b_2 } x w_1} \right] \right>_{{\rm CS}_{\sqrt{b_1 /b_2}}} .
\ee
As we discuss in the next subsection, the limit \eqref{eq:F_limit} of the free energy to rational $b^2$ from the complex plane agrees
with the result directly computed at $b^2 =b_1 /b_2$.

%%%%%%%%%%%%%%%%%%%%%%%
%%%%%%%%%%%%%%%%%%%%%%%
%%%%%%%%%%%%%%%%%%%%%%%
\subsection{A direct computation at rational \boldmath $b^2$}
\label{app:direct}
%%%%%%%%%%%%%%%%%%%%%%%
%%%%%%%%%%%%%%%%%%%%%%%
%%%%%%%%%%%%%%%%%%%%%%%
The double sine function for rational $b^2$ has the following representation \cite{GaroufalidisKashaev}:

\begin{align}
\begin{split}
\log \left[ s_{\sqrt{\frac{b_1}{b_2}}} (z) \right]
= &-\frac{\mathrm{i} \pi z^2}{2}  -\frac{ \mathrm{i}\pi}{24} \left( b^2 + \frac{1}{b^2} \right) 
+\frac{\mathrm{i}}{2 \pi b_1 b_2}{\rm Li}_2 \left[ (-1)^{b_1 +b_2}  \mathrm{e}^{2\pi\sqrt{b_1 b_2}z} \right] \\
& +\left( 1 - \frac{b_1^{-1} +b_2^{-1}}{2} +\frac{\mathrm{i} z}{\sqrt{b_1 b_2}} \right)
\log{\left[ 1 -(-1)^{b_1 +b_2} \mathrm{e}^{2\pi\sqrt{b_1 b_2}z}  \right]} \\
& -\sum_{w_2 =1}^{b_2 -1} \frac{w_2}{b_2}
\log{\left[ 1  +(-1)^{b_1 +b_2} \mathrm{e}^{2\pi \mathrm{i} \frac{b_1}{b_2}(w_2 + 1/2)} \mathrm{e}^{2\pi \sqrt{\frac{b_1}{b_2}} z} \right]} 
+(b_1 \leftrightarrow  b_2).
\end{split}
\end{align}
For our purpose, it is more convenient to rewrite this as
\begin{align}
\begin{split}
\log (s_{\sqrt{\frac{b_1}{b_2}}} (z)) 
= &-\frac{\mathrm{i} \pi z^2}{2} - \frac{\mathrm{i} \pi}{24} \left( b^2 + \frac{1}{b^2} \right) +\frac{\mathrm{i}}{2\pi b_1 b_2}{\rm Li}_2 \left[ (-1)^{b_1 +b_2} \mathrm{e}^{2\pi\sqrt{b_1 b_2}z} \right] 
\\ &+\frac{\mathrm{i}z}{\sqrt{b_1 b_2}} \log{\left[ 1 -(-1)^{b_1 +b_2} \mathrm{e}^{2\pi\sqrt{b_1 b_2}z} \right]} \\
&-\sum_{w_1 =0}^\infty \sum_{w_2 =1}^{b_2 -1}  
\left. \frac{(-1)^w}{w} 
\frac{\mathrm{e}^{\pi \mathrm{i} \frac{b_1}{b_2} w} }{1- \mathrm{e}^{2\pi \mathrm{i} \frac{b_1}{b_2} w}}
\mathrm{e}^{2\pi \sqrt{\frac{b_1}{b_2}} z w} \right|_{w=b_2 w_1 +w_2}
+(b_1 \leftrightarrow  b_2),
\end{split}
\end{align}
where we have used
\begin{align}
\begin{split}
&\sum_{w_1 =0}^\infty \sum_{w_2 =1}^{b_2 -1}  
\left. \frac{(-1)^w}{w} 
\frac{\mathrm{e}^{\pi \mathrm{i} \frac{b_1}{b_2} w} }{1-\mathrm{e}^{2\pi \mathrm{i} \frac{b_1}{b_2} w}}
\mathrm{e}^{2\pi \sqrt{\frac{b_1}{b_2}} z w} \right|_{w=b_2 w_1 +w_2} \\ &= \frac{b_2^{-1}-1}{2}  
\log{\left[ 1-(-1)^{b_1 +b_2}e^{2\pi \sqrt{b_1 b_2} z} \right]}
\\ &+ \sum_{w_2 =1}^{b_2 -1} \frac{w_2}{b_2}
\log{\left[ 1  +(-1)^{b_1 +b_2} \mathrm{e}^{2\pi \mathrm{i} \frac{b_1}{b_2}(w_2 +1/2)} \mathrm{e}^{2\pi \sqrt{\frac{b_1}{b_2}} z} \right]}.
\end{split}
\end{align}
Then, the partition function is given by
\begin{align}
\begin{split}
\tilde{Z} = &\frac{1}{N!}
\int_{-\infty -\frac{\mathrm{i} Q}{4k'}}^{\infty -\frac{\mathrm{i} Q}{4k'}} \mathrm{d}^N x 
\prod_{i<j} \sinh{(\pi b x_{ij})} \sinh{(\pi b^{-1} x_{ij})} 
\prod_{j=1}^N \mathrm{e}^{\mathrm{i} k' \pi x_j^2} \\
&\times \exp \Biggl\{
\frac{\mathrm{i}}{2\pi b_1 b_2}{\rm Li}_2 \left[\mathrm{e}^{2\pi\sqrt{b_1 b_2}(\tilde{\mathrm{M}} - x_j )} \right] 
+\frac{\mathrm{i} (\mathrm{i} Q/2 +\tilde{\mathrm{M}}-x_j )}{\sqrt{b_1 b_2}}  \log{\left[ 1 - \mathrm{e}^{2\pi\sqrt{b_1 b_2}(\tilde{\mathrm{M}}-x_j )}  \right] } \\
&-\sum_{w_1 =0}^\infty \sum_{w_2 =1}^{b_2 -1}  
\left. \frac{1}{w} 
\frac{\mathrm{e}^{2\pi \mathrm{i} \frac{b_1}{b_2} w} }{1-\mathrm{e}^{2\pi \mathrm{i} \frac{b_1}{b_2} w}}
\mathrm{e}^{2\pi \sqrt{\frac{b_1}{b_2}} \left(\tilde{\mathrm{M}}-x_j \right)w} \right|_{w=b_2 w_1 +w_2} +(b_1 \leftrightarrow  b_2)  \Biggr\} .
\end{split}
\end{align}
As in the main text, the $\mathcal{O}(N)$ free energy can be computed by
\begin{align} \label{eq:beforeVEV_rational}
\begin{split}
\tilde{\mathrm{F}} |_{\mathcal{O}(N)} 
= &\left< {\rm Tr}\left\{ \frac{\mathrm{i}}{2\pi b_1 b_2}{\rm Li}_2 \left[\mathrm{e}^{2\pi\sqrt{b_1 b_2}(\mathrm{M} - x_j )} \right]
+ \frac{\mathrm{i}(\mathrm{i} Q/2 + \mathrm{M}-x_j )}{\sqrt{b_1 b_2}}  
\log{\left[ 1 - \mathrm{e}^{2\pi\sqrt{b_1 b_2}(m-x_j )}  \right] } \right\} \right>_{{\rm CS}_b} \\
&- \sum_{w_1 =0}^\infty \sum_{w_2 =1}^{b_2 -1}  
\left. \frac{1}{w} \frac{\mathrm{e}^{2\pi \mathrm{i} \frac{b_1}{b_2} w} }{1-\mathrm{e}^{2\pi \mathrm{i} \frac{b_1}{b_2} w}}
\left< {\rm Tr} \left[ \mathrm{e}^{2\pi \sqrt{\frac{b_1}{b_2}} \left( \mathrm{M}-x \right) w} \right] \right>_{{\rm CS}_b} \right|_{w=b_2 w_1 +w_2}
+(b_1 \leftrightarrow  b_2)  \\
&= \sum_{w=1}^\infty \frac{\mathrm{e}^{2\pi\sqrt{b_1 b_2} \mathrm{M} w}}{w}
\left< {\rm Tr} \left\{ \left[ \frac{\mathrm{i}}{2\pi w b_1 b_2} - \frac{\mathrm{i}( \mathrm{i} Q/2 + \mathrm{M} + x)}{\sqrt{b_1 b_2}} \right] 
\mathrm{e}^{2\pi\sqrt{b_1 b_2} x w} \right\} \right>_{{\rm CS}_b} \\
&- \sum_{w_1 =0}^\infty \sum_{w_2 =1}^{b_2 -1}  
\left. \frac{\mathrm{e}^{2\pi \sqrt{\frac{b_1}{b_2}} \mathrm{M} w}}{w}
\frac{\mathrm{e}^{2\pi \mathrm{i} \frac{b_1}{b_2} w} }{1-\mathrm{e}^{2\pi \mathrm{i} \frac{b_1}{b_2} w}}
\left< {\rm Tr} \left[ \mathrm{e}^{2\pi \sqrt{\frac{b_1}{b_2}} x w} \right] \right>_{{\rm CS}_b} \right|_{w=b_2 w_1 +w_2}
+(b_1 \leftrightarrow  b_2)  .
\end{split}
\end{align}
We see that this result precisely agrees with the one we got in the previous subsection.

\newpage
%%%%%%%%%%%%%%%%%%%%%%%%%%%%%%%%%%%
%%%%%%%%%%%%%%%%%%%%%%%%%%%%%%%%%%%
%%%%%%%%%%%%%%%%%%%%%%%%%%%%%%%%%%%


\addcontentsline{toc}{section}{References}

\begin{thebibliography}{99}
	
	\bibitem{tHooft}
	G.~`t Hooft,
	\newblock {``A Planar Diagram Theory for Strong Interactions},''
	\newblock Nucl. Phys. \textbf{B72} (1974) 461.
	
	\bibitem{KontsevichAiry}
	M.~Kontsevich,
	\newblock {``Intersection Theory on the Moduli Space of Curves and the Matrix Airy Function},''
	\newblock Commun. Math. Phys. \textbf{147} (1992) 1-23.
	
	\bibitem{WittenIntersection}
	E.~Witten,
	\newblock {``Two-Dimensional Gravity and Intersection Theory on Moduli Space},''
	\newblock Surveys in Diff. Geom. \textbf{1} (1991) 243-310.
	
	\bibitem{GrossTaylor}
	D.~J.~Gross and W.~Taylor,
	\newblock {``Two-dimensional QCD is a String Theory},''
	\newblock Nucl. Phys. \textbf{B400} (1993) 181-208
	\newblock [arXiv:hep-th/9301068].
	
	\bibitem{BanksFischlerShenkerSusskind}
	T.~Banks, W.~Fischler, S.~Shenker and L.~Susskind,
	\newblock {``M Theory as a Matrix Model: A Conjecture},''
	\newblock Phys. Rev. \textbf{D55} (1997) 5112-5128
	\newblock [arXiv:hep-th/9610043 ].
	
	\bibitem{MaldacenaHol}
	J.~Maldacena,
	\newblock {``The Large $N$ Limit of Superconformal Field Theories and Supergravity},''
	\newblock Adv. Theor. Math. Phys. \textbf{2} (1998) 231-252
	\newblock [arXiv:hep-th/9711200].
	
	\bibitem{WittenHol}
	E.~Witten,
	\newblock {``Anti de Sitter Space and Holography},''
	\newblock Adv. Theor. Math. Phys. \textbf{2} (1998) 253-291
	\newblock [arXiv:hep-th/9802150].
	
	\bibitem{GKPHol}
	S.~Gubser, I.~Klebanov and A.~Polyakov,
	\newblock {``Gauge Theory Correlators from Noncritical String Theory},''
	\newblock Phys. Lett. \textbf{B428} (1998) 105-114
	\newblock [arXiv:hep-th/9802109].
	
	\bibitem{BerkovitsOoguriVafa}
	N.~Berkovits, H.~Ooguri and C.~Vafa,
	\newblock {``On the World Sheet Derivation of Large $N$ Dualities for the Superstring},''
	\newblock Commun. Math. Phys. \textbf{252} (2004) 259-274
	\newblock [arXiv:hep-th/0310118].
	
	\bibitem{BerkovitsVafa}
	N.~Berkovits and C.~Vafa,
	\newblock {``Towards a Worldsheet Derivation of the Maldacena Conjecture},''
	\newblock JHEP \textbf{0803} (2008) 031 and AIP Conf. Proc. \textbf{1031} (2008) 21-42
	\newblock [arXiv:0711.1799].
	
	\bibitem{GV}
	R.~Gopakumar and C.~Vafa,
	\newblock {``On the Gauge Theory/Geometry Correspondence},''
	\newblock Adv. Theor. Math. Phys. \textbf{3} (1999) 1415-1443
	\newblock [arXiv:hep-th/9811131].
	
	\bibitem{OV}
	H.~Ooguri and C.~Vafa,
	\newblock {``World Sheet Derivation of a Large $N$ Duality},''
	\newblock Nucl. Phys. \textbf{B641} (2002) 3-34
	\newblock [arXiv:hep-th/0205297].
	
	\bibitem{2DPhases}
	E.~Witten,
	\newblock {``Phases of ${\cal N}=2$ Theories in Two Dimensions},''
	\newblock Nucl. Phys. \textbf{B403} (1993) 159-222
	\newblock [arXiv:hep-th/9301042].
		
	\bibitem{WittenChernString}
	E.~Witten,
	\newblock {``Chern-Simons Gauge Theory as a String Theory},''
	\newblock Prog. Math. \textbf{133} (1995) 637-678
	\newblock [arXiv:hep-th/9207094].
	
	\bibitem{GiombiReview}
	S.~Giombi,
	\newblock {``Higher Spin — CFT Duality},''
	\newblock [arXiv:1607.02967].
	
	\bibitem{AganagicCostelloMcNamaraVafa}
	M.~Aganagic, K.~Costello, J.~McNamara and C.~Vafa,
	\newblock {``Topological Chern-Simons/Matter Theories},''
	\newblock [arXiv:1706.09977].
	
	\bibitem{CandelasdelaOssa}
	P.~Candelas and X.~C.~de~la~Ossa,
	\newblock {``Comments on Conifolds},''
	\newblock Nucl. Phys. \textbf{B342} (1990) 246-268.
	
	\bibitem{WittenChernSimons}
	E.~Witten,
	\newblock {``Quantum Field Theory and the Jones Polynomial},''
	\newblock Commun. Math. Phys. \textbf{121} (1989) 351-399.
	
	\bibitem{KnotInvariantsAndTopologicalStrings}
	H.~Ooguri and C.~Vafa,
	\newblock {``Knot Invariants and Topological Strings},''
	\newblock Nucl. Phys. \textbf{B577} (2000) 419-438
	\newblock [arXiv:hep-th/9912123].
	
	\bibitem{LabastidaMarinoVafa}
	J.~Labastida, M.~Mari\~{n}o and C.~Vafa,
	\newblock {``Knots, Links and Branes at Large $N$},''
	\newblock JHEP \textbf{0011} (2000) 007
	\newblock [arXiv:hep-th/0010102].
	
	\bibitem{FramedKnotsAtLargeN}
	M.~Mari\~{n}o and C.~Vafa,
	\newblock {``Framed Knots at Large $N$},''
	\newblock Contemp. Math. \textbf{310} (2002) 185-204
	\newblock [arXiv:hep-th/0108064].
	
	\bibitem{VafaAlgebraicKnots}
	D.~Diaconescu, V.~Shende and C.~Vafa,
	\newblock {``Large N Duality, Lagrangian Cycles, and Algebraic Knots},''
	\newblock Commun. Math. Phys. \textbf{319} (2013) 813-863
	\newblock[arXiv:1111.6533].
	
	\bibitem{MarinoReview}
	M.~Mari\~{n}o,
	\newblock {``Chern-Simons Theory and Topological Strings},''
	\newblock Rev. Mod. Phys. \textbf{77} (2005) 675-720
	\newblock[arXiv:hep-th/0406005].
	
    \bibitem{D3D7us}
     O.~Aharony, A.~Fayyazuddin and J.~M.~Maldacena,
    \newblock {``The Large $N$ Limit of $\mathcal{N}=2$, $\mathcal{N}=1$ Field Theories from Three-branes in F Theory},''
    \newblock JHEP {\bf 9807} (1998) 013
    \newblock [arXiv:hep-th/9806159].

	\bibitem{D3D7}
	A.~Karch and E.~Katz,
	\newblock {``Adding Flavor to AdS/CFT},''
	\newblock JHEP \textbf{0206} (2002) 043
	\newblock [arXiv:hep-th/0205236].
	
	\bibitem{ClossetDumitrescuFestucciaKomargodski1} 
	C.~Closset, T.~T.~Dumitrescu, G.~Festuccia and Z.~Komargodski,
	\newblock {``Supersymmetric Field Theories on Three-Manifolds},''
	\newblock JHEP \textbf{1305} (2013) 017
	\newblock [arXiv:1212.3388].
	
	\bibitem{GeometrySUSYPartition}
	C.~Closset, T.~T.~Dumitrescu, G.~Festuccia and Z.~Komargodski,
	\newblock {``The Geometry of Supersymmetric Partition Functions},''
	\newblock JHEP \textbf{1401} (2014) 124
	\newblock [arXiv:1309.5876].
	
	\bibitem{ClossetDumitrescuFestucciaKomargodski2}
	C.~Closset, T.~T.~Dumitrescu, G.~Festuccia and Z.~Komargodski,
	\newblock {``From Rigid Supersymmetry to Twisted Holomorphic Theories},''
	\newblock Phys. Rev. \textbf{D90} (2014) no. 8, 085006
	\newblock [arXiv:1407.2598].
	
	\bibitem{DumitrescuReview}
	T.~T.~Dumitrescu,
	\newblock {``An Introduction to Supersymmetric Field Theories in Curved Space},''
	\newblock J. Phys. \textbf{A50} (2017) no. 44, 443005
	\newblock [arXiv:1608.02957].
	
	\bibitem{NeitzkeVafa}
	A.~Neitzke and C.~Vafa,
	\newblock {``Topological Strings and Their Physical Applications},''
	\newblock (2004) 77 pp.
	\newblock [hep-th/0410178].
	
	\bibitem{Vonk}
	M.~Vonk,
	\newblock {``A Mini-course on Topological Strings},''
	\newblock (2005) 120 pp.
	\newblock [hep-th/0504147].
	
	\bibitem{KlemmBook}
	A.~Klemm,
	\newblock {``Introduction in Topological String Theory on Calabi-Yau Manifolds},''
	\newblock 126 pp.
	
	\bibitem{WittenTopologicalSigma}
	E.~Witten,
	\newblock {``Topological Sigma Models},''
	\newblock Commun. Math. Phys. \textbf{118} (1988) 411.
	
	\bibitem{WittenTopologicalGravity}
	E.~Witten,
	\newblock {``On the Structure of the Topological Phase of Two-dimensional Gravity},''
	\newblock Nucl. Phys. \textbf{B340} (1990) 281-332.
	
	\bibitem{WittenMirrorManifolds}
	E.~Witten,
	\newblock {``Mirror Manifolds and Topological Field Theory},''
	\newblock in {\it Mirror Symmetry I}, S. T. Yau, ed. (American Mathematical Society, 1998) 121-160
	\newblock [arXiv:hep-th/9112056].
	
	\bibitem{Pestun}
	V.~Pestun,
	\newblock {``Localization of Gauge Theory on a Four-Sphere and Supersymmetric Wilson Loops},''
	\newblock Commun. Math. Phys. \textbf{313} (2012) 71-129
	\newblock [arXiv:0712.2824].
	
	\bibitem{BigMirrorBook}
	K.~Hori, S.~Katz, A.~Klemm, R.~Pandharipande, R.~Thomas, C.~Vafa, R.~Vakil and E.~Zaslow,
	\newblock {``Mirror Symmetry},''
	\newblock American Mathematical Society,
	\newblock 2006,
	\newblock 929 pp.
	
	\bibitem{TopologicalGravityReview}
	R.~Dijkgraaf, H.~Verlinde and E.~Verlinde,
	\newblock {``Notes on Topological String Theory and $2D$ Quantum Gravity},''
	\newblock World Sci. Publ. (1991) 91-156.
	
	\bibitem{BCOVLong}
	M.~Bershadsky, S.~Cecotti, H.~Ooguri and C.~Vafa,
	\newblock {``Kodaira-Spencer Theory of Gravity and Exact Results for Quantum String Amplitudes},''
	\newblock Commun. Math. Phys. \textbf{165} (1994) 311-428
	\newblock [arXiv:hep-th/9309140].
	
	\bibitem{AspinwallMorrison}
	P.~S.~Aspinwall and D.~R.~Morrison,
	\newblock {``Topological Field Theory And Rational Curves},''
	\newblock Commun. Math. Phys. \textbf{151} (1993) 245-262 
	\newblock [arXiv:hep-th/9110048].
	
	\bibitem{OoguriOzYin}
	H.~Ooguri, Y.~Oz and Z.~Yin,
	\newblock {``D-branes on Calabi-Yau Spaces and Their Mirrors},''
	\newblock Nucl. Phys. \textbf{B477} (1996) 407-430
	\newblock [arXiv:hep-th/9606112].
	
	\bibitem{KapustinOrlovPaper}
	A.~Kapustin and D.~Orlov,
	\newblock {``Remarks on A branes, Mirror Symmetry, and the Fukaya Category},''
	\newblock J. Geom. Phys. \textbf{48} (2003) 84
	\newblock [arXiv:hep-th/0109098].
	
	\bibitem{KapustinOrlovLectures}
	A.~Kapustin and D.~Orlov,
	\newblock {``Lectures on Mirror Symmetry, Derived Categories, and D-branes},''
	\newblock Russ. Math. Surveys \textbf{59} (2004) 907
	\newblock [arXiv:math/0308173].
	
	\bibitem{N=2}
	U.~Lindstrom and M.~Zabzine,
	\newblock {``$N=2$ Boundary Conditions for Non-linear Sigma Models and Landau-Ginzburg Models},''
	\newblock JHEP \textbf{0302} (2003) 006
	\newblock [arXiv:hep-th/0209098].
	
	\bibitem{Herbst}
	M.~Herbst,
	\newblock {``On Higher Rank Coisotropic A-branes},''
	\newblock J. Geom. Phys. \textbf{62} (2012) 156-169
	\newblock [arXiv:1003.3771].
	
	\bibitem{DuchampKalka}
	T.~Duchamp and M.~Kalka,
	\newblock {``Deformation Theory for Holomorphic Foliations},''
	\newblock J. Differential Geometry \textbf{14} 317 (1979).
	
	\bibitem{GomezMont}
	X.~Gomez-Mont,
	\newblock {``Transversal Holomorphic Structures},''
	\newblock J. Differential Geometry \textbf{15} 161 (1980).
	
	\bibitem{GirbauHaefligerSundararaman}
	J.~Girbau, A.~Haefliger and D.~Sundararaman,
	\newblock {``On Deformations of Transversely Holomorphic Foliations},''
	\newblock Journal f\"{u}r die reine und angewandte Mathematik \textbf{345} 122 (1983).
	
	\bibitem{BrunellaGhys}
	M.~Brunella and E.~Ghys,
	\newblock {``Umbilical Foliations and Transversely Holomorphic Flows},''
	\newblock J. Differential Geom. \textbf{41} (1995) 1-19.
	
	\bibitem{Brunella}
	M.~Brunella,
	\newblock {``On Transversely Holomorphic Flows I},''
	\newblock Invent. Math. \textbf{126} (1996) 265-279.
	
	\bibitem{Ghys}
	E.~Ghys,
	\newblock {``On Transversely Holomorphic Flows II},''
	\newblock Invent. Math. \textbf{126} (1996) 281-286.
	
	\bibitem{OpenStringBRST}
	A.~Kapustin and Y.~Li,
	\newblock {``Open String BRST Cohomology for Generalized Complex Branes},''
	\newblock Adv. Theor. Math. Phys. \textbf{9} (2005) no.4, 559-574
	\newblock [arXiv:hep-th/0501071].
	
	\bibitem{Kontsevich}
	M.~Kontsevich,
	\newblock {``Enumeration of Rational Curves via Torus Actions},''
	\newblock [arXiv:hep-th/9405035].
	
	\bibitem{GraberPandharipande}
	T.~Graber and R.~Pandharipande, 
	\newblock {``Localization of Virtual Classes},''
	\newblock Invent. Math. \textbf{135} (1999) 487-518
	\newblock [arXiv:alg-geom/9708001].
	
	\bibitem{KatzLiu}
	S.~Katz and M.~Liu,
	\newblock {``Enumerative Geometry of Stable Maps with Lagrangian Boundary Conditions and Multiple Covers of the Disk},''
	\newblock Adv. Theor. Math. Phys. \textbf{5} (2002) 1-49 and Geom. Topol. Monogr. \textbf{8} (2006) 1-47
	\newblock [arXiv:math/0103074].
	
	\bibitem{LiSong}
	Y.~Li and J.~Song,
	\newblock {``Open String Instantons and Relative Stable Morphisms},''
	\newblock Adv. Theor. Math. Phys. \textbf{5} (2002) 67-91
	\newblock [arXiv:hep-th/0103100].
	
	\bibitem{GraberZaslow}
	T.~Graber and E.~Zaslow,
	\newblock {``Open-String Gromov-Witten Invariants: Calculations and a Mirror `Theorem'},''
	\newblock [arXiv:hep-th/0109075].
	
	\bibitem{Saulina}
	N.~Saulina,
	\newblock {``Coisotropic Branes in Toric Calabi-Yau 3-folds},''
	\newblock [arXiv:1410.2340].
	
	\bibitem{AganagicVafa}
	M.~Aganagic and C.~Vafa,
	\newblock {``Mirror Symmetry, D-branes and Counting Holomorphic Discs},''
	\newblock [arXiv:hep-th/0012041].
	
	\bibitem{AganagicKlemmVafa}
	M.~Aganagic, A.~Klemm and C.~Vafa,
	\newblock {``Disk Instantons, Mirror Symmetry and the Duality Web},''
	\newblock Z. Naturforsch. \textbf{A57} (2002) 1-28
	\newblock [arXiv:hep-th/0105045].
	
	\bibitem{TorusKnots} 
	A.~Brini, B.~Eynard and M.~Mari\~{n}o,
	\newblock {``Torus Knots and Mirror Symmetry},''
	\newblock Annales Henri Poincare {\bf 13} (2012) 1873
	\newblock [arXiv:1105.2012].
	
	\bibitem{TopStringsMTheory1}
	R.~Gopakumar and C.~Vafa,
	\newblock {``M-theory and Topological Strings I},''
	\newblock [arXiv:hep-th/9809187].
	
	\bibitem{TopStringsMTheory2}
	R.~Gopakumar and C.~Vafa,
	\newblock {``M-theory and Topological Strings II},''
	\newblock [arXiv:hep-th/9812127].
	
	\bibitem{DiaconescuFloreaGrassi1}
	D.-E.~Diaconescu, B.~Florea and A.~Grassi,
	\newblock {``Geometric Transitions and Open String Instantons},''
	\newblock Adv. Theor. Math. Phys. \textbf{6} (2003) 619-642
	\newblock [hep-th/0205234].
	
	\bibitem{DiaconescuFloreaGrassi2}
	D.-E.~Diaconescu, B.~Florea and A.~Grassi,
	\newblock {``Geometric Transitions, del Pezzo Surfaces and Open String Instantons},''
	\newblock Adv. Theor. Math. Phys. \textbf{6} (2003) 643-702
	\newblock [hep-th/0206163].
	
	\bibitem{FangLiu}
	B.~Fang and C.-C.~M.~Liu,
	\newblock {``Open Gromov-Witten Invariants of Toric Calabi-Yau 3-Folds},''
	\newblock Commun. Math. Phys. \textbf{323} (2013) 285-328
	\newblock [arXiv:1103.0693].
	
	\bibitem{Faber}
	C.~Faber,
	\newblock {``Algorithm for Computing Intersection Numbers on Moduli Spaces of Curves, with an Application to the Class of the Locus of the Jacobians},''
	\newblock [arXiv:math.AG/9706006].
	
	\bibitem{THFSphere}
	H.~Geiges and J.~G.~P\'{o}rez,
	\newblock {``Transversely Holomorphic Flows and Contact Circles
	on Spherical 3-manifolds},''
	\newblock L’Enseignement Mathématique (2) \textbf{62} (2016), 527–567
	\newblock [arXiv:1510.08670].
		
	\bibitem{KapustinWillettYaakov} 
	A.~Kapustin, B.~Willett and I.~Yaakov,
	\newblock {``Exact Results for Wilson Loops in Superconformal Chern-Simons Theories with Matter,}''
	\newblock JHEP {\bf 1003} 089 (2010)
	\newblock [arXiv:0909.4559].
	
	\bibitem{HamaHosomichiLee} 
	N.~Hama, K.~Hosomichi and S.~Lee,
	\newblock {``SUSY Gauge Theories on Squashed Three-Spheres,}''
	\newblock JHEP {\bf 1105} 014 (2011)
	\newblock [arXiv:1102.4716].
	
	\bibitem{NishiokaYonekura} 
	T.~Nishioka and K.~Yonekura,
	\newblock {``On RG Flow of $\tau_{RR}$ for Supersymmetric Field Theories in Three-Dimensions,}''
	\newblock JHEP {\bf 1305} 165 (2013)
	\newblock [arXiv:1303.1522].
		
	\bibitem{Tanaka} 
	A.~Tanaka,
	\newblock {``Comments on Knotted 1/2 BPS Wilson Loops,}''
	\newblock JHEP {\bf 1207} 097 (2012)
	\newblock [arXiv:1204.5975].
	
	\bibitem{LewkowyczMaldacena} 
	A.~Lewkowycz and J.~Maldacena,
	\newblock {``Exact Results for the Entanglement Entropy and the Energy Radiated by a Quark,}''
	\newblock JHEP {\bf 1405} 025 (2014)
	\newblock [arXiv:1312.5682].
	
	\bibitem{WittenNewLook}
	E.~Witten,
	\newblock {``A New Look at the Path Integral of Quantum Mechanics,}''
	\newblock [arXiv:1009.6032].
		
	\bibitem{LensSpaces1}
	N.~Halmagyi, T.~Okuda and V.~Yasnov,
	\newblock {``Large N Duality, Lens Spaces and the Chern-Simons Matrix Model},''
	\newblock JHEP \textbf{0404} (2004) 014
	\newblock [arXiv:hep-th/0312145].
	
	\bibitem{LensSpaces2}
	A.~Brini, L.~Griguolo, D.~Seminara and A.~Tanzini,
	\newblock {``Chern-Simons Theory on $L(p,q)$ Lens Spaces and Gopakumar-Vafa Duality},''
	\newblock J. Geom. Phys. \textbf{60} (2010) 417-429
	\newblock [arXiv:0809.1610].
	
	\bibitem{HoriVafa}
	K.~Hori and C.~Vafa,
	\newblock {``Miror Symmetry},''
	\newblock [arXiv:hep-th/0002222].
	
	\bibitem{CoisotropicNonCommutative}
	M.~Aldi and E.~Zaslow,
	\newblock {``Coisotropic Branes, Noncommutativity, and the Mirror Correspondence},''
	\newblock JHEP \textbf{0506} (2005) 019
	\newblock [arXiv:hep-th/0501247].
	
	\bibitem{TheTopologicalVertex}
	M.~Aganagic, A.~Klemm, M.~Marino and C.~Vafa,
	\newblock {``The Topological Vertex},''
	\newblock Commun. Math. Phys. \textbf{254} (2005) 425-478
	\newblock [arXiv:hep-th/0305132].
	
	\bibitem{VafaSuperstrings}
	C.~Vafa,
	\newblock {``Superstrings and Topological Strings at Large $N$},''
	\newblock J. Math. Phys. \textbf{42} (2001) 2798-2817.
	\newblock [arXiv:hep-th/0008142].
	
	\bibitem{Acharya}
	B.~S.~Acharya,
	\newblock {``On Realizing ${\cal N}=1$ Super Yang-Mills in M theory},''
	\newblock [arXiv:hep-th/0011089].
	
	\bibitem{AtiyahMaldacenaVafa}
	M.~Atiyah, J.~Maldacena and C.~Vafa,
	\newblock {``An M-theory Flop as a Large $N$ Duality},''
	\newblock J. Math. Phys. \textbf{42} (2001) 3209
	\newblock [arXiv:hep-th/0011256].
	
	\bibitem{DijkgraafVafaVerlinde}
	R.~Dijkgraaf, C.~Vafa and E.~Verlinde,
	\newblock {``M-theory and a Topological String Duality},''
	\newblock [arXiv:hep-th/0602087].
	
	\bibitem{AganagicOoguriVafaYamazaki}
	M.~Aganagic, H.~Ooguri, C.~Vafa and M.~Yamazaki,
	\newblock {``Wall Crossing and M-theory},''
	\newblock Publ. Res. Inst. Math. Sci. Kyoto \textbf{47} (2011) 569
	\newblock [arXiv:0908.1194].
	
	\bibitem{DedushWitten}
	M.~Dedushenko and E.~Witten,
	\newblock {``Some Details on the Gopakumar-Vafa and Ooguri-Vafa Formulas},''
	\newblock Adv. Theor. Math. Phys. \textbf{20} (2016) 1-133
	\newblock [arXiv:1411.7108].
	
	\bibitem{FradkinVasiliev1}
	E.~Fradkin and M.~Vasiliev,
	\newblock {``On the Gravitational Interaction of Massless Higher Spin Fields},''
	\newblock Phys. Lett. \textbf{B189} (1987) 89.
	
	\bibitem{FradkinVasiliev2}
	E.~Fradkin and M.~Vasiliev,
	\newblock {``Cubic Interaction in Extended Theories of Massless Higher Spin Fields},''
	\newblock Nucl. Phys. \textbf{B291} (1987) 141.
	
	\bibitem{KP}
	I.~Klebanov and A.~Polyakov,
	\newblock {``AdS Dual of the Critical $O(N)$ Vector Model},''
	\newblock Phys. Lett. \textbf{B550} (2002) 213-219
	\newblock [arXiv:hep-th/0210114].
	
	\bibitem{ABJTriality}
	C.-M.~Chang, S.~Minwalla, T.~Sharma and X.~Yin,
	\newblock {``ABJ Triality: from Higher Spin Fields to Strings},''
	\newblock J. Phys. \textbf{A46} (2013) 214009
	\newblock [arXiv:1207.4485].
	
	\bibitem{ABJQuadrality}
	M.~Honda, Y.~Pang and Y.~Zhu,
	\newblock {``ABJ Quadrality},''
	\newblock JHEP \textbf{1711} (2017) 190
	\newblock [arXiv:1708.08472].
	
	\bibitem{Yangian}
	K.~Costello,
	\newblock {``Supersymmetric Gauge Theory and the Yangian},''
	\newblock [arXiv:1303.2632].
	
	\bibitem{TwistedSupergravityAndItsQuantization}
	K.~Costello and S.~Li,
	\newblock {``Twisted Supergravity and its Quantization},''
	\newblock [arXiv:1606.00365].
	
	\bibitem{CostelloMTheoryOmegaBackground}
	K.~Costello,
	\newblock {``M-theory in the $\Omega$-background and 5-dimensional Non-commutative Gauge Theory},''
	\newblock [arXiv:1610.04144].
	
	\bibitem{HolographyAndKoszulDuality}
	K.~Costello,
	\newblock {``Holography and Koszul Duality: the Example of the M2 Brane},''
	\newblock [arXiv:1705.02500].
	
	\bibitem{CostelloGaiotto}
	K.~Costello and D.~Gaiotto,
	\newblock {``Twisted Holography},''
	\newblock [arXiv:1812.09257].
	
	\bibitem{GaroufalidisKashaev} 
	S.~Garoufalidis and R.~Kashaev,
	\newblock {``Evaluation of State Integrals at Rational Points,}''
	\newblock Commun. Num. Theor. Phys. {\bf 09} no.3, 549 (2015)
	\newblock [arXiv:1411.6062].
	
\end{thebibliography}
\end{document}